\documentclass[reprint,
superscriptaddress,
nofootinbib,
 amsmath,amssymb,
prd,
floatfix,
]{revtex4-2}

\usepackage{mathtools} 
\usepackage{graphicx}
\usepackage{dcolumn}
\usepackage{bm}
\usepackage{hyperref}
\usepackage[x11names]{xcolor}
\hypersetup{
    colorlinks,
    linkcolor={Red4},
    citecolor={Blue3},
    urlcolor={Blue3}
}

\usepackage{xspace}
\usepackage[utf8]{inputenc}
\usepackage{array}   
\newcolumntype{C}{>{$}c<{$}} 
\usepackage[caption=false]{subfig}

\newcommand{\bi}{\begin{itemize}}
\newcommand{\ei}{\end{itemize}}

\newcommand{\ben}{\begin{enumerate}}
\newcommand{\een}{\end{enumerate}} 

\newcommand{\bt}[1]{\begin{table}[tb]\begin{tabular}{#1} \hline\hline  \\[-1.0em]}
\newcommand{\et}[2]{\hline\hline \end{tabular} \caption{#1} \label{#2} \end{table}}

\newcommand{\be}{\begin{equation}}
\newcommand{\ee}{\end{equation}}

\newcommand{\bea}{\begin{eqnarray}}
\newcommand{\eea}{\end{eqnarray}}

\newcommand{\bc}{}

\newcommand{\epem}{\ensuremath{e^+e^-}\xspace}

\newcommand{\pp}{\ensuremath{\pi^+\pi^-}\xspace}

\newcommand{\mev}{\ensuremath{\mathrm{\,Me\kern -0.1em V}}\xspace}
\newcommand{\gev}{\ensuremath{\mathrm{\,Ge\kern -0.1em V}}\xspace}

\newcommand{\mZ}{\ensuremath{M_{Z}}\xspace}

\newcommand{\DaHadZfive}{\ensuremath{\Delta\alpha^{(5)}_{\rm had}(\mZ^2)}\xspace}

\newcommand{\FigChiSqpValue}[7]{
\begin{figure*}[tb]
\centering

\resizebox{0.48\textwidth}{!}{\includegraphics{Sections/Plots/Results_#4/Distrib_chi2_C__LatticeCovarianceMatrix_#7__momentsFit_#5__LowHighMass_RhoOutside_Fit_#2}}
\resizebox{0.48\textwidth}{!}{\includegraphics{Sections/Plots/Results_#4/Distrib_pValue_C__LatticeCovarianceMatrix_#7__momentsFit_#5__LowHighMass_RhoOutside_Fit_#2}} \\

\resizebox{0.48\textwidth}{!}{\includegraphics{Sections/Plots/Results_#4/Distrib_chi2_SW__LatticeCovarianceMatrix_#7__momentsFit_#5__LowHighMass_RhoOutside_Fit_#2}}
\resizebox{0.48\textwidth}{!}{\includegraphics{Sections/Plots/Results_#4/Distrib_pValue_SW__LatticeCovarianceMatrix_#7__momentsFit_#5__LowHighMass_RhoOutside_Fit_#2}} \\

\resizebox{0.48\textwidth}{!}{\includegraphics{Sections/Plots/Results_#4/Distrib_chi2_NoNormalisationFit_C__LatticeCovarianceMatrix_#7__momentsFit_#5__LowHighMass_RhoOutside_Fit_#2}}
\resizebox{0.48\textwidth}{!}{\includegraphics{Sections/Plots/Results_#4/Distrib_pValue_NoNormalisationFit_C__LatticeCovarianceMatrix_#7__momentsFit_#5__LowHighMass_RhoOutside_Fit_#2}}
\caption{Distributions of $\chi^{2}$~(left) and $p$-value~(right), obtained by sampling the bootstrap replicas of the lattice covariance matrix. These figures show the effects of different approaches to determining $\gamma_1$ and its uncertainty {#1-mass region} {#3}. {#6 moment integrals} are considered here, with the {lattice covariance matrix ``#7''}. In the first row, we consider normalization fits using the full covariance matrices for determining the averaging weights--their comparison with the two plots in the following row, is discussed in \app{Appendix:alternate_averaging}. In the middle row, we consider normalization fits with the averaging weights proportional to the inverse of the $\tilde{\gamma}_j$ uncertainties squared, as considered in the present section. In the final row, we set the $\gamma_b$ normalization coefficients to unity, i.e. we perform no rescaling--these are discussed in \sec{sec:testing_lattice_results}. The blue horizontal error bar indicates the asymmetric variance of the distribution, computed on each side of the nominal value~(blue point and blue dashed vertical line). The median~(dashed-dotted black line), mean~(black continuous line), $68.3\%$ and $95.4\%$ quantiles~(coloured bands) of the distribution are also indicated. The number of degrees of freedom in the $\chi^{2}$ calculation is indicated by the dotted pink line. Note that the scales in the last row of plots (corresponding to no rescaling) are very different from those of the two previous rows (in which rescaling is allowed) to account for the fact that the $\chi^2$ and $p$-values with no rescaling are very poor. }

\label{Fig:ChiSqpValue_#2_#5_#7}
\end{figure*}
}
\newcommand{\FigAverageSigma}[7]{
\begin{figure*}[tb]
\centering

\resizebox{0.48\textwidth}{!}{\includegraphics{Sections/Plots/Results_#4/Distrib_average_C__LatticeCovarianceMatrix_#7__momentsFit_#5__LowHighMass_RhoOutside_Fit_#2}}
\resizebox{0.48\textwidth}{!}{\includegraphics{Sections/Plots/Results_#4/Distrib_sigma_C__LatticeCovarianceMatrix_#7__momentsFit_#5__LowHighMass_RhoOutside_Fit_#2}} \\

\resizebox{0.48\textwidth}{!}{\includegraphics{Sections/Plots/Results_#4/Distrib_average_SW__LatticeCovarianceMatrix_#7__momentsFit_#5__LowHighMass_RhoOutside_Fit_#2}}
\resizebox{0.48\textwidth}{!}{\includegraphics{Sections/Plots/Results_#4/Distrib_sigma_SW__LatticeCovarianceMatrix_#7__momentsFit_#5__LowHighMass_RhoOutside_Fit_#2}}

\caption{Distributions of the normalization rescaling factor $\gamma_1$~(left) and its uncertainty propagated from the covariance matrices of the lattice QCD and dispersive results~(right), obtained by sampling the bootstrap replicas of the lattice covariance matrix. These figures show the effects of different approaches to determining $\gamma_1$ and its uncertainty {#1-mass region} {#3}. {#6 moment integrals} are considered here, with the {lattice covariance matrix ``#7''}. In the first row, we consider normalization fits using the full covariance matrices of the $\tilde\gamma_j$ for determining the averaging weights. In the second row, we consider normalization fits with the averaging weights proportional to the inverse of the $\tilde{\gamma}_j$ uncertainties squared.  The blue horizontal line indicates the asymmetric variance of the distribution, computed on each side of the nominal value~(dashed blue vertical line). The median~(dashed-dotted black line), mean~(black continuous line), $68.3\%$ and $95.4\%$ quantiles~(coloured bands) of the distribution are also indicated.}

\label{Fig:AverageSigma_#2_#5_#7}
\end{figure*}
}

\def\nn{\nonumber}
\def\be{\begin{equation}}
\def\ee{\end{equation}}
\def\bea{\begin{eqnarray}}
\def\eea{\end{eqnarray}}

\def\amulohvp{a_\mu^\text{LO-HVP}}
\def\amulohvpwin{a_{\mu,\text{win}}^\text{LO-HVP}}

\def\amupipi{a_\mu^{\pi\pi}}

\def\ctogi{C_{1\gamma I}(t)}
\def\amuwin{a_{\mu,\text{win}}^\text{LO-HVP}}
\def\deltaalphahad{\Delta_\text{had}^{(5)}\alpha}

\def\alat#1{a_{#1}^\text{lat}}
\def\aR#1{a_{#1}^\text{R}}

\def\ndof{\text{\it ndof}}

\def\DaHadQsqFive#1{\ensuremath{\Delta\alpha^{(5)}_{\rm had}(#1)}\xspace}

\def\reff#1{\ref{#1}}

\def\app#1{Appendix~\reff{#1}}
\def\apps#1#2{Appendices~\reff{#1} and \reff{#2}}
\def\eq#1{Eq.~(\reff{#1})}
\def\eqs#1#2{Eqs.~(\reff{#1}) and (\reff{#2})}

\def\eqsss#1#2#3#4{Eqs.~(\reff{#1}), (\reff{#2}), (\reff{#3}) and (\reff{#4})}

\def\eqsint#1#2{Eqs.~(\ref{#1})-(\ref{#2})}

\def\fig#1{Fig.~\reff{#1}}
\def\figs#1#2{Figs.~\reff{#1} and \reff{#2}}

\def\sec#1{Sec.~\reff{#1}}
\def\tab#1{Table~\reff{#1}}

\def\tabss#1#2#3{Tables~\reff{#1}, \reff{#2} and \reff{#3}}
\def\sec#1{Sec.~\reff{#1}}

\def\nn{\nonumber}

\def\Tr{\,\mathrm{Tr}}
\def\mev{\mathrm{Me\kern-0.1em V}}
\def\gev{\mathrm{Ge\kern-0.1em V}}
\def\tev{\mathrm{Te\kern-0.1em V}}
\def\fm{\mathrm{fm}}

\newcommand{\lsim}{ {\
\lower-1.2pt\vbox{\hbox{\rlap{$<$}\lower5pt\vbox{\hbox{$\sim$}}}}\ } }
\newcommand{\gsim}{ {\
\lower-1.2pt\vbox{\hbox{\rlap{$>$}\lower5pt\vbox{\hbox{$\sim$}}}}\ } }

\makeatletter
\DeclareRobustCommand{\text}{%
  \ifmmode\expandafter\text@\else\expandafter\mbox\fi}
\let\nfss@text\text
\def\text@#1{{\mathchoice
  {\textdef@\displaystyle\f@size{#1}}%
  {\textdef@\textstyle\f@size{#1}}%
  {\textdef@\textstyle\sf@size{#1}}%
  {\textdef@\textstyle \ssf@size{#1}}%
  \check@mathfonts
  }%
}
\def\textdef@#1#2#3{\hbox{{%
                    \everymath{#1}%
                    \let\f@size#2\selectfont
                    #3}}}
\makeatother

\def\epemtohad{e^+e^-\to\mbox{hadrons}}
\def\ddalpha{\delta(\deltaalphahad)}
\def\epemtopippimg{e^+e^-\to\pi^+\pi^-(\gamma)}

\newcommand{\ijclab}{\affiliation{IJCLab, Université Paris-Saclay et CNRS/IN2P3, Orsay, France}}
\newcommand{\lpnhe}{\affiliation{LPNHE, Sorbonne Université, Université Paris Cité, CNRS/IN2P3, Paris, 75252, France}}
\newcommand{\marseille}{\affiliation{Aix Marseille Univ, Université de Toulon, CNRS, CPT, IPhU, Marseille, France}}
\newcommand{\wuppertal}{\affiliation{Department of Physics, University of Wuppertal, D-42119 Wuppertal, Germany}}
\newcommand{\juelich}{\affiliation{Jülich Supercomputing Centre, Forschungszentrum Jülich, D-52428 Jülich, Germany}}
\newcommand{\budapest}{\affiliation{Institute for Theoretical Physics, Eötvös University, H-1117 Budapest, Hungary}}
\newcommand{\penn}{\affiliation{Department of Physics, Pennsylvania State University, University Park, PA, USA}}

\begin{document}


\title{Hadronic vacuum polarization: comparing lattice QCD and data-driven results in systematically improvable ways}

\author{Michel Davier}
\ijclab
\author{Zoltán Fodor}
\penn
\wuppertal
\budapest
\juelich
\author{Antoine Gérardin}
\marseille
\author{Laurent Lellouch}
 \email{laurent.lellouch@cnrs.fr}
 \homepage{https://www.cpt.univ-mrs.fr/$\sim$lellouch}
 \marseille
 \author{Bogdan Malaescu}
 \email{malaescu@in2p3.fr}
 \homepage{http://lpnhe.in2p3.fr/malaescu}
\lpnhe
\author{Finn M. Stokes}
\juelich
\author{Kálmán K. Szabó}
\wuppertal
\juelich
\author{Balint C. Toth}
\wuppertal
\author{Lukas Varnhorst}
\wuppertal
\author{Zhiqing Zhang}
\ijclab

\date{\today}

\begin{abstract}
The precision with which hadronic vacuum polarization (HVP) is obtained determines how accurately important observables, such as the muon anomalous magnetic moment, $a_\mu$, or the low-energy running of the electromagnetic coupling, $\alpha$, are predicted.
The two most precise approaches for determining HVP are: dispersive relations combined with $e^+e^-\to\text{hadrons}$ cross-section data, and lattice QCD.
However, the results obtained in these two approaches display significant tensions, whose origins are not understood. 
Here we present a framework that sheds light on this issue and---if the two approaches can be reconciled---allows them to be combined. 
Via this framework, we test the hypothesis that the tensions can be explained by modifying the R-ratio in different intervals of center-of-mass energy $\sqrt{s}$.
As ingredients, we consider observables that have been precisely determined in both approaches. These are the leading hadronic contributions to $a_\mu$, to the so-called intermediate window observable and to the running of $\alpha$ between spacelike virtualities $1\,\gev^2$ and $10\,\gev^2$ (for which only a preliminary lattice result exists).
Our tests take into account all uncertainties and correlations, as well as uncertainties on uncertainties in the lattice results.
Among our findings, the most striking is that results obtained in the two approaches can be made to agree for all three observables by modifying the $\rho$ peak in the experimental spectrum.
In particular, we find that this requires a common $\sim 5\%$ increase in the contributions of the peak to each of the three observables.
This finding is robust against the presence or absence of one of the constraining observables.
However, such an increase is much larger than the uncertainties on the measured R-ratio.
We also discuss a variety of generalizations of the methods used here, as well as the limits in the information that can be extracted from the R-ratio via a finite set of observables.

\end{abstract}

\maketitle


\section{Introduction}

A virtual, propagating photon polarizes the vacuum into quarks and gluons. Known as hadronic vacuum polarization (HVP), this effect is important when processes involving electromagnetism at the quantum level are studied with high precision. 
For a photon of small virtuality, predicting this effect requires being able to describe the strong interaction in its nonperturbative regime. 
At present, there are two approaches for making precise predictions of this polarization. 
These are based, on the one hand, on large-scale, numerical simulations in lattice quantum chromodynamics (QCD); and on the other, on the exploitation of $\epemtohad$ data.

HVP has been recently the center of much attention, because of its importance in the standard model prediction of the anomalous magnetic moment of the muon, $a_\mu$. This quantity is currently being measured at the Fermi National Accelerator Laboratory (FNAL)~\cite{Abi:2021gix} and was previously measured at the Brookhaven National Laboratory (BNL)~\cite{Bennett:2006fi}. 
While the data-driven determination of the leading-order (LO) HVP contribution to $a_\mu$ \cite{Davier:2019can,Keshavarzi:2019abf,Aoyama:2020ynm}, $\amulohvp$, yields a standard model prediction $4.2\sigma$ below the measurement of the total $a_\mu$ \cite{Abi:2021gix}, the most precise lattice calculation, of this contribution \cite{Borsanyi:2020mff}, reduces this difference to $1.5\sigma$, with comparable uncertainties \cite{Vbwgm2_LL_202105}.
As discussed below, and in Ref.~\cite{Borsanyi:2020mff}, this reduction in the tension between prediction and experiment comes at the expense of being $2.0\sigma$ above the data-driven prediction. 
Moreover, for the so-called intermediate-window contribution to $\amulohvp$~\cite{Blum:2018mom}, this excess rises to $3.8\sigma$ (\cite{Borsanyi:2020mff} and below), and even $4\sigma$ for a weighted average of independent lattice determinations of this quantity \cite{Borsanyi:2020mff,Ce:2022kxy,ExtendedTwistedMass:2022jpw,Blum:2023qou,Bazavov:2023has} (see \sec{sec:testing_lat}).

The above discussion does not take into account the measurement of the $\epemtopippimg$ cross section, from threshold to $1.2\,\gev$, reported by the CMD-3 collaboration in the preprint \cite{CMD-3:2023alj}.
This is because an explanation has not yet been found for the fact that the cross section which they obtain is significantly larger than all previous, modern measurements and, in particular, the one published by the same collaboration in Ref.~\cite{CMD-2:2006gxt}.

Another place where HVP plays an important role is in the scale-dependence of the electromagnetic coupling, $\alpha$. 
The fact that the lattice predicts a larger value for $\amulohvp$ has an impact on the value of $\alpha$ at the scale of the $Z$-boson mass, $M_Z$.
This has been investigated in Refs.~\cite{Crivellin:2020zul,Keshavarzi:2020bfy,deRafael:2020uif,Malaescu:2020zuc}. 
The overall conclusion is that the lattice excess in $a_\mu$, alone, does not imply a change in $\alpha(M_Z^2)$ large enough to have an impact on precision electroweak tests.
This is confirmed by direct lattice calculations of the running of $\alpha$ \cite{Borsanyi:2017zdw,Borsanyi:2020mff,Ce:2022eix}. 
In particular, Ref.~\cite{Borsanyi:2020mff} suggests that the faster running of $\alpha$ observed in lattice calculations is concentrated for spacelike values of momentum scale below a few $\gev^2$.

Beyond signaling tensions between the lattice and data-driven approaches for specific physical observables, the discrepancies discussed above also contain information about the agreement or disagreement between the primary quantities used to compute these HVP related observables in each of the two approaches. 
On the lattice this primary observable is the quark electromagnetic-current, two-point function with vanishing three-momentum, computed as a function of Euclidean time $t$. 
In the data-driven approach, it is instead the cross section for $e^+e^-$ annihilation into hadrons, measured as a function of center-of-mass (c.o.m.) energy $\sqrt{s}$, normalized by the tree-level cross section for $e^+e^-\to \mu^+\mu^-$ in the massless limit, i.e. the R-ratio $R(s)$.~\footnote{The determination of $R(s)$ actually mixes experimental measurements and perturbative QCD predictions, depending on the $\sqrt{s}$ range, as discussed in \app{sec:R-ratio_obs}. In the following it will generally be referred to as ``experimental R-ratio", and more specifically as ``measured R-ratio" for the $\sqrt{s}$ regions where measurements are being used.}

To help pinpoint the possible sources of the disagreement between the two approaches, and possibly correct them, one needs to be able to make sharp statements about the agreement or disagreement between the lattice correlation function and the experimental R-ratio in different regions of $t$ and $\sqrt{s}$. 
Because the current correlator is proportional to the Laplace transform of $\sqrt{s}R(s)$ \cite{Bernecker:2011gh}, it is straightforward to determine that correlator, and any observable that can be obtained from it, once $R(s)$ is measured. 
Thus, it is relatively simple to identify regions of $t$ in which the lattice and data-driven approaches disagree, possibly pointing to a problem with lattice computations at those length scales. 

Determining the R-ratio in specific intervals of $\sqrt{s}$ from a lattice computation of the current correlator requires performing an inverse Laplace transform. Such lattice results will have uncertainties and be obtained at a finite number of points, making this inverse a notoriously ill-posed problem. 

A number of interesting methods have been advocated for relating the two approaches. 
For instance, Ref.~\cite{Hansen:2019idp} proposes a modification of the Backus-Gilbert method that involves reconstructing, not the spectral function itself, but rather the spectral function convoluted, at individual values of $\sqrt{s}$, with a Gaussian whose narrowness is limited by the statistical and systematic uncertainties on the lattice results. 
A first, application of this method was performed in Ref.~\cite{Alexandrou:2022tyn}. 
In Ref.~\cite{Colangelo:2018mtw} the authors propose a physics- and data-constrained dispersive representation of the pion electromagnetic form factor. Then, in Ref.~\cite{Colangelo:2020lcg} they use it to study the consequences of a value of $\amulohvp$, such as the one obtained on the lattice in Ref.~\cite{Borsanyi:2020mff}, on important observables impacted by this form factor.
Another approach proposes to gain information about the R-ratio in different regions of $\sqrt{s}$, via well-chosen linear combinations of so-called time-window observables computed on the lattice \cite{Colangelo:2022vok}.
Yet another approach proposes to use spectral-width sumrules to constrain $R(s)$, in narrow regions of $\sqrt{s}$, using the lattice correlator~\cite{Boito:2022njs}.

More generally, a variety of methods have been developed for extracting real-time properties~(e.g.\ spectral functions) from Euclidean correlation functions computed in lattice QCD. 
These include approaches based on: the Backus-Gilbert method with various regularizations \cite{Brandt:2015sxa,Hansen:2017mnd,Bulava:2019kbi,Bonanno:2023ljc,Frezzotti:2023nun}; smeared spectral functions using Chebyshev polynomials~\cite{Bailas:2020qmv}; the maximum entropy method (MEM)~\cite{Nakahara:1999vy,Nakahara:1999bm,Asakawa:2000tr,Rothkopf:2011db,Rothkopf:2011ef,Asakawa:2020hjs,Rothkopf:2020qqt}; a Bayesian reconstruction approach (BR method)~\cite{Burnier:2013nla,Rothkopf:2016luz}; Bayesian GrHMC~(for retarded propagator Gr employing Hamiltonian Monte-Carlo techniques)~\cite{Cyrol:2018xeq}; sumrules or other parametrizations~\cite{Li:2020xrz,Li:2020fiz,Li:2020ejs,Xiong:2022uwj}, also combined with Bayesian approaches~\cite{Gubler:2010cf}; a modified lattice correlator which emphasizes different parts of the spectral function at large times~\cite{Bruno:2020kyl}; recent developments based on special properties of certain conformal maps \cite{Bergamaschi:2023xzx}; Gaussian processes~\cite{Horak:2021syv}; and machine learning~\cite{Offler:2019eij,Offler:2021fmg,Kades:2019wtd,laanait2019exascale,Chen:2021giw,Wang:2021jou,Wang:2021cqw,Shi:2022yqw,Lechien:2022ieg,Boyda:2022nmh,Buzzicotti:2023qdv}.~\footnote{Some of these methods have also been applied in the context of the reconstruction of parton distribution functions or of distribution amplitudes of hadrons, from lattice QCD~\cite{Chambers:2017dov,Karpie:2019eiq,Candido:2023nnb}.} 
Comparisons of various approaches, together with their regularization, have also been performed~\cite{Dudal:2013yva,Tripolt:2018xeo,Rothkopf:2022ctl,Rothkopf:2022fyo}.

Here we propose to begin more modestly in terms of R-ratio reconstruction. However, we do build a quantitative measure of comparison between the two approaches into our formalism and include it in our results.

To provide a relevant comparison between lattice and data-driven results, the lattice results used must have subpercent uncertainties.
At such precisions, leading-order strong-isospin-breaking and QED corrections are required.
Moreover, if one is interested in probing the spectral function above the $c\bar c$ threshold, one must include the contributions of at least the four lightest quark flavors.
For information above the $b\bar b$ threshold, a fifth quark flavour must be included.
Of course, all continuum and infinite-volume limits must be taken in a controlled fashion.
In addition, as the ranges of $\sqrt{s}$ that one is interested in reconstructing become narrower, the function with which the lattice results must be convoluted becomes more and more oscillatory. 
Thus, not only do the uncertainties have to be small, but also the statistical and systematic correlations between the lattice quantities have to be well known.

In the present paper we focus on the results published in Ref.~\cite{Borsanyi:2020mff}, which satisfy all of the above requirements.
These comprise $\amulohvp$ and the intermediate-time window observable, $\amulohvpwin$.
We also call upon the preliminary result for the hadronic contribution to the running of $\alpha$ between spacelike virtualities $1\,\gev^2$ and $10\,\gev^2$, $\ddalpha\equiv \deltaalphahad(-10\,\gev^2)-\deltaalphahad(-1\,\gev^2)$ \cite{Borsanyi:2020mff}, which provides interesting complementary information. 

Here we use a limited form of the more general approach we propose, focusing on different intervals of $\sqrt{s}$ in the $\epemtohad$ spectrum, such as the $\rho$-peak region, the low-mass region below that peak, different high-mass regions, etc. 
The idea is to test the extent to which the lattice results are consistent with a modification of the spectrum that leads to a common rescaling of the observables of interest in the chosen region, and to determine the required rescaling factor. 
The simplest allowed modification would consist of directly rescaling the experimental R-ratio within the chosen region by that same factor.
This would be a rather crude distortion of the experimental spectral function that can be viewed as a first approximation to a more physical modification.  
In fact, because the constraints that we consider pertain only to integrals of the R-ratio, there remains a significant amount of freedom in the shape of the corresponding modification.
Therefore, many other modifications are possible, some of which may actually be physical.
This point is discussed in more detail in \app{sec:general_model}.

Beyond complete and precise lattice results for the HVP quantities discussed above, our proposed reconstruction and comparison approach also requires state-of-the-art determinations of the contributions to those quantities from different $\sqrt{s}$-intervals in the data-driven approach, including a reliable quantification of correlations. 
For this we use the DHMZ methodology implemented in the HVPTools software~\cite{Davier:2010rnx,Davier:2010nc,Davier:2017zfy,Davier:2019can}.

Another feature of our analysis is that we include ``uncertainties on the uncertainties" for the lattice results. 
That is, we determine statistical and systematic uncertainties on the covariance matrix between the different lattice observables considered. 
This is important, because reconstruction methods are very sensitive to uncertainties and correlations in the input data, a corollary of the ill-posed nature of inverse methods. 
The same should eventually be done for the correlations between the corresponding data-driven observables. 
This is left for future work.

The remainder of the paper is organized as follows. \sec{sec:HVP_corr} briefly presents the lattice correlation function and how it can be obtained from the R-ratio. 
The observables that we use in the comparison of the lattice and data-driven approaches are defined in \sec{sec:observables}, including the window observables put forward in Ref.~\cite{Blum:2018mom}. 
In \sec{sec:comparison_methodology}, we introduce the comparison methodology for testing the lattice correlation function with data-driven results. 
More importantly, we present our method for determining the size of the modification of the experimental R-ratio, in a given $\sqrt{s}$-interval, that would be required to reconcile the lattice and data-driven values of the observables of interest.
\sec{Sec:Results} presents results of applications of these methods to the comparison of the two approaches.
This is followed by our conclusions in \sec{sec:conclusion}, in which we also summarize the main results of our study.

In addition, we provide a number of appendices. 
\app{sec:R-ratio_obs} presents the data-driven determination of the relevant observables and of their correlations and \app{sec:lat_obs}, a summary of the lattice determination of these quantities, including uncertainties on the covariance matrix.
In \app{Appendix:Chi2Methodology} we discuss, in more detail, the method used in this paper for determining the size of the possible modifications of the experimental R-ratio and its possible pitfalls.
\app{Appendix:alternate_averaging} is dedicated to showing the stability of this method with respect to the averaging procedure which it calls upon.
\app{app:beyond_rescaling} presents various generalizations of this method, which can provide increasingly refined reconstructions of the R-ratio required to accommodate lattice results, as more of these results become available and/or their precision increases.
Finally, in \app{app:more_moments}, some of the limitations on the number of observables that can be studied in the data-driven approach are discussed.

\section{The HVP correlator from the lattice and the R-ratio}
\label{sec:HVP_corr}

Recent lattice calculations of leading-order (LO) HVP effects are based on the following Euclidean-time $t$, quark, electromagnetic-current two-point function with vanishing three-momentum and averaged over spatial components~\cite{Bernecker:2011gh}:
\be
\label{eq:correlator}
C(t)=\frac1{3e^2}\sum_{i=1}^3\int d^3x\,\left\langle J_i(\vec{x},t)J_i(0)\right\rangle
\ .\ee
Here, $e>0$ is the unit of electric charge, $\frac{J_\mu}e=\frac23 \bar u\gamma_\mu u - \frac13 \bar d\gamma_\mu d -\frac13 \bar s\gamma_\mu s + \frac23 \bar c\gamma_\mu c - \frac13 \bar b\gamma_\mu b + \frac23 \bar t\gamma_\mu t$ and the angle brackets stand for the QCD + QED expectation value up to and including order $e^2$. For the quantities considered here, the top-quark contribution is either negligible or excluded by definition. However, all other flavors are required. Thus, the lattice results that we use \cite{Borsanyi:2020mff} are obtained including the first five quark flavors. 

The two-point function of \eq{eq:correlator} is not generically studied in the data-driven approach, because quantities of phenomenological interest are usually obtained directly from the R-ratio. The latter is defined in terms of the cross-section for $e^+e^-$ annihilation into hadrons, and is a function of the c.o.m.\ energy, $\sqrt{s}$:
\be
\label{eq:R-ratio}
R(s)\equiv\frac{\sigma(e^+e^-(s)\to\text{hadrons}(+\gamma))}{4\pi\alpha^2(s)/(3s)}
\ ,
\ee
where the denominator is the Born cross section for $e^+e^-\to \mu^+\mu^-$ in the massless limit.
In that denominator, it is the electromagnetic coupling at c.o.m.\ energy squared, $s$, that is used, to remove unwanted vacuum polarization effects in the numerator. 
In particular, in comparisons of $R(s)$ and $C(t)$, this means that it is the one-photon-irreducible part of the latter that is relevant. 
We denote it $\ctogi$. 
Moreover, the cross section in the numerator of \eq{eq:R-ratio} includes the final-state radiation (FSR) of photons. 
While these photons imply that effects of higher order in $\alpha$ are included, they are kept to obtain an infrared-safe cross section at next-to-leading order (NLO) in $\alpha$.
Moreover, in applications, the contributions of these photons can consistently be taken into account in higher-order calculations. 

From $R(s)$, it is straightforward to obtain $\ctogi$, after invoking the optical theorem that relates the $\epemtohad$ cross section to the spectral function associated with $\ctogi$. 
The result is proportional to the Laplace transform of $\sqrt{s}R(s)$~\cite{Bernecker:2011gh}:
\be
\label{eq:laplace}
\ctogi=\frac1{24\pi^2}\int_0^\infty ds\,\sqrt{s}R(s)\, e^{-|t|\sqrt{s}} 
\ .\ee
The integral diverges at short distances like $t^{-3}$, up to logarithms. However, in physical quantities, $\ctogi$ is multiplied by weights that vanish faster than $t^3$ as $t\to 0$.

\section{Observables of interest}
\label{sec:observables}

In principle, one could make a direct comparison of $\ctogi$ as a function of Euclidean time $t$, computed on the lattice and in the data-driven approach, via \eq{eq:laplace}. However, $\ctogi$ has not yet been computed on the lattice, as a function of $t$, in the continuum and infinite-volume limits at the physical mass point. Here, to compare the two approaches, we focus on derived quantities, computed on the lattice with high precision, that are of direct phenomenological interest or that contribute to such quantities. 

The LO-HVP contribution to the anomalous magnetic moment of the muon can be obtained from $\ctogi$ via~\cite{Bernecker:2011gh}:
\be
\label{eq:amulohvp}
\amulohvp=\alpha^2\int_0^\infty dt\,K(t)\ctogi
\ ,\ee
where the kernel is
\be
\label{eq:kernel}
K(t)=\int_0^\infty\frac{dQ^2}{m_\mu^2}\,\omega\left(\frac{Q^2}{m_\mu^2}\right)\left[t^2-\frac4{Q^2}\sin^2\left(\frac{t\sqrt{Q^2}}2\right)\right]
\ ,\ee
$\omega(r)=[r+2-\sqrt{r(r+4)}]^2/\sqrt{r(r+4)}$ and $\alpha$ is the fine-structure constant in the Thomson limit. 
In fact, it is a discretized version of \eq{eq:amulohvp} that is the basis of the lattice calculation of $a_\mu^\text{LO-HVP}$.

When the data-driven approach is used, $a_\mu^\text{LO-HVP}$ is computed directly from $R(s)$ as an integral over $s$~(see \app{sec:R-ratio_obs}). 
A comparison with lattice results gives a first means of confronting the two approaches. 
However, in the case of a disagreement, it is important to understand more precisely how different Euclidean times, $t$, in $\ctogi$ computed on the lattice, contribute to this tension. Indeed, different length scales on the lattice are subject to different statistical and systematic uncertainties, and knowing which ones agree or disagree with the data-driven approach may help point to aspects of the lattice calculations that require more attention. 

To shed light on this issue, particularly useful quantities are the so-called Euclidean-time window contributions to $\amulohvp$ \cite{Blum:2018mom,Colangelo:2022vok}. These are restrictions, of the integral in \eq{eq:amulohvp}, to time-intervals of interest, using a smoothed-out Heaviside function that limits edge effects when dealing with discrete times, as is the case on the lattice. This function is taken to be $\theta(t;\Delta) \equiv \frac12\left[1+\tanh\left(t/\Delta\right)\right]$, 
where $\Delta>0$ determines the time-range over which the function transitions from 0 to 1. The desired, restricted integrals are then defined as:
\be
\label{eq:amuwin}
\amuwin(t_i,t_f)=\alpha^2\int_0^\infty dt\,W(t;t_i,t_f,\Delta)K(t)\ctogi
\ ,\ee
where the window function, $W(t;t_i,t_f,\Delta)$, takes on different forms depending on the values of $t_i$ and $t_f$:
\be
\label{eq:window_fns}
W(t;t_i,t_f,\Delta)=\left\{
\begin{array}{lr}
\multicolumn{2}{r}{1,\ \text{when } t_i=0 \;\&\; t_f=\infty}\ , \\
1-\theta(t-t_f;\Delta),  & \\
\multicolumn{2}{r}{\text{when } 0=t_i<t_f<\infty}\ , \\
\theta(t-t_i;\Delta)-\theta(t-t_f;\Delta), & \\
\multicolumn{2}{r}{\text{when } 0<t_i<t_f<\infty}\ ,\\
\theta(t-t_i;\Delta), & \\
\multicolumn{2}{r}{\text{when } 0<t_i<t_f=\infty} \ , 
\end{array}
\right.\ee
where we take $\Delta=0.15\,\fm$, as advocaated in \cite{Blum:2018mom}
Note that $W(t;t_i,t_f,\Delta)$ is defined such that $\amuwin(0,\infty)=\amulohvp$.
Also, in the following, we call $\amulohvpwin$ the Euclidean-time window contribution to $\amulohvp$ from the so-called ``intermediate'' time interval $[0.4,1.0]\,\fm$, i.e.\ \eqs{eq:amuwin}{eq:window_fns} with $t_i=0.4\,\fm$ and $t_f=1.0\,\fm$. Here we use a version of \eq{eq:kernel} in which the upper limit of the integral is $Q^2_\text{max}=3\,\gev^2$ for reasons explained in Ref.~\cite{Borsanyi:2017zdw}.

These time-window observables are particularly useful, because they are straightforward to compute in both the lattice and data-driven approaches. 

On the lattice, their computation requires modifying the kernel in \eq{eq:amulohvp} used to compute $a_\mu^\text{LO-HVP}$, that is using an appropriate discretization of \eq{eq:amuwin}. 
Of course, the difficulty of the continuum and infinite-volume extrapolations, as well as the quality of the statistical signal, will depend acutely on the chosen window. 

With the R-ratio, one simply substitutes \eq{eq:laplace} into \eq{eq:amuwin}, yielding:
\be
\label{eq:amuwin_R-ratio}
\amuwin(t_i,t_f)=\left(\frac{\alpha m_\mu}{3\pi}\right)^2\int_0^\infty \frac{ds}{s^2}\hat K(s;t_i,t_f)R(s)
\ ,\ee
with
\be
\label{eq:khats}
\hat K(s;t_i,t_f)=\frac{3s^{5/2}}{8m_\mu^4}\int_0^\infty dt\,e^{-t\sqrt{s}}W(t;t_i,t_f,\Delta)K(t)
\ .\ee
A first quantitative comparison of the two approaches can proceed through the comparison of individual windows, as was done in Refs.~\cite{Borsanyi:2020mff,Colangelo:2022vok}.

In addition to time-window observables, one can further compare the lattice and data-driven approaches via the HVP function, $\hat\Pi(Q^2)$, at different values of the spacelike momentum $q$, with $Q^2=-q^2\ge0$ \cite{Borsanyi:2017zdw,Ce:2022eix}.
Alternatively, one can consider the difference of this function at two values of $Q^2$, to isolate an interval of interest. 
This has already be done in Ref.~\cite{Borsanyi:2020mff}. 
Instead of directly considering $\hat\Pi(Q^2)$, which is convention dependent, here we focus on the hadronic running of the electromagnetic coupling in the on-shell scheme, to which it is related. Thus, we work here with the hadronic running of the fine structure constant in the five-flavor theory in the spacelike region, $\deltaalphahad(-Q^2)$. In terms of the lattice correlation function of \eq{eq:correlator}, this quantity is given by
\bea
\deltaalphahad(-Q^2)& = & e^2\int_0^\infty dt\left[t^2-\frac4{Q^2}\sin^2\left(\frac{t\sqrt{Q^2}}2\right)\right]\nn\\
&&\qquad\qquad\qquad\times\ctogi
\label{eq:deltaalphalat}
\eea
and, in terms of the R-ratio, by
\be
\deltaalphahad(-Q^2) = \frac{\alpha Q^2}{3\pi}\int_0^\infty ds\, \frac{R(s)}{s(s+Q^2)}
\ .
\label{eq:deltaalphaRratio}
\ee

So as to simplify notations in the following, we denote quantities computed on the lattice, $\alat{j}$, and their counterparts computed in the data-driven approach, $\aR{j}$. 
Here the $j$ indexes different moment integrals, such as the total $\amulohvp$, window contributions to $\amulohvp$, as well as the running of $\alpha$ between two spacelike virtualities.
\section{Comparison methodology}
\label{sec:comparison_methodology}

In the following two subsections, we present in more detail the methodologies that we will use to compare and test the lattice and data-driven approaches in \sec{Sec:Results}.

\subsection{Testing the lattice with R-ratio results}
\label{sec:testing_lat}

Using the notation introduced in the last paragraph of the previous section, the first step is to compare the observables of interest, $\alat{j}$ and $\aR{j}$, one by one, and to determine their level of compatibility.
Depending on how localized these observables are in Euclidean time or in c.o.m.\ energy, one can isolate distance, virtuality or energy scales that may be problematic in each approach. 
Since a more effective way of isolating possibly problematic regions of the experimental R-ratio is presented below in \sec{sec:testing_Rratio}, the observables $a^\text{lat/R}_j$ that we focus on  are chosen to emphasize features of the lattice correlator.
Detailed results for the comparison of individual observables obtained from the lattice and the experimental R-ratio, are given in \sec{sec:testing_lattice_results}.

To go beyond a comparison of individual observables, we propose a more global comparison of the two approaches that includes not just one observable at a time, but many.
Thus, we define the following $\chi^2$ function:
\bea
\chi^2 & = & \sum_{j,k}\left[\alat{j} - a_j\right][C^{-1}_\text{lat}]_{jk}\left[\alat{k} - a_k\right]\nn\\
&& + \sum_{j,k}\left[\aR{j} - a_j\right][C^{-1}_\text{R}]_{jk}\left[\aR{k} - a_k\right]\ ,
  \label{eq:chisqwin}
\eea
where $j,k$ run over the observables considered in the comparison and where the $a_j$ are parameters. 
This function requires knowledge of the covariance matrices, $C_\text{lat/R}$, between the observables obtained in each approach. 
Correlations between lattice and experimental R-ratio results are assumed to vanish, which is the case in the examples considered below.
This more global measure of the compatibility of the two approaches is also important because it serves as a baseline for the improvements in the agreement brought by modifications of the experimental R-ratio considered in \sec{sec:testing_Rratio}.

For absolute uncertainties~\footnote{This can be generalized for also treating relative uncertainties that are scaled with the fitted quantities, e.g.\ by explicitly including the corresponding dependencies on the fitted quantities in the $\chi^2$ definition and/or using an iterative procedure~\cite{DAgostini:1993arp,Blobel:2003wa}~(see also \sec{sec:testing_Rratio} and \app{Appendix:Chi2Methodology}).},
the minimum of the $\chi^2$ of \eq{eq:chisqwin}, as a function of the parameters $a_i$, is attained at 
\bea
a_i &=& \sum_{j,k}\Bigl[ \Bigl(C^{-1}_\text{lat} + C^{-1}_\text{R} \Bigr)^{-1} \Bigr]_{ij}
\left[ [C^{-1}_\text{lat}]_{jk} \alat{k} \right.\nn\\ 
& & \left. \qquad + [C^{-1}_\text{R}]_{jk}  \aR{k} \right]\ ,
  \label{eq:amuwinmin}
\eea
and takes the minimum value, 
\be
\chi^2_\text{min} = \sum_{j,k}\left[ \alat{j} - \aR{j} \right]
[ \left(C_\text{lat} + C_\text{R} \right)^{-1} ]_{jk}\left[ \alat{k} - \aR{k}\right]\ .
  \label{eq:chisqwinMin}
\ee

We take the $p$-value associated with the minimum value of $\chi^2$ and the number of degrees of freedom ($\ndof$) to be a measure of the overall agreement of the two approaches. 
Here, the $\ndof$ is simply the number of observables considered in the comparison. 
This measure of agreement makes sense only if $C_\text{lat/R}$ are well known, because the latter determine how much independent information is available in the lattice and R-ratio inputs and because the $p$-value obtained will depend strongly on them. 
This $p$-value may dilute some of the tensions observed in individual observables. 
However, in the presence of strong correlations, its value will better reflect the significance of having many observables disagree.
Detailed results for this combined comparison of lattice and experimental R-ratio observables, via \eq{eq:chisqwinMin}, are given in \sec{sec:testing_lattice_results}.

Note that if the $p$-value corresponding to this minimum is acceptable, then the values of $a_j$ at the minimum of $\chi^2$ correspond to a weighted average of the experimental R-ratio and lattice integrals.
However, this averaging constrains the shape of the underlying spectral functions only in a very limited fashion.
These aspects are discussed in more detail in  \app{app:beyond_rescaling}.

\subsection{Testing the experimental R-ratio with lattice results: rescaling}
\label{sec:testing_Rratio}

While fully reconstructing the R-ratio from the HVP correlator computed on the lattice is an ill-posed problem, interesting information about $R(s)$ can still be derived. 

The method proposed here is applicable to any observable, $a_j$, related to the HVP.
It is meant to help isolate c.o.m.\ energy regions, in the experimental R-ratio, that may be responsible for tensions with the chosen lattice observables.
Thus, we split up the observables obtained from the R-ratio, via e.g.\ \eqs{eq:amuwin_R-ratio}{eq:deltaalphaRratio}, into contributions from different $\sqrt{s}$ intervals, $I_b$:
\be
\label{eq:amuwin_split}
\aR{j}=\sum_b\aR{jb}
\ .\ee
Here the sum over $b$ (i.e.\ over all of the intervals) covers the full support of $R(s)$. 
In the case of $\amulohvp$, of window observables and of the running of $\alpha$ between two spacelike momenta, the $\aR{jb}$ are obtained by restricting the integrals over $s$ in \eqs{eq:amuwin_R-ratio}{eq:deltaalphaRratio}, respectively, to the interval $I_b$. 

When computing the Euclidean-time window observables, the edges of the corresponding intervals are smoothed out to account for the fact that, at finite lattice spacing, the density of points in $\ctogi$ is not that large. This is not necessary for $R(s)$, because the density of measured points in $s$ is significant, allowing us to consider sharp intervals $I_b$, with a interpolation between the two points bracketing each boundary of the interval.

As discussed in the Introduction, in this paper we investigate whether a tension between lattice and data-driven results can be explained by a change in the experimental R-ratio that is consistent with a rescaling of the contributions, from one or more intervals in $\sqrt{s}$, to the observables of interest. 
Thus, we wish to solve the following system of equations for $\gamma_b$:
\be
\label{eq:rescaleR} 
\alat{j} = \sum_{b} \gamma_b \,\aR{jb}
\ ,\ee
taking into account all uncertainties and correlations among the lattice results, on the one hand, and the experimental R-ratio ones, on the other. 
If a given $\delta_b \equiv (\gamma_b - 1)$ is set to zero {\em a priori}, the R-ratio integrals remain unchanged in the corresponding interval. 

Here we consider the special case where a subset of the $\sqrt{s}$-intervals, $A$, give contributions to the observables that are rescaled by a common factor $\gamma$ while, for the complementary subset of intervals, $B$, the contributions are kept unchanged~\footnote{In practice, for the cases we consider here, these subsets contain one or two $\sqrt{s}$-intervals.}. 
In such a case \eq{eq:rescaleR} becomes
\be
\label{eq:rescaleRoneNorm} 
\alat{j} = \sum_{b \in A} \gamma \,\aR{jb} + \sum_{b \in B} \aR{jb}
\ ,\ee
which implies,
\be
\label{eq:rescaleRoneNormConstr} 
\gamma = \frac{ \alat{j} - \sum_{b \in B} \aR{jb} }{ \sum_{b \in A} \,\aR{jb} } \equiv \tilde{\gamma}_j
\ .\ee
With more than one independent observable, $a_j$, this becomes an over-constrained system, typically characterized by a $\chi^2$ function:
\be
\chi^2 = \sum_{j,k}\left[  \gamma -  \tilde{\gamma}_j \right]
\left[ \left(C^{\tilde{\gamma}}_\text{lat} + C^{\tilde{\gamma}}_\text{R} \right)^{-1} \right]_{jk}\left[ \gamma -  \tilde{\gamma}_k \right]\ ,
  \label{eq:chisqAveragefit}
\ee
where $C^{\tilde{\gamma}}_\text{lat}$ and $C^{\tilde{\gamma}}_\text{R}$ are the covariance matrices of the $\tilde{\gamma}_j$ rescaling coefficients, originating from the lattice and the dispersive uncertainties respectively.
They are obtained through a linear propagation of uncertainties, from the lattice QCD and experimental R-ratio $a_j$ integrals, through \eq{eq:rescaleRoneNormConstr}. 

The rescaling factor $\gamma$ optimizing the constraints from \eq{eq:rescaleRoneNormConstr} can be determined by minimizing the $\chi^2$ function from \eq{eq:chisqAveragefit} with respect to $\gamma$. 
This minimization yields an average of the input $\tilde{\gamma}_j$ values, with weights given by the inverse of the sum of covariance matrices that appears in \eq{eq:chisqAveragefit}.
Although very commonly used, this approach can yield biased results~(see e.g.\ Ref.~\cite{DAgostini:1993arp} and the discussion in \app{Appendix:Chi2Methodology}).
In particular, such biases are caused by the uncertainties on the uncertainties and on their correlations, present both for the experimental measurements used in the dispersive approaches and for the lattice QCD results.
In order to avoid these undesirable effects, we employ the weighted average of \eq{Eq:chi2minimum}, in which we set the off-diagonal elements of $C^{\tilde{\gamma}}_\text{lat}$ and $C^{\tilde{\gamma}}_\text{R}$ to zero, followed by a full propagation of the uncertainties with their correlations~(see \app{Appendix:Chi2Methodology} for a detailed discussion).

We also consider a method based on a generalization of the $\chi^2$ of \eq{eq:chisqwin}, instead of the weighted averages discussed above.
This method also allows us to study similar modifications of the experimental R-ratio.
It is discussed in \app{sec:multiple_rescaling_fits}. 

\begin{figure*}[t]
\centerline{\includegraphics[width=0.45\linewidth]{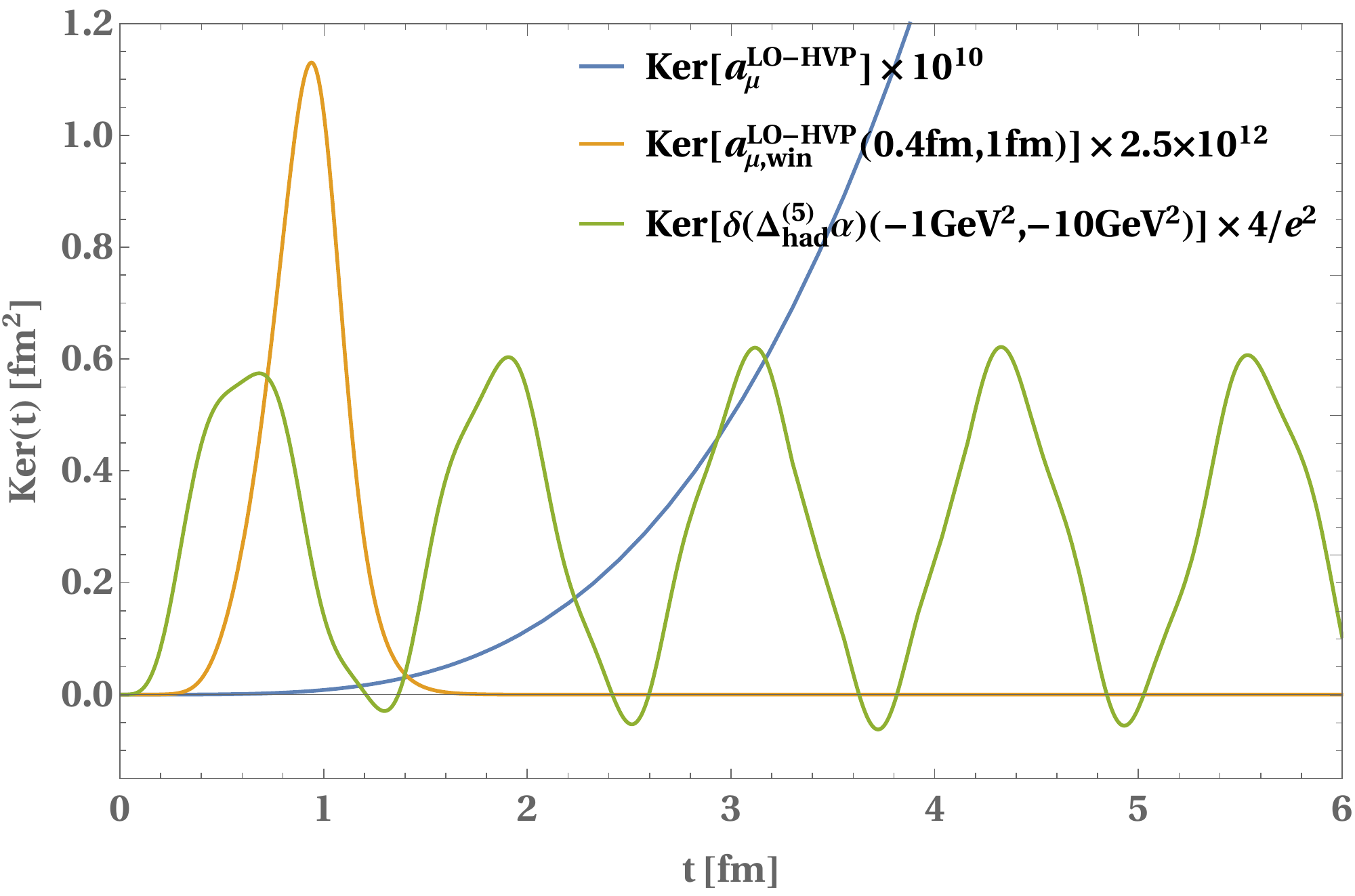}\hspace{0.05\linewidth}
\includegraphics[width=0.45\linewidth]{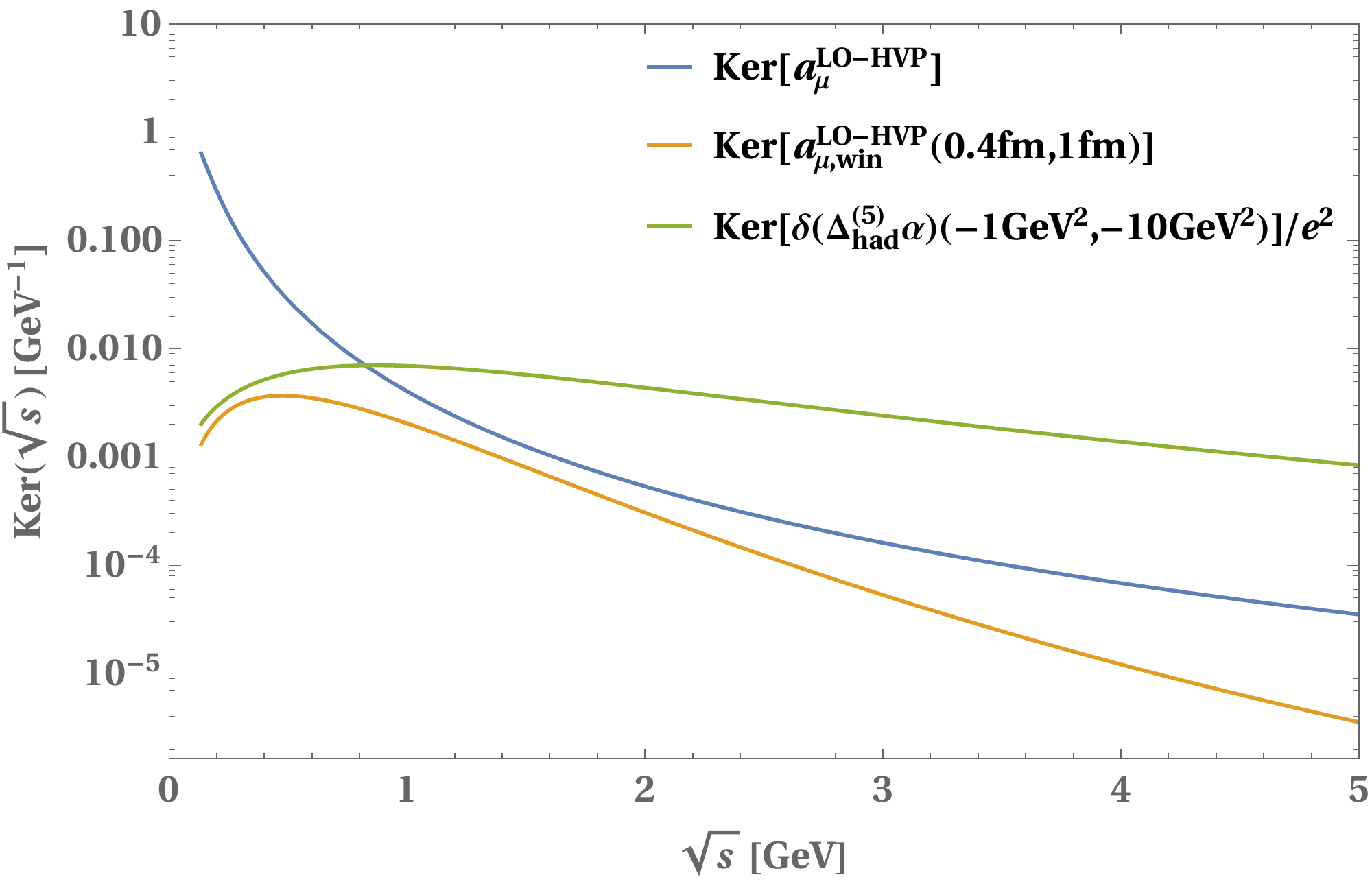}}
\caption{\label{fig:support}Kernels for $\amulohvp$, $\amulohvpwin$ and $\ddalpha$: as a function of Euclidean time (left panel); of c.o.m.\ energy, $\sqrt{s}$ (right panel). Some of the kernels are rescaled for better visibility. $\amulohvpwin$ is a contribution to $\amulohvp$ so that its kernel (orange curve) is a contribution to $\amulohvp$'s kernel (blue curve), peaked around $0.9\,\fm$. }
\end{figure*}

\subsection{Possible extensions of the methodology}
\label{sec:testing_Rratio_beyond}

In \app{app:beyond_rescaling}, we consider comparisons and tests of the lattice and data-driven approaches that go beyond the rescaling of the experimental R-ratio integrals in a $\sqrt{s}$-interval, by a single, common parameter $\gamma$. Moreover, in \app{app:more_moments} we consider the possible advantages and limitations associated with the use of additional observables in the data-driven approach.

\section{Comparison results}
\label{Sec:Results}

\subsection{Testing the lattice with R-ratio results}
\label{sec:testing_lattice_results}

As a first step, we compare individually $\amulohvp$, the intermediate-time window observable $\amulohvpwin$ and 
$\ddalpha$. 
The kernels for these quantities are shown as functions of $t$ and $\sqrt{s}$ in \fig{fig:support}. 
As is well known, the kernels for $\amulohvp$ is highly peaked for small values of $\sqrt{s}$ and large values of $t$. 
While the Euclidean-time kernel for $\amulohvpwin$ is strongly localized in $t$ in the interval ranging approximately from $0.35$ to $1.51\,\fm$,~\footnote{This interval is the range of $t$ values within which the kernel for $\amulohvpwin$ takes on values greater than $1\%$ of its maximum value. The same exercise on the window function of \eq{eq:window_fns}, with $t_i=0.4\,\fm$ and $t_f=1\,\fm$,  gives the interval $[0.05,1.35]\,\fm$. Thus, the region of the lattice correlator that is probed via $\amulohvpwin$ is shifted upwards relative to the one given by the window function alone.} its c.o.m.\ counterpart emphasizes a $\sqrt{s}$ region from threshold to around $3.2\,\gev$.~\footnote{This upper limit is the value of $\sqrt{s}$ below which the kernel takes on values larger than $1\%$ of its maximum value.}
Finally, the kernels for $\ddalpha$ extends over all $t$, with an oscillation of period $\sim 1\,\fm$, and over a large range of $\sqrt{s}$, going from threshold to well beyond $5\,\gev$.
The upshot is that the observables which we consider here probe the behavior of the lattice correlator $\ctogi$ at very different distance scales while the c.o.m.\ energies at which the R-ratio is probed have more overlap.
Thus, the comparison of these observables tells us more about specific features of the lattice correlator than they do about those of the timelike spectrum.

\begin{figure*}[tb]
\centerline{\includegraphics[width=0.45\linewidth]{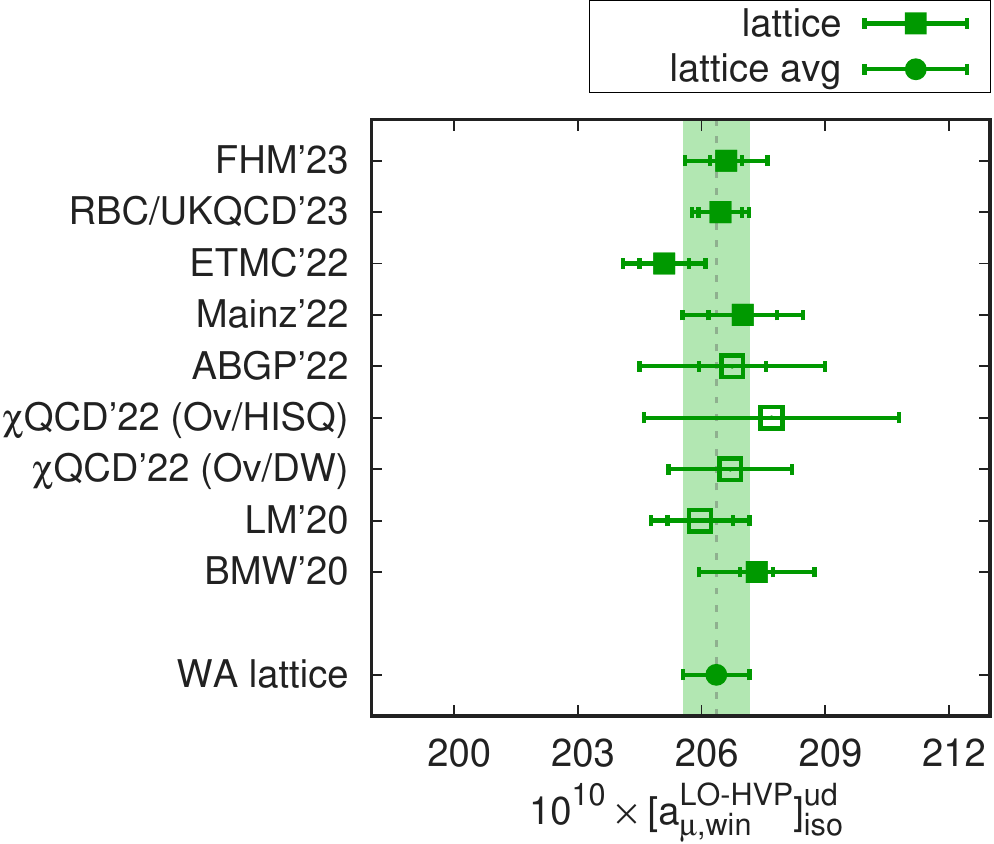}
\hspace{0.05\linewidth}
\includegraphics[width=0.45\linewidth]{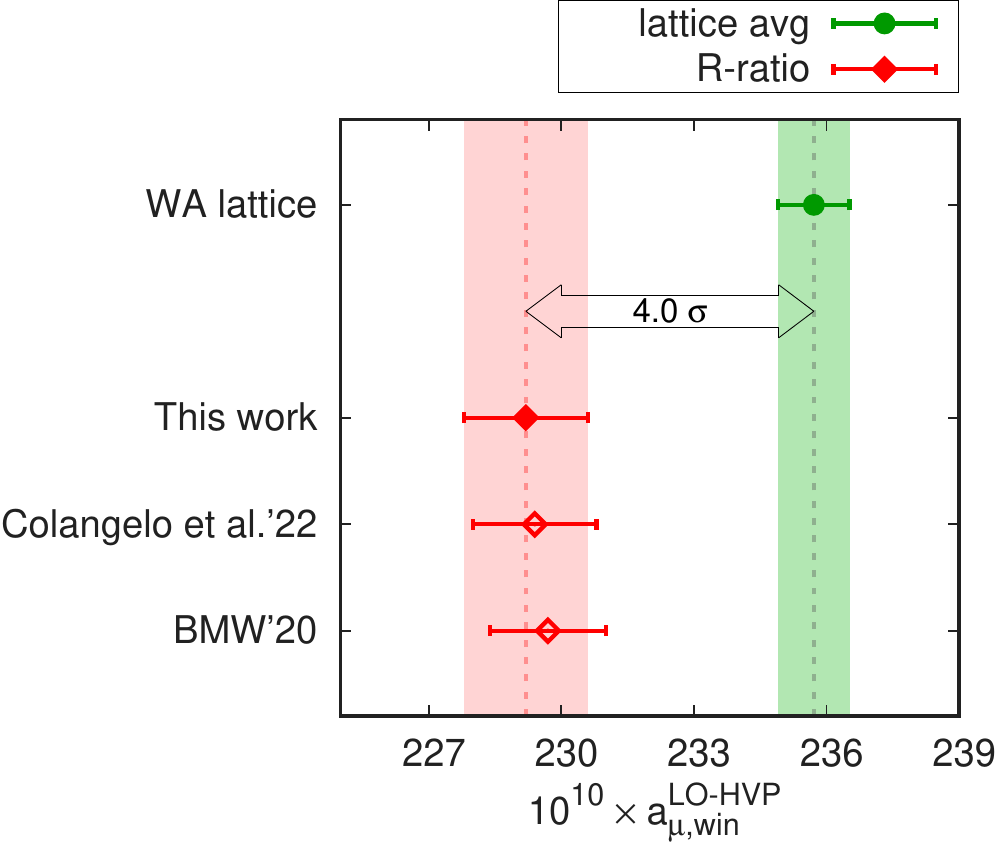}}
\caption{\label{fig:win_cmp}
Comparison of lattice and data-driven results for the contribution $\amulohvpwin$ to $\amulohvp$, from the Euclidean time interval $[0.4,1.0]\,\fm$ (see \eqs{eq:amuwin}{eq:window_fns} and subsequent text). 
Left panel: Comparison of lattice results for the isospin-symmetric, $u$ and $d$ connected contribution to the intermediate-window observable, noted $[\amulohvpwin]_\text{iso}^{ud}$ (green squares). For each group, only the most recent results are shown. BMW'20~\cite{Borsanyi:2020mff}, LM'20~\cite{Lehner:2020crt}, ABGP'22~\cite{Aubin:2022hgm} and FHM'23~\cite{Bazavov:2023has} are obtained with different varieties of staggered fermions; Mainz'22~\cite{Ce:2022kxy} with $O(a)$-improved Wilson, ETMC'22~\cite{ExtendedTwistedMass:2022jpw}  with twisted-mass and RBC/UKQCD'23~\cite{Blum:2023qou} with domain-wall fermions; $\chi$QCD'22~\cite{Wang:2022lkq} with overlap valence quarks on either HISQ or domain-wall configurations. 
The filled squares correspond to fully independent results while the open ones are obtained using subsets of configurations from other calculations. 
LM'20 relies on a subset of the configurations used in FHM'23 and $\chi$QCD'22 on subsets of those used in FHM'23 and RBC/UKQCD'23.
The green band corresponds to a weighted average of the fully-independent (filled squares) lattice results for $[\amulohvpwin]_\text{iso}^{ud}$.
The mean is performed without correlations in the determination of the weights while the uncertainty propagation assumes a $100\%$ correlation between the total systematic error of the different calculations.
The resulting correlated $\chi^2/\ndof$ is $2.3/4$.
Right panel: Comparison of the weighted average of lattice results (green filled circle and band) with a number of R-ratio results for the intermediate-window observable (red diamonds), including the one determined in this paper (filled red diamond and band). 
The weighted average for $\amulohvpwin$ is obtained from that for $[\amulohvpwin]_\text{iso}^{ud}$, to which we have added all other quark, QED and strong-isospin contributions from Ref.~\cite{Borsanyi:2020mff}. The resulting average is $\amulohvpwin=(235.7\pm0.8)\times 10^{-10}$. 
The difference of this lattice average with the data-driven one of the present work is $4.0\sigma$, shown as a horizontal arrow. ``WA" stands for ``world average".}
\end{figure*}

The lattice \cite{Borsanyi:2020mff} and data-driven results for the observables of interest are shown in \tab{tab:observables}, together with their pairwise differences. 
While the preliminary lattice \cite{Borsanyi:2020mff} and the data-driven results for $\ddalpha$ are compatible within $1.4\sigma$, there is a $2.0\sigma$ tension between the results for $\amulohvp$. 
For $\amulohvpwin$, the tension is much more significant, at $3.8\sigma$. 
This number rises to $4.0\sigma$ when the lattice result of Ref.~\cite{Borsanyi:2020mff} is replaced by a weighted average of independent lattice determinations of this quantity~\cite{Borsanyi:2020mff,Ce:2022kxy,ExtendedTwistedMass:2022jpw,Blum:2023qou,Bazavov:2023has}. 
All of these points are illustrated in \fig{fig:win_cmp}.
This large discrepancy points to a significant inconsistency between the approaches that must be resolved before giving a standard model prediction for $\amulohvp$. 

Because the window observable probes intermediate Euclidean-time distances, its computation is particularly well suited to the lattice approach. 
The observable is significantly less noisy statistically than $\amulohvp$, less sensitive to finite-volume effects, and less sensitive to short-distance discretization effects, making its computation in lattice QCD particularly reliable.

In principle, the disagreement with the data-driven approach could be due to a common feature in the lattice calculations of \fig{fig:win_cmp} that would lead to an overlooked systematic error. 
For instance, one may worry that all the calculations are based on the time-momentum representation (TMR) of the current-current correlator given in \eq{eq:correlator}.
However, Ref.~\cite{Chao:2022ycy} explicitly checked that the connected contributions of the $ud$ and $s$ quarks to the window observable, obtained using a Lorentz-covariant coordinate-space representation (CCS), agree with those computed in the TMR to the level of $0.7\%$, in simulations with  pion and kaon masses of $\sim 350\,\mev$ and $\sim 450\,\mev$, respectively. 
Alternatively, one may question the universality of the continuum limit. 
However, as discussed in the caption of \fig{fig:win_cmp}, the lattice results for $[\amulohvpwin]^{ud}_\text{iso}$, included in the average~\cite{Borsanyi:2020mff,Ce:2022kxy,ExtendedTwistedMass:2022jpw,Blum:2023qou,Bazavov:2023has} are obtained with different discretizations, thus testing this universality.
Moreover, two of these analyses were blinded \cite{Blum:2023qou,Bazavov:2023has}, strengthening this test.
There are four additional calculations of the leading $u$ and $d$ quark contributions \cite{Blum:2018mom,Lehner:2020crt,Wang:2022lkq,Aubin:2022hgm}.
All nine calculations agree, as shown in the left panel of \fig{fig:win_cmp}.

\begin{table*}[t]
    \centering
        \caption{Comparison of lattice and data-driven results for the three main observables of interest in this paper. Each row contains the results and comparisons for the observable described in the first column. The second column provides the corresponding lattice result from Ref.~\cite{Borsanyi:2020mff} and the third, the data-driven one computed here, using the methods of Ref.~\cite{Davier:2019can}. The fourth column displays the absolute difference between the two results, the fifth their relative difference, the sixth their difference in units of combined standard deviations and the last column, the corresponding $p$-value. The values in parentheses correspond to the total uncertainties of the corresponding quantities. Note that the powers of 10 in the observable column only apply to columns two through four. The three remaining columns have units specified in the corresponding column label.}
    \label{tab:observables}
    \begin{tabular}{lcccccc}
    \hline\hline
      Observable &  lattice \cite{Borsanyi:2020mff} & data-driven & diff. & \%\ diff. & $\sigma$ & $p$-value [\%]\\
      \hline
    $\amulohvp\times 10^{10}$ & $707.5(5.5)$ & $694.0(4.0)$ & $13.5(6.8)$ & $1.9(1.0)$ & $2.0$ & $4.7$\\
    $\amulohvpwin\times 10^{10}$ & $236.7(1.4)$& $229.2(1.4)$ & $7.5(2.0)$ & $3.2(0.8)$  & $3.8$ & $0.01$\\
    $\left[\deltaalphahad(-10\,\gev^2)-\deltaalphahad(-1\,\gev^2)\right]\times 10^4$ & $48.67(0.32)$\footnote{This result's continuum limit does not include the logarithmically-enhanced discretization uncertainties discovered subsequently in Ref.~\cite{Ce:2021xgd}, nor was this quantity the focus of Ref.~\cite{Borsanyi:2020mff}. However, we include it in the present study to illustrate how using a quantity which is complementary to $\amulohvp$ and $\amulohvpwin$ can provide important information. }
    & $48.02(0.32)$ & $0.65(0.45)$ & $1.3(0.9)$ & $1.4$ & $15.$\\
    \hline\hline
    \end{tabular}
\end{table*}

From the perspective of the R-ratio as a function of $\sqrt{s}$, the range covered by the intermediate window is more spread out. 
Though it is not exclusively emphasized, this region includes the $\rho$ peak, in which some disagreement between the different measurements is apparent \cite{Davier:2019can,CMD-3:2023alj}. 
However, the difference between the KLOE~\cite{KLOE:2008fmq,KLOE:2010qei,KLOE:2012anl} and the BABAR~\cite{BaBar:2012bdw} measurements of the $\epemtopippimg$ cross section is taken into account as an additional systematic uncertainty, as described in Ref.~\cite{Davier:2019can}. 
So, unless a large, unknown systematic uncertainty is present in the measurements of the $\epemtohad$ cross section \cite{KLOE:2008fmq,KLOE:2010qei,KLOE:2012anl,BaBar:2012bdw}, the strong tension with the lattice calculation of the intermediate window cannot be explained within the data-driven approach. 
The very recent measurement of the $\epemtopippimg$ cross section by the CMD-3 collaboration \cite{CMD-3:2023alj} helps bring the data-driven results for $\amulohvp$ and $\amulohvpwin$ in line with the lattice results for those quantities. 
However, it does so at the expense of a dramatic tension with previous measurements that is not yet understood. 

As a preamble to the following discussion about the consequences, for the lattice correlator, of the tensions and agreement in the values of $\amulohvp$, $\amulohvpwin$ and $\ddalpha$ computed in the two approaches, it is important to remember that the current correlator is a smooth, monotonically decreasing function of $t$ that behaves as $t^{-3}$ for small $t$, up to logarithms, and falls off exponentially with exponent $2M_\pi t$ at large $t$ in infinite volume.

While the significance of the difference between the lattice and data-driven determinations of $\amulohvp$ is smaller than in $\amulohvpwin$, the absolute difference itself is about twice as large.
Thus, about half the discrepancy in $\amulohvp$ must come from the value of the lattice correlator for $t$ below $\sim 0.35\,\fm$ and/or above $\sim 1.51\,\fm$.
An important caveat is that the uncertainty on this second half of the discrepancy is around $100\%$ (see \tab{tab:observables}), making this contribution to the discrepancy not significantly different from $0$.

Combining this observation with the results for $\ddalpha$, which agree in the two approaches, is more complicated.
Indeed, the kernel of this observable is oscillatory in $t$ and has support at all times.
However, the kernel of $\ddalpha$ has significant overlap with the one of $\amulohvpwin$.
This suggests that $\ddalpha$ will receive an enhancement similar to the one of $\amulohvpwin$, from the region delimited by the kernel of this window observable.
Of course, that will not be true if all of the enhancement in $\amulohvpwin$ comes from times $1.2\,\fm\lsim t\lsim 1.5\,\fm$, where the running-of-$\alpha$ kernel is suppressed.
Assuming, for the moment, that the enhancement of $\ddalpha$ is present, agreement on this quantity in the two approaches then requires that the lattice correlator be suppressed for some values of $t$ outside the window.

This discussion leads to a situation where $\amulohvp$ suggests that the lattice correlator is enhanced for $t$ below $\sim 0.35\,\fm$ and/or above $\sim 1.51\,\fm$, while 
$\ddalpha$ suggests the opposite.
To try to reconcile the two statements it is useful to look at contributions to each of the two observables from the short-distance (SD) window (defined via the window function of \eq{eq:window_fns}, with $t_i=0\,\fm$ and $t_f=0.4\,\fm$), the intermediate-distance (ID) one (denoted simply as window in this paper, and with $t_i=0.4\,\fm$ and $t_f=1\,\fm$), and the long-distance (LD) one (with  $t=1\,\fm$ to $\infty$).
Using the data from Ref.~\cite{Keshavarzi:2018mgv} for a rough estimate of the ratio of the contributions of the SD:ID:LD windows, we find 10\%:33\%:57\%\ for $\amulohvp$ and 70\%:29\%:1\%\ for $\ddalpha$.

Thus, if the excess in $\ddalpha$ from the ID window were to be compensated by a suppression of the correlator in the LD window alone, which contributes only 1\%, this suppression would have to be significant. 
However, such a significant suppression would also reduce $\amulohvp$, since the contribution of the LD window to that observable is dominant.
To compensate that suppression would require a significant enhancement of the lattice correlator in the SD window, because that window only contributes at the level of 10\% to $\amulohvp$.
But such an SD enhancement would make the lattice determination of $\ddalpha$ significantly larger than that from the experimental R-ratio, given that the SD window is responsible for about 70\% of the value of that observable.
This discussion suggests that the SD window is not responsible for the additional enhancement in the lattice computation of $\amulohvp$ over that of $\amulohvpwin$. It also suggests that an LD suppression of the lattice correlator cannot be responsible for mitigating a possible enhancement due to the ID window contribution to $\ddalpha$.

To summarize, we know for sure that the lattice correlator is enhanced, in the ID window region, compared to the one obtained from the data-driven approach. In addition, the arguments made above suggest that the lattice correlator is smaller in the SD window and larger in the LD one. 

\medskip

We now consider combined comparisons of the lattice and experimental R-ratio results for the above observables, via the minimization of the $\chi^2$ function defined in \eq{eq:chisqwin}.
Beyond giving a more global measure of the compatibility of the two approaches, the resulting $p$-values also provide baselines against which to compare those obtained when we study possible rescalings of integrals of the R-ratio in chosen $\sqrt{s}$-intervals, in the following section.
These comparisons require knowledge of the covariance matrices, among the observables considered, in each of the lattice and data-driven approaches.
While the correlations between $\amulohvp$ and $\amulohvpwin$, determined in lattice QCD, are obtained as described in \app{sec:lat_obs}, the determination of the running of $\alpha$ is impacted by uncertainties that are, to good approximation, independent of those of the former.
On the other hand, for the dispersive approach, the correlations among the uncertainties of the contributions to all moment integrals are derived as discussed in \app{sec:R-ratio_obs}.

We first consider a combined comparison of $\amulohvp$ and $\amulohvpwin$. The resulting $\chi^2/\ndof$ range from $14.4^{+3.0}_{-2.1}/2$ to $18.8^{+2.0}_{-1.7}/2$ when the two most extreme of the four systematic variations of the lattice covariance matrices, given in \app{sec:lat_obs}~\footnote{These two results correspond to the covariance matrices labelled 2 and 3, respectively.}, are considered.
The central values correspond to the nominal lattice sample. 
The uncertainties are statistical, obtained from the $\pm 1\sigma$ quantile deviations from the median of the bootstrap distributions for the lattice covariance matrices~
\footnote{\label{foot:lat_stat} As discussed in \app{sec:lat_obs}, it is important to note that the statistical covariance of the lattice data is determined from only $48$ binned samples, which is smaller than the number of independent samples in the full dataset. However, as seen below, the statistical uncertainties on the lattice covariance matrix that arise due to this conservative binning are sufficiently small for the purposes of the present study. In future work, we will aim to improve this evaluation.}. 
Half the difference between the central values, i.e. $4.4/2=2.2$, can be viewed as the lattice systematic uncertainty on these results. 
It is of comparable size to the statistical one. 
Unlike what has been just discussed for the lattice results, statistical and systematic uncertainties on the uncertainties associated with the data-driven approach are not yet included here, but should be in the future.

The corresponding $p$-values are $7^{+10}_{-6}\times 10^{-4}$ and $8^{+13}_{-6}\times 10^{-5}$, respectively. 
These are very small and significantly smaller than the $p$-value for $\amulohvp$ alone in \tab{tab:observables}.  
Nevertheless, the first of the above $p$-values can be as much as 17 times larger than the one in given in \tab{tab:observables} for $\amulohvpwin$ and the second as much as a factor of 5 smaller, within one standard deviation. 
In any case, the probability that the lattice and data-driven results for $\amulohvp$ and $\amulohvpwin$ agree simultaneously is very small.

The disagreement reduces some when $\ddalpha$ is added to the mix. In the lattice calculation, which is still preliminary, this quantity is negligibly correlated with $\amulohvp$ and $\amulohvpwin$. While here the $\chi^2$ remain essentially unchanged, ranging from $14.4^{+3.0}_{-2.1}$ to $18.8^{+2.0}_{-1.7}$, the $\ndof$ increases from $2$ to $3$, leading to improved $p$-values that range from $2.3^{+4.0}_{-1.9}\times 10^{-3}$ and $3.0^{+4.2}_{-2.1}\times 10^{-4}$. 
That is, the inclusion of $\ddalpha$ and, to a lesser extent, $\amulohvp$ is observed to dilute the disagreement between the lattice and data-driven approaches for $\amulohvpwin$.
This dilution is expected, because the disagreement between the two approaches is much smaller for both $\ddalpha$  and $\amulohvp$.
However, in the presence of strong correlations, this combined measure will better reflect the significance of having agreement or disagreement among many observables.

Very similar results are obtained in the approach of \sec{sec:testing_Rratio}, when the $\gamma$ rescaling coefficient in \eq{eq:rescaleR} is set to unity. 
This is not surprising because, with $\gamma=1$, \eqs{eq:chisqwinMin}{eq:chisqAveragefit} indicate that the $\chi^2$ and $\ndof$ are the same in the two approaches, up to possible nonlinearities in the propagation of uncertainties.
In particular, the minimal/maximal $\chi^2/\ndof$ values are stable, within $0.2$ or less, with respect to the choice of the split in various $\sqrt{s}$ regions.
When sampling the $\chi^2/\ndof$ values for the various bootstrap replicas of the lattice covariance matrix, with either two or three moment integrals, the median of the resulting distribution is close to the nominal values~(within $0.15$ or less), while the variance and $\pm 1~\sigma$ quantiles of the distribution are smaller than $1.5$ and the $\pm 2~\sigma$ quantiles are smaller than $3.0$~(similar to the uncertainty values given earlier in this section). 
It is hence observed that the level of the tension between the dispersive- and lattice-based moment integrals is relatively stable with respect to the choice of the split in various $\sqrt{s}$ regions and/or statistical and systematic variations of the lattice covariance matrix. 

Of course, other quantities related to hadronic vacuum polarization, that are computed in the lattice and data-driven approaches, can also be compared individually or added with correlations to the $\chi^2$, to further sharpen the comparison of the two approaches. 

\subsection{Testing the experimental R-ratio with lattice results: rescaling}
\label{sec:testing_R-ratio_results}

\begin{table*}[t]
    \centering
        \caption{Results for various rescaling scenarios, using two observables ($\amulohvp$ and $\amulohvpwin$) or three (adding $\ddalpha$), as indicated in column 1. 
    In these scenarios it is assumed that the integrals of the experimental R-ratio in the $\sqrt{s}$-interval $I_1$ are rescaled by a common $\gamma_1 \equiv 1+\delta_1$. 
    This interval is given in column 2 and the {\it nominal} rescaling percentage $\delta_1$~(obtained through the weighted average discussed in \sec{sec:testing_Rratio}, using the {\it nominal} values of the lattice covariance matrices obtained in \app{sec:lat_obs}), in column 4, with the corresponding {\it nominal} uncertainty propagated from the covariance matrices of the lattice QCD and dispersive results~($C^{\tilde{\gamma}}_\text{lat}$ and $C^{\tilde{\gamma}}_\text{R}$) indicated between ``$()$''.
    Column 3 indicates which of the four systematic uncertainty variations of the lattice covariance matrices give the best and worst fit qualities.
    Column 5 gives the {\it nominal} $\chi^2/\ndof$ for the scenario~(obtained by injecting $\gamma_1$ into \eq{eq:chisqAveragefit}) and column 6, the corresponding $p$-value.
    The values indicated between ``$[]$'' correspond to the $\pm 1 \sigma$ quantiles based on the lattice bootstrap replicas discussed in \app{sec:lat_obs} (see also footnote~\ref{foot:lat_stat}).
    The last column indicates the shift  on \DaHadZfive induced by the $\delta_1$ rescaling. }
    \label{tab:rescalingAverage}
    \resizebox{\textwidth}{!}{
    \begin{tabular}{cCCCCCC}
    \hline\hline
    Number of observables & I_1\;[\gev] & {\rm Lat. cov.} &  \delta_1 & \chi^2/\ndof  & p\text{-value} & \delta_1 \times \DaHadZfive[I_1] \times 10^{4} \\
    \hline
    2 & [\sqrt{s_\text{th}},0.63] & 0 & 15.9 (5.3)[^{+ 0.9}_{- 0.8}]\% & 10.0 [^{+ 2.4}_{- 1.9}]/1 & 0.16 [^{+ 0.31}_{- 0.13}]\%    & 0.80 \\  
    2 & [\sqrt{s_\text{th}},0.63] & 3 & 17.4 (5.7)[^{+ 0.6}_{- 0.5}]\% & 17.4 [^{+ 2.2}_{- 1.9}]/1 & 0.003 [^{+ 0.010}_{- 0.004}]\% & 0.88 \\  
    2 & [0.63,\infty[             & 0 &  3.1 (0.9)[^{+ 0.05}_{- 0.05}]\% & 0.9 [^{+ 0.1}_{- 0.1}]/1 & 34.6 [^{+ 3.2}_{- 3.2}]\%     & 8.49 \\  
    2 & [0.63,\infty[             & 3 &  3.2 (0.9)[^{+ 0.02}_{- 0.02}]\% & 1.3 [^{+ 0.1}_{- 0.1}]/1 & 25.2 [^{+ 2.8}_{- 2.2}]\%     & 8.71 \\  
    \hline
    3 & [\sqrt{s_\text{th}},0.63] & 0 & 16.4 (5.4)[^{+ 0.9}_{- 0.7}]\% & 10.6 [^{+ 2.2}_{- 1.7}]/2 & 0.49 [^{+ 0.73}_{- 0.36}]\%    & 0.83 \\  
    3 & [\sqrt{s_\text{th}},0.63] & 3 & 17.9 (5.8)[^{+ 0.6}_{- 0.5}]\% & 17.8 [^{+ 2.1}_{- 1.9}]/2 & 0.013 [^{+ 0.038}_{- 0.016}]\% & 0.91 \\  
    3 & [0.63,\infty[             & 0 &  2.5 (0.7)[^{+ 0.08}_{- 0.07}]\% & 3.8 [^{+ 0.6}_{- 0.5}]/2 & 14.7 [^{+ 4.2}_{- 4.0}]\%     & 6.68 \\  
    3 & [0.63,\infty[             & 3 &  2.6 (0.7)[^{+ 0.04}_{- 0.04}]\% & 5.3 [^{+ 0.5}_{- 0.4}]/2 & 7.0 [^{+ 1.8}_{- 1.6}]\%      & 6.96 \\  
    \hline
    \hline
    2 & [\sqrt{s_\text{th}},0.96] & 0 &  3.7 (1.1)[^{+ 0.1}_{- 0.1}]\% & 2.8 [^{+ 0.5}_{- 0.4}]/1 & 9.3 [^{+ 2.8}_{- 2.5}]\%        & 1.32 \\  
    2 & [\sqrt{s_\text{th}},0.96] & 3 &  3.9 (1.1)[^{+ 0.06}_{- 0.06}]\% & 4.4 [^{+ 0.4}_{- 0.5}]/1 & 3.5 [^{+ 1.4}_{- 1.0}]\%      & 1.39 \\  
    2 & [0.96,\infty[             & 0 &  9.4 (2.6)[^{+ 0.04}_{- 0.04}]\% & 0.09 [^{+ 0.01}_{- 0.009}]/1 & 77.0 [^{+ 1.2}_{- 1.3}]\% & 22.59 \\  
    2 & [0.96,\infty[             & 3 &  9.5 (2.5)[^{+ 0.02}_{- 0.02}]\% & 0.12 [^{+ 0.01}_{- 0.01}]/1 & 72.9 [^{+ 1.5}_{- 1.1}]\%  & 22.75 \\  
    \hline
    3 & [\sqrt{s_\text{th}},0.96] & 0 &  3.8 (1.1)[^{+ 0.09}_{- 0.09}]\% & 3.1 [^{+ 0.4}_{- 0.3}]/2 & 21.7 [^{+ 4.3}_{- 4.2}]\%     & 1.36 \\  
    3 & [\sqrt{s_\text{th}},0.96] & 3 &  4.0 (1.1)[^{+ 0.06}_{- 0.05}]\% & 4.5 [^{+ 0.4}_{- 0.4}]/2 & 10.7 [^{+ 3.2}_{- 2.3}]\%     & 1.42 \\  
    3 & [0.96,\infty[             & 2 &  3.5 (1.3)[^{+ 0.2}_{- 0.2}]\% & 10.9 [^{+ 2.2}_{- 1.6}]/2 & 0.43 [^{+ 0.54}_{- 0.30}]\%    & 8.35 \\  
    3 & [0.96,\infty[             & 3 &  3.7 (1.3)[^{+ 0.1}_{- 0.1}]\% & 14.1 [^{+ 1.5}_{- 1.2}]/2 & 0.089 [^{+ 0.083}_{- 0.052}]\% & 8.91 \\ 
    \hline
    \hline
    2 & [\sqrt{s_\text{th}},1.1]  & 0 &  3.3 (1.0)[^{+ 0.1}_{- 0.1}]\% & 2.2 [^{+ 0.4}_{- 0.3}]/1 & 13.4 [^{+ 3.2}_{- 2.9}]\%       & 1.40 \\  
    2 & [\sqrt{s_\text{th}},1.1]  & 3 &  3.4 (1.0)[^{+ 0.05}_{- 0.04}]\% & 3.5 [^{+ 0.3}_{- 0.4}]/1 & 6.3 [^{+ 1.9}_{- 1.3}]\%      & 1.46 \\  
    2 & [1.1,\infty[              & 0 & 14.1 (3.9)[^{+ 0.07}_{- 0.08}]\% & 0.1 [^{+ 0.02}_{- 0.02}]/1 & 70.9 [^{+ 1.6}_{- 1.6}]\%   & 33.01 \\  
    2 & [1.1,\infty[              & 3 & 14.3 (3.8)[^{+ 0.04}_{- 0.04}]\% & 0.2 [^{+ 0.02}_{- 0.02}]/1 & 65.8 [^{+ 1.8}_{- 1.4}]\%   & 33.31 \\ 
    \hline
    3 & [\sqrt{s_\text{th}},1.1]  & 0 &  3.4 (1.0)[^{+ 0.07}_{- 0.07}]\% & 2.4 [^{+ 0.3}_{- 0.3}]/2 & 30.3 [^{+ 4.5}_{- 4.4}]\%     & 1.44 \\  
    3 & [\sqrt{s_\text{th}},1.1]  & 3 &  3.5 (1.0)[^{+ 0.04}_{- 0.04}]\% & 3.5 [^{+ 0.3}_{- 0.3}]/2 & 17.8 [^{+ 3.8}_{- 2.8}]\%     & 1.49 \\  
    3 & [1.1,\infty[              & 2 &  3.5 (1.4)[^{+ 0.2}_{- 0.2}]\% & 13.0 [^{+ 2.9}_{- 2.0}]/2 & 0.15 [^{+ 0.27}_{- 0.12}]\%    & 8.14 \\  
    3 & [1.1,\infty[              & 3 &  3.7 (1.4)[^{+ 0.1}_{- 0.1}]\% & 17.1 [^{+ 1.9}_{- 1.6}]/2 & 0.019 [^{+ 0.027}_{- 0.014}]\% & 8.70 \\ 
    \hline
    \hline
    2 & [\sqrt{s_\text{th}},1.8]  & 0 &  2.9 (0.8)[^{+ 0.1}_{- 0.1}]\% & 1.7 [^{+ 0.3}_{- 0.2}]/1 & 19.8 [^{+ 3.4}_{- 3.2}]\%       & 1.63 \\ 
    2 & [\sqrt{s_\text{th}},1.8]  & 3 &  3.1 (0.9)[^{+ 0.03}_{- 0.03}]\% & 2.5 [^{+ 0.2}_{- 0.3}]/1 & 11.3 [^{+ 2.4}_{- 1.8}]\%     & 1.69 \\  
    2 & [1.8,\infty[              & 0 & 31.8 (9.1)[^{+ 0.6}_{- 0.6}]\% & 1.5 [^{+ 0.2}_{- 0.2}]/1 & 21.4 [^{+ 3.4}_{- 3.2}]\%       & 70.17 \\  
    2 & [1.8,\infty[              & 3 & 32.9 (9.2)[^{+ 0.4}_{- 0.3}]\% & 2.3 [^{+ 0.2}_{- 0.2}]/1 & 12.8 [^{+ 2.5}_{- 1.8}]\%       & 72.62 \\  
    \hline
    3 & [\sqrt{s_\text{th}},1.8]  & 0 &  3.0 (0.9)[^{+ 0.05}_{- 0.05}]\% & 1.7 [^{+ 0.2}_{- 0.2}]/2 & 43.7 [^{+ 4.9}_{- 5.2}]\%          & 1.65 \\  
    3 & [\sqrt{s_\text{th}},1.8]  & 3 &  3.1 (0.9)[^{+ 0.03}_{- 0.03}]\% & 2.5 [^{+ 0.2}_{- 0.3}]/2 & 28.6 [^{+ 4.5}_{- 3.5}]\%          & 1.70 \\  
    3 & [1.8,\infty[              & 2 &  3.5 (1.7)[^{+ 0.2}_{- 0.1}]\% & 15.1 [^{+ 3.6}_{- 2.4}]/2 & 0.052 [^{+ 0.130}_{- 0.046}]\%      & 7.79 \\  
    3 & [1.8,\infty[              & 3 &  3.7 (1.7)[^{+ 0.08}_{- 0.08}]\% & 20.3 [^{+ 2.4}_{- 2.0}]/2 & 0.0039 [^{+ 0.0081}_{- 0.0034}]\% & 8.18 \\  
    \hline
    \hline
    2 & [\sqrt{s_\text{th}},3.0]  & 0 &  2.8 (0.8)[^{+ 0.06}_{- 0.06}]\% & 1.5 [^{+ 0.2}_{- 0.2}]/1 & 22.0 [^{+ 3.4}_{- 3.2}]\%      & 2.03 \\  
    2 & [\sqrt{s_\text{th}},3.0]  & 3 &  2.9 (0.8)[^{+ 0.03}_{- 0.03}]\% & 2.3 [^{+ 0.2}_{- 0.2}]/1 & 13.2 [^{+ 2.5}_{- 1.9}]\%      & 2.10 \\  
    2 & [3.0,\infty[              & 0 & 70.5 (22.4)[^{+ 3.6}_{- 3.2}]\% & 7.8 [^{+ 1.7}_{- 1.4}]/1 & 0.51 [^{+ 0.66}_{- 0.35}]\%     & 143.42 \\  
    2 & [3.0,\infty[              & 3 & 76.9 (23.9)[^{+ 2.4}_{- 2.2}]\% & 13.4 [^{+ 1.6}_{- 1.5}]/1 & 0.025 [^{+ 0.052}_{- 0.024}]\% & 156.38 \\  
    \hline
    3 & [\sqrt{s_\text{th}},3.0]  & 0 &  2.7 (0.8)[^{+ 0.05}_{- 0.05}]\% & 1.7 [^{+ 0.3}_{- 0.2}]/2 & 43.1 [^{+ 5.8}_{- 6.1}]\%          & 1.97 \\  
    3 & [\sqrt{s_\text{th}},3.0]  & 3 &  2.8 (0.8)[^{+ 0.03}_{- 0.03}]\% & 2.7 [^{+ 0.3}_{- 0.3}]/2 & 26.3 [^{+ 4.3}_{- 3.7}]\%          & 2.04 \\  
    3 & [3.0,\infty[              & 2 &  4.2 (2.4)[^{+ 0.09}_{- 0.08}]\% & 16.0 [^{+ 3.9}_{- 2.6}]/2 & 0.033 [^{+ 0.094}_{- 0.030}]\%    & 8.59 \\  
    3 & [3.0,\infty[              & 3 &  4.3 (2.4)[^{+ 0.06}_{- 0.05}]\% & 21.7 [^{+ 2.7}_{- 2.2}]/2 & 0.0020 [^{+ 0.0049}_{- 0.0018}]\% & 8.80 \\  
    \hline
    \hline
    2 & [0.63,0.92]               & 0 &  4.8 (1.4)[^{+ 0.1}_{- 0.1}]\% & 1.7 [^{+ 0.3}_{- 0.2}]/1 & 19.6 [^{+ 3.4}_{- 3.2}]\%                    & 1.42 \\  
    2 & [0.63,0.92]               & 3 &  4.9 (1.4)[^{+ 0.06}_{- 0.05}]\% & 2.5 [^{+ 0.2}_{- 0.3}]/1 & 11.2 [^{+ 2.4}_{- 1.8}]\%                  & 1.47 \\  
    2 & [\sqrt{s_\text{th}},0.63]\cup[0.92,\infty[ & 0 &  6.2 (1.8)[^{+ 0.1}_{- 0.1}]\% & 1.6 [^{+ 0.2}_{- 0.2}]/1 & 20.4 [^{+ 3.4}_{- 3.2}]\%   & 15.33 \\  
    2 & [\sqrt{s_\text{th}},0.63]\cup[0.92,\infty[ & 3 &  6.5 (1.8)[^{+ 0.07}_{- 0.07}]\% & 2.4 [^{+ 0.2}_{- 0.2}]/1 & 11.9 [^{+ 2.4}_{- 1.8}]\% & 15.88 \\ 
    \hline
    3 & [0.63,0.92]               & 0 &  4.9 (1.4)[^{+ 0.08}_{- 0.08}]\% & 1.8 [^{+ 0.2}_{- 0.2}]/2 & 40.2 [^{+ 4.0}_{- 4.1}]\%                     & 1.45 \\  
    3 & [0.63,0.92]               & 3 &  5.0 (1.4)[^{+ 0.05}_{- 0.05}]\% & 2.6 [^{+ 0.2}_{- 0.2}]/2 & 27.9 [^{+ 4.0}_{- 3.1}]\%                     & 1.50 \\  
    3 & [\sqrt{s_\text{th}},0.63]\cup[0.92,\infty[ & 0 &  3.4 (1.1)[^{+ 0.2}_{- 0.1}]\% & 9.0 [^{+ 1.8}_{- 1.3}]/2 & 1.1 [^{+ 1.1}_{- 0.7}]\%       & 8.30 \\  
    3 & [\sqrt{s_\text{th}},0.63]\cup[0.92,\infty[ & 3 &  3.6 (1.1)[^{+ 0.08}_{- 0.08}]\% & 12.4 [^{+ 1.3}_{- 1.1}]/2 & 0.21 [^{+ 0.18}_{- 0.12}]\% & 8.80 \\  
    \hline\hline
    \end{tabular}
    }
\end{table*}

\begin{figure*}[tb]
\centerline{\includegraphics[width=0.85\linewidth]{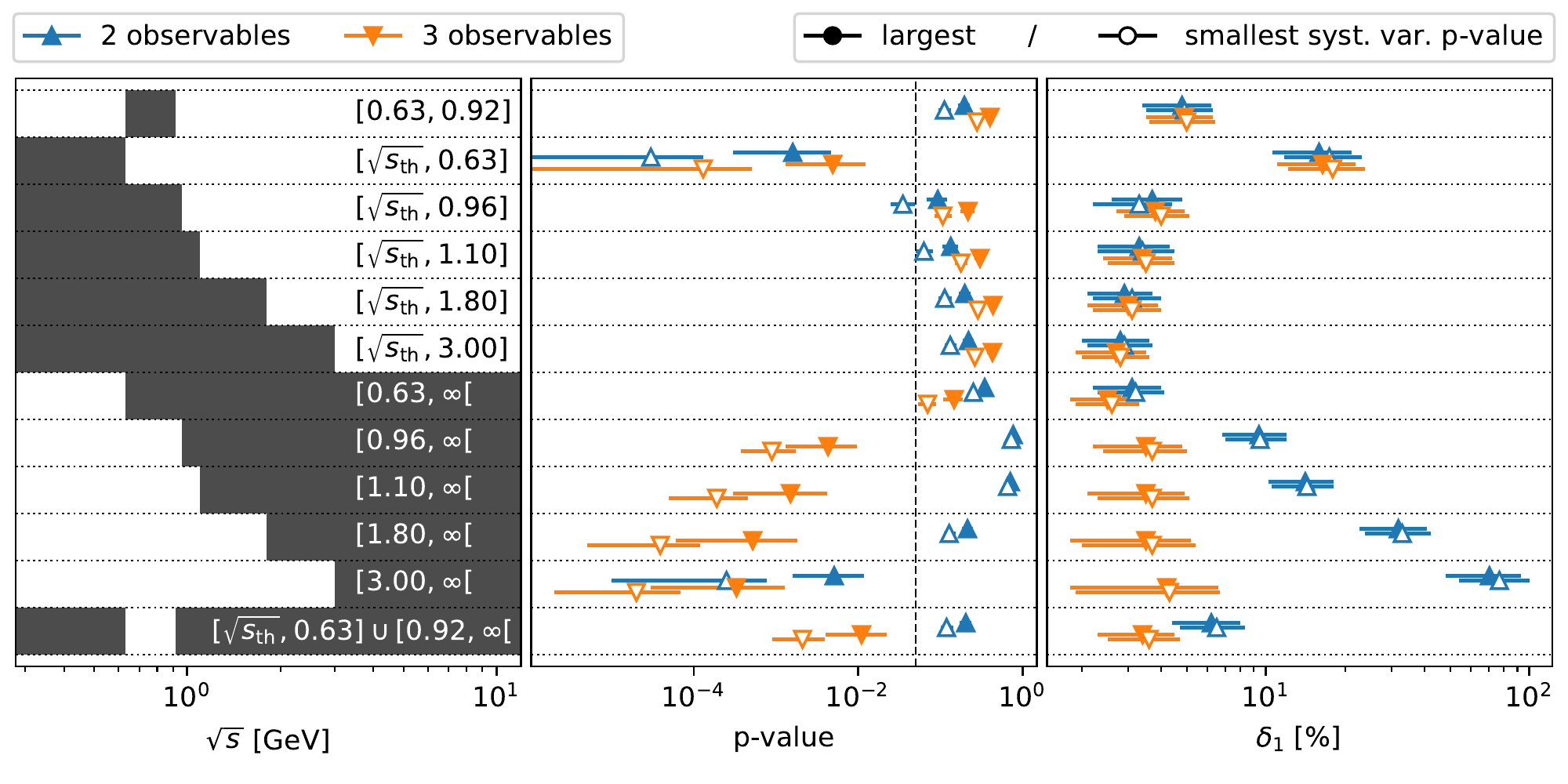}}
\caption{\label{fig:tab2_fig}
Plot illustrating \tab{tab:rescalingAverage}. The first panel displays, in dark grey, the various $\sqrt{s}$-intervals considered here. The first row corresponds to the $\rho$-peak. In the second panel, we plot the $p$-values that indicate the compatibility of the rescaling hypothesis, in the given interval, with the lattice and data-driven results for $\amulohvp$ and $\amulohvpwin$ (2 observables, blue triangles) and with the results for the additional observable $\ddalpha$ (3 observables, orange upside-down triangles). These $p$-values are obtained from the $\chi^2$ defined in \eq{eq:chisqAveragefit}, using the appropriate $\ndof$. The error bars are the statistical uncertainties resulting from those on the lattice covariance matrices determined in \app{sec:lat_obs} (see also footnote~\ref{foot:lat_stat}). The corresponding uncertainty distributions are plotted in \figs{Fig:ChiSqpValue_0_2_0}{Fig:ChiSqpValue_1_2_3} (for the third and eleventh row of this plot respectively). The filled/empty points correspond to the largest/smallest $p$-value obtained from the four systematic variations of the lattice covariance matrix of Eqs.~\ref{eq:latcov1}-\ref{eq:latcov4}. The vertical dashed line corresponds to 5\%. We consider solutions with $p$-values above and close to that line to be compatible with the rescaling hypothesis in the corresponding interval, and those far below to be incompatible. The rightmost panel displays the rescaling percentages, $\delta_1$, corresponding to the $p$-values on the same row that have the same plot symbol. Here the uncertainties are the leading ones, given by the nominal values of the lattice and R-ratio covariance matrices, i.e.\ not including the small, additional uncertainty associated with the statistical uncertainty on the lattice covariance matrices.
}
\end{figure*}

In this section we present the results obtained when comparing the lattice QCD and dispersive results, employing the methodology discussed in \sec{sec:testing_Rratio} and \app{Appendix:Chi2Methodology}.
With this methodology, we determine the size of modifications to the experimental R-ratio that lead to a rescaling of the contributions, from a given $\sqrt{s}$-interval, to the observables of interest.
As in the previous subsection, the study is performed for either two moment integrals~($\amulohvp$ and $\amulohvpwin$) or three of them~(i.e.\ including, in addition, $\ddalpha$).
Results are given for various scenarios, i.e.~$\sqrt{s}$ regions where the dispersive integrals are computed and the rescaling factors are determined.
These include $\sqrt{s}$ regions below or above $0.63$, $0.96$, $1.1$, $1.8$ and $3.0$~GeV respectively, as well as rescaling inside or outside the $[0.63,0.92]$~GeV interval.
While the first set of intervals covers various low- and high-mass regions, with different contributions to the dispersive integrals and different methodologies used for deriving them, the last one the region dominated by the $\rho$-resonance peak and the region complementary to it.
The intervals are displayed in the left panel of \fig{fig:tab2_fig}.
All uncertainties and correlations are accounted for, as discussed in the previous subsection.
There we needed those associated with the moment integrals of the R-ratio over the full range of $\sqrt{s}$.
Here we need those corresponding to the contributions to moment integrals from the various $\sqrt{s}$-intervals, whose determinations are also described in \app{sec:R-ratio_obs}.

\tab{tab:rescalingAverage} and \fig{fig:tab2_fig} also present the results for the rescaling factor, $\gamma_1 \equiv 1+\delta_1$, for the various $\sqrt{s}$ regions $I_1$ discussed in the previous paragraph, using either two or three moment integrals.
The results are determined via the averages defined by the $\chi^2$ function of \eq{eq:chisqAveragefit}, in which $\gamma$ is replaced by $\gamma_1$ and with the weights of the input $\tilde{\gamma}_j$ values obtained by setting all correlations to zero, for reasons discussed in \app{Appendix:alternate_averaging}.
Out of the four systematic uncertainty variations of the lattice covariance matrices discussed in \app{Sec:UncOnLattCov}, results are presented here for the two which induce the smallest/largest $\chi^2$ values in the fit for each of the intervals.
Indeed, depending on the fit configuration, the smallest $\chi^2$ values are generally obtained for matrices ``$0$'' and ``$2$'', while the largest ones for matrix ``$3$''.

More specifically, \tab{tab:rescalingAverage} and \fig{fig:tab2_fig} report the {\it nominal} results~(obtained using the {\it nominal} lattice covariance matrices) for: the rescaling percentage $\delta_1$ with its uncertainty propagated from the covariance matrices of the lattice QCD and dispersive results~\footnote{It has also been checked that non-linear effects in the uncertainty propagation are small, by comparing the outcome of the uncertainty propagation when performing $+1\sigma$ variations, on the one hand, and $-1\sigma$ variations, on the other.}
in \tab{tab:rescalingAverage}; the $\chi^2/\ndof$ obtained by injecting $\gamma_1$ into \eq{eq:chisqAveragefit}; the corresponding $p$-value.

Using either two or three moment integrals, the rescaling of the integrals of the R-ratio by a common factor, in the $\sqrt{s}$ regions below $0.96$, $1.1$, $1.8$ or $3.0$~GeV, allows a good agreement between the lattice and R-ratio approaches.
In those scenarios, the rescalings range from approximately 2 to 5\%, the lower number corresponding to the larger intervals.
In addition, a rescaling by around 5\%\ in the $\rho$-peak interval, $[0.63,0.92]$~GeV, also restores the agreement.
However, the tension persists when the rescaling is performed in the region below $0.63$~GeV.

When using only the two observables, $\amulohvp$ and $\amulohvpwin$, agreement is also restored through rescalings above either $0.63$, $0.96$, $1.1$, $1.8$~GeV, or outside the $[0.63,0.92]$~GeV interval, while tensions persist for a rescaling performed above $3.0$~GeV.

Adding the third observable, $\ddalpha$, increases the constraints and tensions arise for a rescaling in any of the intervals that do not include the $\rho$-peak. 

It is worth noting that the values of the rescaling percentage $\delta_1$ for the various configurations presented in \tab{tab:rescalingAverage} are much larger than the (sub-)percent-level uncertainties of the experimental R-ratio in the corresponding $\sqrt{s}$ regions.
Therefore, such shifts are not to be interpreted as a plausible way of solving the tension between lattice QCD and the dispersive approaches, but rather as a way of comparing various hypotheses for the possible source(s) of the tension.

\FigChiSqpValue{in the low}{0}{{below $0.96\,\gev$}}{2moments0,96GeV}{2}{Two}{0}

\FigAverageSigma{in the low}{0}{{below $0.96\,\gev$}}{2moments0,96GeV}{2}{Two}{0}

\FigChiSqpValue{in the high}{1}{{above $3.0\,\gev$}}{2moments3,0GeV}{2}{Two}{3}

\FigAverageSigma{in the high}{1}{{above $3.0\,\gev$}}{2moments3,0GeV}{2}{Two}{3}

Also given in \tab{tab:rescalingAverage} are the $\pm 1 \sigma$ statistical quantiles obtained for $\delta_1$, for the $\chi^2/\ndof$ and for the corresponding $p$-value, when considering their distributions obtained from the bootstrap replicas of the lattice covariance matrices computed in \app{Appendix:alternate_averaging} (see also footnote~\ref{foot:lat_stat}). 
Examples of such distributions are shown in 
Figs.~\ref{Fig:ChiSqpValue_0_2_0}--\ref{Fig:AverageSigma_1_2_3}.
In the case of \fig{fig:tab2_fig}, which summarizes our results, those uncertainties are shown only for the $p$-values.

For $\delta_1$, in many scenarios the $\pm 1 \sigma$ quantiles based on the lattice bootstrap replicas  are smaller than the primary uncertainty propagated from the covariance matrices of the lattice QCD and dispersive results, by about an order of magnitude or more.
The changes associated with the four systematic variations of the lattice covariance matrix, are also small, though somewhat larger than the bootstrap ones.
This is not necessarily surprising because they are uncertainties on the primary uncertainty.
In fact, the latter are of only a few percent.
Furthermore, the same statistical and systematic uncertainties on the $\chi^2/\ndof$ and on the corresponding $p$-values do not change the fact that a given fit can be considered ``good" or ``bad".
These observations indicate that the statistical and systematic variations of the lattice covariance matrix do not change the conclusions qualitatively.

As mentioned in the previous subsection, the same bootstrap resampling can present some slight differences between the mean, median and nominal values of the corresponding distributions.
In particular, the asymmetric, non-Gaussian tails~(present especially for the $p$-values) impact the mean values~(see e.g.\ \figs{Fig:ChiSqpValue_0_2_0}{Fig:ChiSqpValue_1_2_3}).
Still, these differences are well within the $\pm 2\sigma$ quantiles~(most often within the $\pm 1\sigma$ quantiles) and do not impact the conclusions of the study.
The asymmetric tails also induce some differences between the uncertainties obtained from asymmetric variances~\footnote{These variances are computed from the values in the distribution that are above and below the nominal one.}, those from the $\pm 1\sigma$ quantiles and those from the $\pm 2\sigma$ ones (divided by 2).
However, here also the conclusions of the study are not impacted by the slight differences among these three estimates of uncertainties.

In order to display the implications of observed normalization shifts for EW precision observable (EWPO) fits,
in \tab{tab:rescalingAverage} we indicate the impact of the $\delta_1$ rescaling on \DaHadZfive.~\footnote{While the uncertainty of the $\delta_1$ rescaling factors is determined from the fits, we do not attempt a correlated propagation to (the uncertainty of) the corresponding shift of \DaHadZfive, nor a simple rescaling of the uncertainty of the latter. This follows the rescaling approach discussed in Ref.~\cite{Malaescu:2020zuc}. There it was pointed out that, if such rescaling is necessary for the nominal values of the hadronic spectra, it is not obvious how the corresponding uncertainties should be computed.}
As expected, the largest impacts on \DaHadZfive are observed for modifications of the R-ratio in the high-$\sqrt{s}$ intervals.
However, while such modifications may have reasonable $p$-values with constraints from only $\amulohvp$ and $\amulohvpwin$, when one also includes the hadronic contribution to the five-flavor running of the electromagnetic coupling, between $-10$ and $-1~{\rm GeV}^2$, the fit quality becomes very poor and the common rescaling of the contributions to the observables, in these high-$\sqrt{s}$ intervals, is strongly reduced too.
Thus, in the presence of the $\ddalpha$ constraint, the corresponding shifts on $ $ will not significantly impact the conclusions of EWPO fits.

As they are for the comparisons of \sec{sec:testing_lattice_results}, the conclusions of the studies performed here are stable against the statistical and systematic variations of the lattice covariance matrix. 
As shown in \app{Appendix:alternate_averaging}, they are also stable with respect to the averaging methodology that is employed.
Moreover, very similar results are obtained using the $\chi^2$ approach to rescaling described in \app{sec:multiple_rescaling_fits}.

\section{Summary of results and conclusions}
\label{sec:conclusion}

We have presented a framework that enables a comparison of the primary ingredients that are used for determining HVP in the lattice QCD and data-driven approaches.
The framework makes use of observables computable in both formalisms.
These primary ingredients are: for the lattice approach, the zero three-momentum, current-current correlator of \eq{eq:correlator}, studied as a function of Euclidean time, $t$; for the data-driven approach, the experimental R-ratio studied as a function of timelike c.o.m.\ energy, $\sqrt{s}$ (\eq{eq:R-ratio}).
The observables are integrals over $t$ and $\sqrt{s}$, weighted with appropriately chosen kernels. We alternatively call these observables or moment integrals.
 
The kernels can be chosen to be localized on the $t$-axis, as are the $\amulohvp$ time windows proposed in Ref.~\cite{Blum:2018mom}.
Then by comparing the corresponding observables obtained in the two approaches, one can aim to isolate distance scales in the lattice correlator that may lie at the origin of the observed disagreement in the values, for instance, of $\amulohvp$ and $\amulohvpwin$.
Such comparisons are also interesting for HVP observables that have limited support in photon virtuality.
These are typically related to the Adler function and the hadronic running of the electromagnetic coupling.
They provide a complementary way to isolate relevant scales in the lattice correlator.
More generally, our method applies to any observable related to HVP.

Our framework also allows us to gain information about the R-ratio in chosen regions of c.o.m.\ energy, from these same observables.
The main challenge here comes from the fact that this requires solving a notoriously ill-posed, inverse problem, that arises because the lattice correlator is a Laplace transform of the R-ratio (\eq{eq:R-ratio}).
It also comes from the fact that observables which are localised in Euclidean time are much less so in c.o.m.\ energy.
We solve the second problem by partitioning the $\sqrt{s}$-axis from threshold to infinity into intervals, and computing the contributions of each such interval to the observables of interest.
The first challenge --- the reconstruction of a continuous function from a finite number of its weighted integrals does not have, a priori, a unique solution --- is addressed by limiting the information that we try to extract while monitoring the possible pitfalls associated with such a reconstruction.

In the present work, we focus on simple scenarios. 
We consider three observables related to HVP: $\amulohvp$,  $\amulohvpwin$ and the hadronic running of the electromagnetic coupling between spacelike virtualities $1\,\gev^2$ and $10\,\gev^2$, $\ddalpha$. 
This choice is motivated by the fact that these quantities have been determined on the lattice to the sub-percent level, with all relevant corrections \cite{Borsanyi:2020mff}. 
It is important to note that the result for $\ddalpha$ is not the main focus of Ref.~\cite{Borsanyi:2020mff} and should be understood as preliminary.
Nevertheless, we include it because it imposes complementary constraints on the spectral function and demonstrates the flexibility of our approach.

Then, we use these observables to:
\begin{itemize}
    \item [I.] isolate regions of $t$ in which the lattice correlator leads to values of the observables that are inconsistent with those obtained in the data-driven approach;
    \item[II.] determine intervals of $\sqrt{s}$ in which a change in the experimental R-ratio could explain the observed discrepancies between the lattice and R-ratio approaches, by allowing a common rescaling of the the contributions from these intervals to the observables of interest.
\end{itemize}
Because our approach is based on the minimization of $\chi^2$ functions, it allows us to quantify the compatibility of the observables in the two approaches, as well as the consistency of the observables with our hypothesis that a change of the experimental R-ratio, in a particular $\sqrt{s}$-interval, could explain the observed discrepancies.
Clearly, the very simple model of item II tells us little about the shape of the deformations of the timelike spectral function that may be required.
Moreover, it does not incorporate, {\em a priori}, physical constraints, such as those imposed by chiral perturbation theory at small c.o.m.\ energies, by perturbative QCD at large ones, or by analyticity and positivity at all energies.
However, it is not necessarily incompatible with some of them either.
In addition, it can help isolate $\sqrt{s}$-intervals in which the experimental R-ratio may have to be modified to explain the lattice results.

Within a given approach, the observables obtained are not independent. 
Thus, it is critical to take into account correlations between them. 
This is necessary to correctly quantify the compatibility of the lattice and data-driven observables with one another. 
It is also particularly important when considering the inverse problem, because its solutions are very sensitive to these correlations. 
In fact, because of the exponential enhancement of fluctuations, it is also important to consider the stability of the results against uncertainties on the corresponding covariance matrices. 
We take into account all uncertainties and correlations and, for the lattice results, we also consider the effect of uncertainties on the corresponding covariance matrices.
The consideration of such an effect, in the data-driven approach, is left for future work.

Using the methods described in \sec{sec:testing_lat} we find the results presented and discussed in \sec{sec:testing_lattice_results}. 
They represent our attempt to isolate regions in the lattice correlator, on the Euclidean time axis, that do not agree with the data-driven approach.

As already observed in Ref.~\cite{Borsanyi:2020mff}, the lattice result for $\amulohvp$ is $2.1\sigma$ above the one obtained from the data-driven approach (corresponding to a $p$-value of $5\times 10^{-2}$), the one for $\amulohvpwin$, $3.8\sigma$ above ($p$-value of $1\times 10^{-4}$), while the preliminary lattice result for $\ddalpha$ is only $1.4\sigma$ above ($p$-value of $0.15$). 
Moreover, the lattice result for the leading, isospin-limit, up-and-down quark contribution to the window observable has been independently confirmed by nine collaborations \cite{Borsanyi:2020mff,Ce:2022kxy,ExtendedTwistedMass:2022jpw,Blum:2023qou,Bazavov:2023has}, of which two also confirm the results for other contributions \cite{Ce:2022kxy,Blum:2023qou}.
This increases the tensions with the data-driven results to a conservative $4\sigma$.

The discrepancies discussed in the previous paragraph are confirmed when correlations are taken into account via \eq{eq:chisqwin}.
Considering first $\amulohvp$ and $\amulohvpwin$ together we find $p$-values which range approximately from $2\times 10^{-5}$ to $1.7\times 10^{-3}$, taking into account the maximum systematic variation and $1\sigma$ statistical fluctuation in the lattice covariance matrices.
These are substantially smaller than the one for $\amulohvp$ alone.

Adding to the comparison the preliminary lattice result for $\ddalpha$, and its data-driven counterpart, reduces the overall tension to some degree.
The $p$-values now range from approximately $10^{-4}$ to $6\times 10^{-3}$, still making an overall agreement very unlikely.

Combing all of this information with the shape of the observable's kernels, in \sec{sec:testing_lattice_results} we argue that our results point to an excess of the lattice correlator in the intermediate-time range from $0.4$ to $1.5\,\fm$, a probable excess at time separations above $1.5\,\fm$ and a possible suppression in the short-time range below $0.4\,\fm$.

Then, we turn to the implications that the comparison of the two approaches may have on the experimental R-ratio.
Again, the tests considered take into account all systematic errors and correlations, as well as uncertainties on uncertainties in the lattice results.

Using the averaging procedure of \sec{sec:testing_Rratio}, we find that the lattice predictions, for $\amulohvp$ and the corresponding $\amulohvpwin$ (but excluding $\ddalpha$), can be reproduced within uncertainties via a modification of the experimental R-ratio in various intervals of $\sqrt{s}$.
More specifically, if the lower end of these intervals begins at threshold, the intervals must include the $\rho$ peak.
Indeed, for intervals whose upper bound is anywhere between $\sqrt{s}=0.96$ to $3.0\,\gev$, the modifications have to be such that they increase the contributions to both the data-driven $\amulohvp$ and $\amulohvpwin$ in those intervals by approximately $2$ to $5\%$, the lower value corresponding to the larger intervals.
These conclusions allow scenarios, such as those advocated in \cite{Colangelo:2022xfy}, in which the R-ratio would have to receive modifications above $1\,\gev$ to explain the lattice results for $\amulohvp$ and $\amulohvpwin$.

Interestingly, a modification of only the $\rho$-peak region is also compatible with the lattice results, but requires around a 5\% increase of the contributions to the two observables from that region.
It is worth noting that such an increase is comparable to the one claimed by CMD-3 in a very similar region \cite{CMD-3:2023alj}. 
However, without having worked out the correlations between CMD-3 and the other $\epemtohad$ measurements, it is impossible to make a more quantitative statement.
Moreover, in the presence of large, unexplained discrepancies between measurements, and thus of large unknown systematics, one cannot claim to control such correlations and uncertainties.

Somewhat surprisingly, a scenario in which the experimental R-ratio is modified in an interval from threshold to infinity from which the $\rho$ peak is excluded is also permitted, with $p$-values above $\sim 10\%$ in the two-observable case--i.e.\ beginning right below the $c\bar c$ resonance region.
Moreover, modifications of the experimental R-ratio, in intervals that extend to infinity and begin at $\sqrt{s}$ as small as $0.63\,\gev$ and as large as $1.8\,\gev$, are also allowed.
In that case, the contributions to the data-driven $\amulohvp$ and  $\amulohvpwin$ must be increased anywhere between approximately $2\%$ to $40\%$.
On the other hand, an explanation of the discrepancy between the two approaches via a modification of the experimental R-ratio only above $3\,\gev$ is excluded with high probability.

In order to see if some of the above scenarios can be eliminated, we consider adding the observable $\ddalpha$.
This observable has a footprint along the $\sqrt{s}$-axis that is significantly more extended towards larger c.o.m.\ energies than those of $\amulohvp$ and $\amulohvpwin$.
We find that the addition of this observable eliminates, with significant probability, all scenarios in which the modification of the experimental R-ratio excludes the $\rho$ peak.
Nevertheless, the scenario including only the $\rho$-peak interval, and the one in which the interval begins below the $\rho$ peak and extends to infinity, both survive this additional constraint.
Moreover, these possible modifications of the experimental R-ratio imply a rescaling of the data-driven results, for all three observables, by a common factor between about $1.02$ and $1.06$.
In particular, the scenario including only the $\rho$ peak is barely changed by the addition of the constraint from $\ddalpha$, making it a robust explanation for the discrepancy between the lattice and data-driven approaches.

Not surprisingly, very similar conclusions are obtained using the fitted-moment formalism given in \app{sec:multiple_rescaling_fits}. 
There, however, the natural interpretation of the results is slightly different.
In that approach, the contributions of different $\sqrt{s}$-intervals to the various observables become fit parameters, as are the rescaling factors.
Thus, under the assumption that the proposed rescaling scenario is correct, as should also be verified by the corresponding $\chi^2/\ndof$, the fitted contributions to the observables can be understood as a combination of lattice and data-driven results for these contributions.

It is important to remember that all of the modifications to the experimental R-ratio discussed above are not consistent with the uncertainties of the timelike spectrum.
This is made obvious by the very small $p$-values, discussed above, that correspond to the minimum of the $\chi^2$ of \eq{eq:chisqwin}, which measures the combined compatibility, of the lattice and data-driven results, for $\amulohvp$, $\amulohvpwin$ and $\ddalpha$.

As discussed in \app{app:beyond_rescaling}, the methods presented above can be generalized to include more and more information about HVP from the lattice and data-driven approaches.
This additional information will allow us to refine our understanding of the discrepancies between the two approaches, both in terms of the Euclidean correlation function and the experimental R-ratio.
Our methods can also be generalized to include more theoretical constraints, such as those discussed earlier in this section.
Moreover, the methods presented in \app{app:beyond_rescaling} supply means of obtaining combinations of lattice and data-driven results for various quantities of interest related to HVP.
These combinations will only be as valid as the model used to reconcile the two approaches is well motivated.

Finally, in \app{app:more_moments} we argue that the number of linearly independent moments that can be obtained with reasonable precision, from the experimental R-ratio, is less than 10 for the types of observables considered here, which should be computable with sub-percent precision in the lattice approach.

\section*{Acknowledgements}

MD, BM and ZZ acknowledge their fruitful collaboration with Andreas Hoecker. The authors also warmly thank Jérôme Charles and Eduardo de Rafael for enlightening discussions. This work received funding from the French National Research Agency under contract ANR-22-CE31-0011, from the Excellence Initiative of Aix-Marseille University -- A*Midex, a French ``Investissements d’Avenir" program under grant AMX-18-ACE-005, and from grants ERC-MUON, NW21-024-A and BMBF-05P21PXFCA. The authors gratefully acknowledge the Gauss Centre for Supercomputing (GCS) e.V. (www.gauss-centre.eu) and GENCI (www.genci.fr/en) (grant 502275) for providing computer time on the GCS supercomputers SuperMUC-NG at Leibniz Supercomputing Centre in München, HAWK at the High Performance Computing Center in Stuttgart and JUWELS at Forschungszentrum Jülich, as well as on the GENCI supercomputers Joliot-Curie/Irène Rome at TGCC and Jean-Zay V100 at IDRIS. 

\appendix

\section{Data-driven determination of the observables and their correlations}
\label{sec:R-ratio_obs}

Computing the contributions to the moments integrals from the low energy region, below $1.8$~GeV, requires combining the spectra measured by various experiments.
These measurements are performed either through a scan of the energy in the centre-of-mass of the collider, as is the case of e.g.\ the CMD-2 and SND experiments, or using the ISR technique, in the case of KLOE and BABAR (see e.g.\ Refs.~\cite{Aulchenko:2006dxz,CMD-2:2006gxt,Achasov:2006vp,KLOE:2008fmq,KLOE:2010qei,KLOE:2012anl,BaBar:2009wpw,BaBar:2012bdw} for the $\pi^+\pi^-$ channel, dominant in the case of $\amulohvp$).
In the latter case, selecting events with a hard photon emitted from the initial state allows such studies to cover a broad range of invariant masses for the hadronic final state system.
Actually, BABAR covers the full energy range of interest, due to selecting events with the hard photon reconstructed in the detector and to the high energy in the centre-of-mass of the collider.
While the KLOE $'08$~\cite{KLOE:2008fmq} and KLOE $'12$~\cite{KLOE:2012anl} measurements consider events with the hard ISR photon along the beam, the KLOE $'10$ measurement uses large-angle ISR photons~\cite{KLOE:2010qei}.
Furthermore, in measurements where the ISR luminosity is evaluated in-situ based on $\epem \rightarrow \mu^+\mu^-$ events~\cite{BaBar:2009wpw,BaBar:2012bdw,KLOE:2012anl}, several systematic uncertainties are reduced.
Still, among the currently available results, only the BABAR study uses the reconstructed extra photons~(performing a measurement reconstructed at ``NLO'' and inclusive of any number of additional photons), necessary at the sub-percent level precision aimed for here.
The differences between the sampled event topologies, the detector technologies, and especially the analysis methodologies, result in systematic tensions between the BABAR and KLOE measurements~\cite{Davier:2017zfy,Davier:2019can}, but also between the KLOE measurements themselves~(see Ref.~\cite{Aoyama:2020ynm} and references therein).
Since these tensions are still to be understood and fixed, they require special care in the current combinations.

The channel-by-channel combinations in the DHMZ approach~(implemented in the HVPTools software)~\cite{Davier:2010rnx,Davier:2010nc,Davier:2017zfy,Davier:2019can} are performed in fine energy ranges, using a spline-based procedure.
It takes into account the information on the uncertainties and their correlations between bins/points and between experiments.
The DHMZ procedure also accounts for correlations between the various different exclusive channels, when summing their contributions, which induces a necessary enhancement of the total uncertainty.
Furthermore, the local tensions between the measurements are taken into account based on the computation of a $\chi^2/\ndof$ in fine energy ranges, used for the rescaling of the uncertainties.
However, in the presence of systematic tensions, as the ones between the BABAR and the KLOE measurements in the $\pi\pi$ channel, the local rescaling of the uncertainties is not sufficient.
For this reason, in the most recent DHMZ study~\cite{Davier:2019can}, these tensions are taken into account by treating half of the difference between integrals without either BABAR or KLOE as an extra uncertainty.
This yields the dominant uncertainty on $\amulohvp$, amounting to $2.8 \cdot 10^{-10}$.
The DHMZ approaches for accounting for correlations between channels and for the BABAR--KLOE tension in the $\pi\pi$ channel, have also been adopted for the result of the Theory Initiative White Paper~\cite{Aoyama:2020ynm}.

The procedure described above is used to derive the experimental values of ${\rm R}_{e^+e^- \to \textrm{hadrons}}({\rm s})$, together with its covariance matrix, up to $1.8$~GeV.
It is completed with contributions from inclusive experimental measurements in the range between $3.7$ and $5$~GeV, combined using the same methodology as above, with contributions from narrow resonances, as well as from perturbative QCD for the continuum~\cite{Davier:2019can}.
An uncertainty accounting for possible duality violation effects is also included.
A direct integration of the resulting ${\rm R}_{e^+e^- \to \textrm{hadrons}}({\rm s})$ on the full mass range, together with a linear propagation of the uncertainties with their correlations, yields the moment integrals with the kernels of interest here, together with their covariance matrix.

\section{Determination of lattice observables and of their correlations}
\label{sec:lat_obs}
The lattice QCD results used in this work were obtained in Ref.~\cite{Borsanyi:2020mff}.
In order to calculate the two-point function $C(t)$ defined in \eq{eq:correlator}, we need to evaluate the current correlator $\left\langle J_{\mu}(x)J_{\bar{\mu}}(\bar{x})\right\rangle$ in our lattice regularization.
We define the regularized current correlator as the second differential of the partition function with respect to an external photon field
\begin{equation}
    \left\langle J_{\mu}(x)J_{\bar{\mu}}(\bar{x})\right\rangle \equiv \frac{\delta^2 \log Z}{\delta A_{\mu,x}^\mathrm{ext} \delta A_{\bar{\mu},\bar{x}}^\mathrm{ext}} {\Bigg|}_{A^\mathrm{ext}=0}\,.
\end{equation}
Thus, we use the conserved current at source and sink.
The correlator can be evaluated as
\begin{align}
    \left\langle J_{\mu}(x)J_{\bar{\mu}}(\bar{x})\right\rangle / e^2 = \biggl\langle\sum_f \,& C^{\mathrm{conn}}_{\mu,x,\bar{\mu},\bar{x}}(m_f, e q_f) \nonumber\\ 
    + \,& C^{\mathrm{disc}}_{\mu,x,\bar{\mu},\bar{x}}\biggr\rangle\,,
\end{align}
plus a contact term that gives no contribution to the observables of interest here. $C^{\mathrm{conn}}_{\mu,x,\bar{\mu},\bar{x}}$ and $C^{\mathrm{disc}}_{\mu,x,\bar{\mu},\bar{x}}$ arise respectively from the connected and disconnected Wick contractions of the quark fields and are given by
\begin{align}
    C^{\mathrm{conn}}_{\mu,x,\bar{\mu},\bar{x}}(m_f, e q_f) &= -\frac{1}{4} \Tr\Big(M_f^{-1} D_\mu \!\left[i P_x V_U e^{ie q_f A}\right]\nonumber\\
    &\qquad\cdot M_f^{-1} D_{\bar{\mu}}\!\left[i P_{\bar{x}} V_U e^{ie q_f A}\right]\Big)\,,
\end{align}
\begin{align}
    C^{\mathrm{disc}}_{\mu,x,\bar{\mu},\bar{x}} &= \sum_{f,\bar{f}} I_{\mu,x}(m_f, e q_f) I_{\bar{\mu},\bar{x}}(m_{\bar{f}}, e q_{\bar{f}})\,,
\end{align}
where $\Tr$ is the spin-color trace, $P_x$ is a projection operator that sets all sites other than $x$ to zero, $M_f^{-1}$ is the propagator for a quark of mass $m_f$ on a gauge background $V_U e^{ie q_f A}$, and
\begin{equation}
    I_{\mu,x}(m_f, e q_f) \equiv \frac{1}{4} \Tr\!\left(M_f^{-1} D_\mu\!\left[i P_x V_U e^{ie q_f A}\right]\right)\,.
\end{equation}

These calculations are performed gauge-ensemble by gauge-ensemble.
The SU(3) configurations are generated at the isospin-symmetric point, where $m_u=m_d$ and $e=0$. Leading-order QED and strong-isospin-breaking (SIB) corrections are then added as perturbations.

The leading QED corrections are obtained as follows. For each SU(3) gauge configuration, a stochastic U(1) photon field is generated with the non-compact, $\text{QED}_L$ action. 
The U(1) fields are exponentiated into link variables that we multiply with the corresponding SU(3) links.
The covariant Dirac operator is then inverted on each of the corresponding gauge configurations and the resulting quark propagators are appropriately contracted.
The first and second derivatives of the contractions, with respect to the unit of electric charge $e$, are obtained by computing the contractions for $+e$ and $-e$ and by taking the appropriate finite differences. 
The expressions for the first and second derivatives of the fermion determinant, in terms of quark propagators, are computed analytically. 
For the second derivative, around 2000 photon fields are used. 

These first and second derivatives are appropriately combined, with the $e=0$ contractions as well, to obtain all $O(e^2)$ contractions relevant for the current correlator.
The desired correlators are then obtained by averaging the various contractions over the SU(3) and U(1) gauge configurations of each ensemble.

At leading order, the fermion determinant does not contribute SIB corrections. The valence contributions are obtained by computing the first derivative of the contractions, with respect to light-quark mass, and then multiplying it by the $u$-and-$d$ quark mass difference. As above, the desired correlators are then obtained by averaging the various contractions over the SU(3) and U(1) gauge configurations of each ensemble.

In order to obtain physically relevant values, we perform global fits to the lattice spacing, quark mass, and electric charge dependence of $\amulohvp$ and $\amuwin{j}$.
These fits take the form
\begin{align}\label{eq:globalfit}
    Y = \,&A + B X_l + C X_s\nonumber\\
    &+ D X_{\delta m} + E X_{vv} + F X_{vs} + G X_{ss}\,,
\end{align}
where the first line corresponds to a definition of the isospin-symmetric limit, and the second line gives the isospin breaking effects.
The $X_l$ and $X_s$
variables describe the deviation from the physical light and strange mass
\begin{align}
    X_l &= \frac{M_{\pi_0}^2}{M_\Omega^2} - {\Bigg[\frac{M_{\pi_0}^2}{M_\Omega^2}\Bigg]}_{\mathrm{ph}}\,,\\
    X_s &= \frac{M_{K_\chi}^2}{M_\Omega^2} - {\Bigg[\frac{M_{K_\chi}^2}{M_\Omega^2}\Bigg]}_{\mathrm{ph}}\,,
\end{align}
where ``ph'' indicates the physical value.
The remaining $X$ variables measure the distance to the isospin-symmetric limit
\begin{align}
    X_{\delta m} &= \frac{\Delta M_K^2}{M_\Omega^2}\,,\\
    X_{vv} &= e_v^2\,,\\
    X_{vs} &= e_v e_s\,,\\
    X_{ss} &= e_s^2\,,
\end{align}
where $e_v$ and $e_s$ are the valence and sea electric charges, respectively.
The meson masses are defined as
\begin{align}
    M_{K_\chi}^2 &\equiv \frac{1}{2} \left(M_{K_0}^2 + M_{K_+}^2 - M_{\pi_+}^2\right)\,,\\
    \Delta M_{K}^2 &\equiv M_{K_0}^2 - M_{K_+}^2\,.
\end{align}

The coefficients $A, B, \ldots$ in \eq{eq:globalfit} are specific to the observable $Y$.
They can depend on the lattice spacing, and also on the $X$ variables defined above, in particular we use:
\begin{align}
    A &= A_0 + A_2 \left[a^2 {\left(\alpha_s(1/a)\right)}^n\right]\nonumber\\
      &\quad+ A_4 {\left[a^2 {\left(\alpha_s(1/a)\right)}^n\right]}^2
      + A_6 {\left[a^2 {\left(\alpha_s(1/a)\right)}^n\right]}^3 \,, \label{eq:fit_model_A}\\
    B &= B_0 + B_2 a^2\,,\\
    C &= C_0 + C_2 a^2\,,\\
    D &= D_0 + D_2 a^2 + D_4 a^4 + D_l X_l + D_s X_s\,,\\
    E &= E_0 + E_2 a^2 + E_4 a^4 + E_l X_l + E_s X_s\,,\\
    F &= F_0 + F_2 a^2\,,\\
    G &= G_0 + G_2 a^2\,.
\end{align}

For any ensemble, $a$ is given by the ratio of the $\Omega$ mass, in lattice units, to its experimental value.
For the $A$ coefficient of observables which give a large contribution to the final result, we include powers $n$ of the strong coupling $\alpha_s(1/a)$ in the lattice-spacing dependence~\cite{Husung:2019ytz}. We take both the commonly used $n=0$ (corresponding to the usual polynomial expansion in $a^2$), and $n=3$, as suggested in Ref.~\cite{Husung:2020}.
For the strong coupling, we use its four-flavor, $\overline{\text{MS}}$ value at scale $1/a$. We determine this value from the world average value of $\alpha_s(M_Z)$~\cite{ParticleDataGroup:2020ssz}, by running down from $M_Z$ to $1/a$ in five-loop perturbation theory~\cite{Herzog:2017ohr}, taking into account four-loop threshold corrections~\cite{Schroder:2005hy} at the $b$-quark mass that are given in Ref.~\cite{ParticleDataGroup:2020ssz}.

The choice of $n$ is included as a source of systematic uncertainty in the analysis described below.
In addition, we vary the polynomial order of the coefficients as an additional source of uncertainty.
As required by the data, we take the A coefficient to be linear, quadratic, or, in some cases cubic in $a^2\alpha_s(1/a)^n$.
For all other coefficients, the lattice spacing dependence is assumed to be a function of $a^2$ only. In the $D$ and $E$ coefficients, up to quadratic dependencies are used, in all other cases only a linear one is needed.
Depending on the fit qualities, some of these parameters will be set to zero.

We perform simulations at six different values of the lattice spacing, at a range of quark masses scattered around the isospin-symmetric point.
We exclude zero or more of the coarsest lattice spacings from the analysis, and the resulting variation in the global fit provides another source of systematic uncertainty.

\subsection{Estimating the covariance matrix}

The results of this lattice calculation are subject to both statistical uncertainties arising from the Monte Carlo sampling of the QCD gauge fields, and systematic uncertainties arising from choices made during the analysis.
We estimate the variance and covariance arising from both types of uncertainty by using a combination of statistical resampling, and histogramming of systematic variations~\cite{Durr:2008zz,Borsanyi:2014jba,Borsanyi:2020mff}.

The statistical covariance matrix is computed through jackknife resampling of the two-point, electromagnetic current-current correlator $C(t)$.
The gauge field configurations are computed through a Markov chain algorithm that induces an auto-correlation between subsequent configurations.
We suppress this effect by introducing a blocking procedure, where the individual configurations are grouped together into $N_J$ blocks and then the average of $C(t)$ on each block is used in the resampling procedure instead of the value on each configuration.
To simplify the analysis, we take $N_J$ to be the same on all ensembles. Specifically, we take $N_J=48$ which gives typical blocks of 100 or more configurations, larger than the autocorrelation time of the topological charge (which is $\sim 20$ on the finest ensembles).

Systematic uncertainties arise from a number of choices that must be made throughout the analysis.
For example, the choice of appropriate fit ranges, the experimental value of the $\Omega$-baryon mass used to set the scale, or the particular fit form used to perform the global fit.
We call the global fit resulting from a particular set of these choices an analysis.
In order to quantify the systematic uncertainties, we need to estimate the variance of the final value across all analyses arising from reasonable choices.
To do this we follow the strategy used in Ref.~\cite{Borsanyi:2020mff}.

As the full value of $\amulohvp$ and the window observable $\amulohvpwin$ are evaluated with similar techniques on the same set of gauge configurations, a number of choices in the analyses for $\amulohvp$ and $\amulohvpwin$ are similar and hence introduce correlations. Therefore, the procedure described in Ref.~\cite{Borsanyi:2020mff} had to be adapted to take into account these correlations. $\ddalpha$ is assumed to have negligible correlations with $\amulohvp$ and $\amulohvpwin$ because it is mostly determined by contributions from shorter distances. 

The systematic uncertainties affecting any given analysis can be arranged into two groups: those that induce correlations, and those that induce no correlation.
The systematics that induce no correlation are further subdivided into those that enter with a flat weight, and those that are weighted with the Akaike Information Criterion:
\be
\mathrm{AIC} \propto \exp\left[-\frac{1}{2} \left(\chi^2 + 2 n_\mathrm{par} - n_\mathrm{data} \right)\right]\,,
\ee
where the $\chi^2$, the number of fit parameters $n_\mathrm{par}$ and the number of data points $n_\mathrm{data}$ describe the global fit corresponding to the given analysis.
These relative AIC weights are employed between analyses differing in the choice of fit functions or in the lattice spacings included. For other systematics, including all systematics that induce correlations, a flat weight is employed.
The choices that are assumed to introduce correlations are: the fit ranges in the extraction of masses from correlation functions, the experimental input for the $\Omega$ mass, the taste improvement time-ranges, and the choice of $t_c$ in the bounding method for the extraction of $\amulohvp$.
Details of this procedure and explanation of the individual contributions to the systematic uncertainties can be found in Ref.~\cite{Borsanyi:2020mff}. 
In the following, we denote with $A(\phi,\psi^\text{flat},\psi^\text{AIC})$ the result for some observable $A$, either $\amulohvp$ or $\amulohvpwin$, with a specific set of systematics. 
Here, $\phi$ is a label for a set of specific choices that will induce correlations between $\amulohvp$ and $\amulohvpwin$ and the labels $\psi^\text{flat}$ and $\psi^\text{AIC}$ refer to choices that are made independently and are weighted either by a flat weight or by an AIC weight. In this notation, the histogram for a single observable $A$ can be expressed as
\begin{multline}
H(A) = \sum_{\phi,\psi^\text{flat},\psi^\text{AIC}}
w(\phi,\psi^\text{flat},\psi^\text{AIC}) \times \\
\mathcal N_1(A,\overline A(\phi,\psi^\text{flat},\psi^\text{AIC}), 
\sigma_{A(\phi,\psi^\text{flat},\psi^\text{AIC})})
\end{multline}
where $\mathcal N_1(x,\mu,\sigma)$ is the probability density function of the normal distribution with mean $\mu$ and standard deviation $\sigma$. $\overline A(\phi,\psi^\text{flat},\psi^\text{AIC})$ refers to the mean value of the observable $A$ in the analysis with the systematic variation indicated in parentheses. In the same sense, $\sigma_{A(\phi,\psi^\text{flat},\psi^\text{AIC})}$ corresponds to the statistical uncertainty on this analysis determined from the jackknife samples. The weight function
\be
w(\phi,\psi^\text{flat},\psi^\text{AIC}) = \frac{\operatorname{AIC}(\phi,\psi^\text{flat},\psi^\text{AIC})}{ \sum_{\psi^\text{AIC}} \operatorname{AIC}(\phi,\psi^\text{flat},\psi^\text{AIC}) }
\ee
ensures a correct relative weight between the analyses. Here, $\operatorname{AIC}(\cdots)$ denotes the AIC weight of the indicated analysis.

This procedure can be generalized to the study of the correlations between $\amulohvp$ and $\amulohvpwin$ by introducing a two-dimensional histogram:
\begin{multline}
H(A, B) =
\sum_{\substack{\phi,\psi_{A}^\text{flat},\psi_{A}^\text{AIC},\\
\psi_{B}^\text{flat},\psi_{B}^\text{AIC}}}
w_{A}(\phi,\psi_{A}^\text{flat},\psi_{A}^\text{AIC}) \times \\
w_{B}(\phi,\psi_{B}^\text{flat},\psi_{B}^\text{AIC}) \times \\
\mathcal N_2(A,B, \overline{A}(\phi,\psi_{A}^\text{flat},\psi_{A}^\text{AIC}), \\
\overline{B}(\phi,\psi_{B}^\text{flat},\psi_{B}^\text{AIC}), \\
C_\text{stat}^{(1)}(\phi,\psi_{A}^\text{flat},\psi_{A}^\text{AIC},\psi_{B}^\text{flat},\psi_{B}^\text{AIC}))\,.
\label{eq:hist_2d}
\end{multline}
Here, $C_\text{stat}^{(1)}$ refers to the statistical covariance matrix determined from the jackknife samples for each individual combination of analyses and $\mathcal N_2$ is the bivariate normal distribution with this covariance matrix. $A$ and $B$ refer to either $\amulohvp$ and $\amulohvpwin$, or individual flavor and contraction contributions to them.

We treated the systematic error introduced by the taste correction as a special case---both $\amulohvp$ and $\amulohvpwin$ are affected by this systematic in a correlated way. 
For $\amulohvp$, four time-ranges for taste improvements to the correlation function are used, starting at the Euclidean times $0.4 \, \text{fm}$, $0.7 \, \text{fm}$, $1.0 \, \text{fm}$, and $1.3 \, \text{fm}$. But since the window observable $\amulohvpwin$ is not affected by the taste improvement in the last range, for $\amulohvpwin$ the two last ranges give the same results.
Therefore, using all four ranges for both observables would bias the window observable. To avoid this but still retain a measure of correlation, we considered 12 systematic combinations where, for each matrix, we fixed the range of the taste correction to $\amulohvp$ and $\amulohvpwin$. The starting times for the taste correction ranges in all 12 cases are shown in \tab{tbl:sliding_ranges}.
\begin{table}
 \caption{\label{tbl:sliding_ranges}
 Starting time for the taste corrections in the 12 systematic combinations used to estimate the correlation introduced by the taste correction. The first column contains the number of the systematic combination, the second column contains the starting time for the taste correction in $\amulohvp$ and the last column contains the starting time for the taste correction for $\amulohvpwin$.}
 \centering
 \begin{tabular}{ccc}
 \hline
 \hline
  \#\ syst. comb. & $t^{\text{start}}_{\amulohvp}/\text{fm}$ & $t^{\text{start}}_{\amulohvpwin}/\text{fm}$ \\
  \hline
  1 & 0.4 & 0.4 \\
  2 & 0.4 & 0.4 \\
  3 & 0.4 & 0.4 \\
  4 & 0.7 & 0.4 \\
  5 & 0.7 & 0.7 \\
  6 & 0.7 & 0.7 \\
  7 & 1.0 & 0.7 \\
  8 & 1.0 & 0.7 \\
  9 & 1.0 & 1.0 \\
  10 & 1.3 & 1.0 \\
  11 & 1.3 & 1.0 \\
  12 & 1.3 & 1.0\\
  \hline
  \hline
 \end{tabular}
\end{table}
Then, we averaged the 12 covariance matrices.

\begin{figure}
    \centering
    \subfloat[light contribution\label{fig:hist2d:light}]{
    \includegraphics[width=0.4\textwidth]{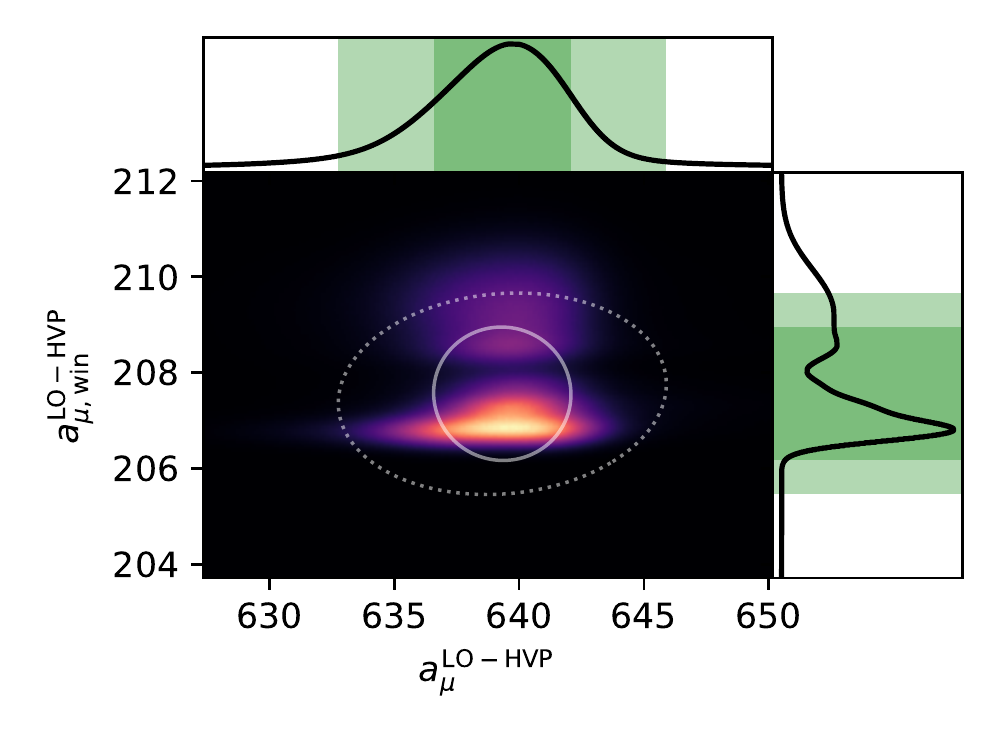}}\\\subfloat[strange contribution]{
    \includegraphics[width=0.4\textwidth]{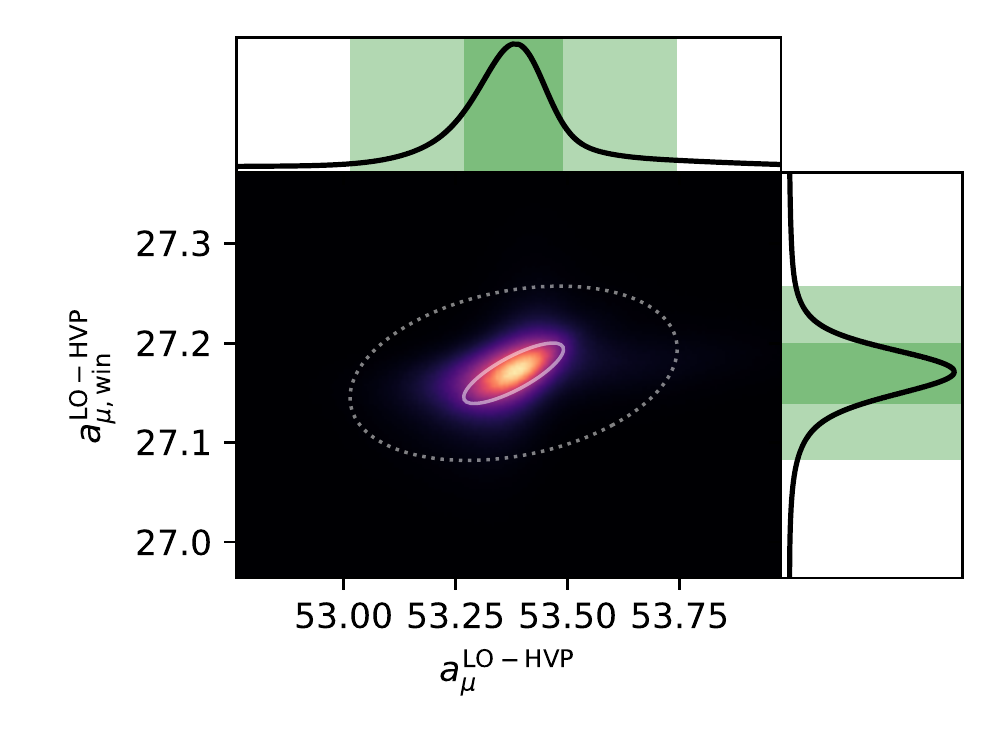}}\\
    \subfloat[disconnected contribution]{
    \includegraphics[width=0.4\textwidth]{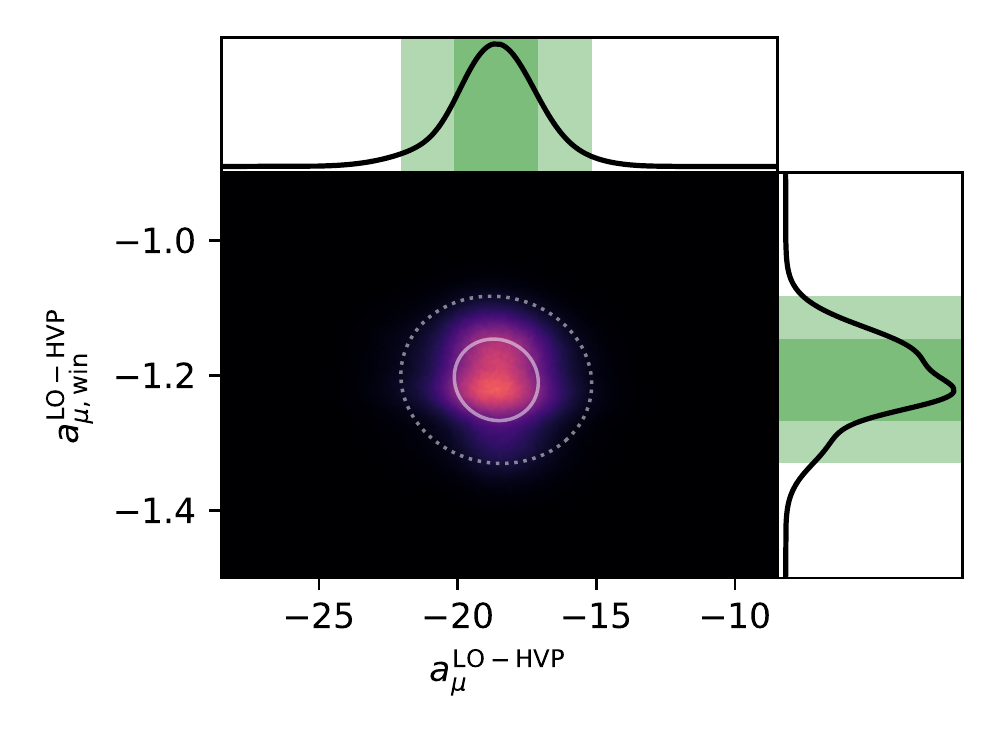}}
    \caption{Histograms for the estimation of the total, statistical and systematic lattice covariance matrices for the connected $ud$ (top), the connected $s$ (middle) and disconnected (bottom) contributions to $a^\text{LO-HVP}_{\mu}$ and $a^\text{LO-HVP}_{\mu,\text{win}}$. 
    These are obtained via the specialization of \eq{eq:hist_2d} to these individual contributions, as described in the text.
    Each of the three plots has a heat map in its center, illustrating the two-dimensional covariance distribution of the total, combined uncertainty. 
    As usual, colors from white to blue represent higher to lower probability densities. 
    These densities are calculated using the random sampling discussed in the text.
    The solid and dashed ellipses in these maps indicate the one and two sigma errors determined from the one and two sigma quantiles.
    In the upper and right panels around the heat maps, histograms of the marginalized distributions are shown. 
    These are calculated exactly without random sampling. 
    The two green bands show the one and two sigma uncertainties determined by the one and two sigma quantiles.}
    \label{fig:hist2d}
\end{figure}

As there is a large number of choices with many of these inducing a correlation, explicitly evaluating \eq{eq:hist_2d}, as a function of $\amulohvp$ and $\amulohvpwin$, is unfeasible. 
Instead, we prepared three one-dimensional histograms for the distributions of $X = \amulohvp$, $Y = \amulohvpwin$ and $Z = \amulohvp + \amulohvpwin$. 
From these three histograms, we determined the median and the interquantile distances $d^\sigma_X$, $d^\sigma_Y$ and $d^\sigma_Z$. 
Here, $\sigma$ indicates the number of standard deviations the interquantile distances would correspond to, in the case of normal distributions. Using these results, we construct the covariance matrices as
\be
C = 
\begin{bmatrix}
 \frac{(d_X^\sigma)^2}{\sigma^2} & \frac{(d_Z^\sigma)^2 - (d_X^\sigma)^2 - (d_Y^\sigma)^2}{2\sigma^2} \\
 \frac{(d_Z^\sigma)^2 - (d_X^\sigma)^2 - (d_Y^\sigma)^2}{2\sigma^2} & \frac{(d_Y^\sigma)^2}{\sigma^2}
\end{bmatrix}.
\ee
To avoid any bias from the details of the histogram binning, we considered a large number of 2000 bins per histogram. 
The upper and lower bounds of the histogram were chosen such that they do not influence the results. These values can be found in \tab{tbl:hist_bounds}.
\begin{table*}[htpb!]
\centering
\caption{\label{tbl:hist_bounds}
Upper and lower bounds of the histograms used for determining the lattice covariance matrices in different flavor channels.}
 \begin{tabular}{ccccc}
 \hline
 \hline
 Channel & Lower bound $a^\text{LO-HVP}_{\mu,\text{chan}}$ & Upper bound $a^\text{LO-HVP}_{\mu,\text{chan}}$ & Lower bound $a^\text{LO-HVP}_{\mu,\text{chan,win}}$ & Upper bound $a^\text{LO-HVP}_{\mu,\text{chan,win}}$\\
 \hline
  Light & 600 & 680 & 190 & 230 \\
  Strange & 51 & 55 & 25 & 29 \\
  Disconnected & $-28.5$ & $-8.5$ & $-1.5$ & $-0.9$\\
 \hline
 \hline
 \end{tabular}
\end{table*}
The resulting histograms for $X$ and $Y$ are plotted as black curves on the upper and on the right sides of the heat maps of \fig{fig:hist2d}.
For illustrative purposes, we also calculated two-dimensional histograms with a smaller number of 1000 bins in each direction. 
These two-dimensional histograms are also shown \fig{fig:hist2d}.
The bins were filled by sampling 5 random points for each summand in \eq{eq:hist_2d}. 
Note that the random sampling is only included in the visualization and has no impact on the end result.
It is noteworthy that in the case of the light quark contribution to the window quantity, a double peak structure appears. This structure is due to the variation of the parameter $n=0,3$ in \eq{eq:fit_model_A}. The effect of the non-gaussianity of the distribution introduced by this structure is taken into account by looking at both the $\sigma$ and $2\sigma$ quantiles.

In order to separate the systematic uncertainties and correlations from the statistical ones, we repeat, in analogy to the procedure in Ref.~\cite{Borsanyi:2020mff}, the full procedure while replacing $C$ with $\lambda C$ in \eq{eq:hist_2d}. 
We choose $\lambda = 2$. Then, we solve the equations
\begin{align}
    C &= C_\text{stat} + C_\text{syst} \\
    C_{\lambda} &= \lambda C_\text{stat} + C_\text{syst}
\end{align}
for $C_\text{stat}$ and $C_\text{syst}$. Here, $C_\text{stat}$ and $C_\text{syst}$ are the statistical and systematic covariance matrices of $\amulohvp$ and $\amulohvpwin$, extrapolated to $a=0$ and interpolated to the physical mass point.

\subsection{Systematic and statistical uncertainties on the covariance matrix}
\label{Sec:UncOnLattCov}
The specific value taken by the estimated covariance matrix can have an important effect on the results described in the main paper, in particular on those for $\chi^2$, for the corresponding $p$-values, as well as those resulting from the inverse problem.
As such, we found it important to quantify the uncertainties in this estimate. We identified two major sources of systematic uncertainty in our covariance estimate: non-normal distributions, and correlation of continuum extrapolations.
Each of these sources is described below, along with our approach to estimating the associated uncertainty.

Our jackknife estimate of the statistical covariance matrix for each analysis has itself a statistical uncertainty. We estimate this uncertainty by applying an approximate second-order resampling technique. We observe that the jackknife bias in $\amulohvp$ and $\amulohvpwin$ is small. Hence, we can approximate the true bootstrap distribution by bootstrapping the reconstructed samples $X_i = N_J \bar{X} - (N_J - 1) \bar{X}_i$, where $\bar{X}$ is the ensemble mean and $\bar{X}_i$ is the $i$th jackknife. We construct $N_B=1000$ bootstraps in this way, and then on each bootstrap we compute the jackknife statistical covariance and perform the histogramming technique described above to obtain the full covariance matrix. In this way we compute the bootstrapped distribution of our final covariance matrix.

To obtain a covariance from the histogram, we make the assumption that the histogram is normally distributed.
Under this assumption, the one-sigma quantiles described above will fall one standard deviation below and above the median.
Any deviation away from the normal distribution will hence introduce a systematic uncertainty.
One example of such a non-normal distribution is the double-peak structure observed in \fig{fig:hist2d:light}.
This double peak arises from the two-point systematic related to the power $n$ of the strong coupling in the description of the lattice spacing dependence of $a^\text{LO-HVP}_{\mu,\text{win},l}$ (i.e.\ the connected light ($ud$) quark contribution to the window observable) given in \eq{eq:fit_model_A}.
It is an artefact of how we sample the errors, and not intrinsic to the underlying distribution, but it still introduces an uncertainty in how the covariance is defined.
To quantify this uncertainty, we also consider the covariance matrix obtained by considering the two-sigma quantiles to be two standard deviations below and above the median.
If the distribution were normal, the covariance matrix obtained in each of these ways would be identical.
We take the difference between these two determinations of the covariance matrix as a two-point systematic.

In producing the histograms we allow the form of the continuum fit to vary independently for $\amulohvp$ and $\amulohvpwin$, as there is no a priori reason that the fit forms should be the same.
Hence, by construction we obtain no correlation between systematic uncertainties arising from the continuum limit of each quantity.
However, since $\amulohvp$ can be considered to be the sum of $W = \amulohvpwin$ and its complement $C$, it is reasonable to surmise that there may be some correlation between their continuum limits.
To estimate this potential correlation, we consider taking the continuum limits of $W$ and $C$ independently, and adding them together to obtain $T = W + C = a_{\mu}$.
In this scenario, let $W$ have variance $(dW)_c^2$ associated with the continuum limit, and similarly for $C$. Then the variance of $T$ is
\begin{equation}
    (dT)^2 = (dT)_o^2 + (dW)_c^2 + (dC)_c^2 \,,
\end{equation}
where $(dT)_o^2$ is the variance of $T$ due to uncertainty sources other than the continuum extrapolation.
Under these assumptions, the covariance matrix between $T$ and $W$ is
\begin{equation}
    C_{TW} = C^o_{TW} + \begin{bmatrix}
        (dW)_c^2 + (dC)_c^2 & (dW)_c^2 \\
        (dW)_c^2 & (dW)_c^2
    \end{bmatrix}\,,
\end{equation}
where $C^o_{TW}$ describes the covariance not associated with the continuum extrapolation. 
In practice, since the continuum limit of the combined $T = W + C$ is taken instead, the $C$ component may interfere and partially spoil the correlation.
As such we expect the true covariance to lie somewhere between this case, and the case where this contribution is neglected.
We take as an additional two-point systematic covariance matrices including either 0\% or 100\% of this estimated continuum limit contribution.

The connected contribution from valence charm quarks was computed in Ref.~\cite{Budapest-Marseille-Wuppertal:2017okr}, and the data was not available to perform the covariance analysis described here.
To address this, we assume that the correlation between $a^\text{LO-HVP}_{\mu,c}$ and $a^\text{LO-HVP}_{\mu,c,\text{win}}$ lies somewhere between 0\% and 100\%, and treat the difference between these two extremes as an additional uncertainty.
The charm contribution is a very small part of the total variance so this uncertainty is small.
It is closely related to the uncertainty from the correlation of continuum extrapolations and as such is treated as part of the same two-point systematic.

Finally, the finite-size effects computed in Ref.~\cite{Borsanyi:2020mff} have a large contribution from the continuum limit of dedicated large-volume lattice studies, so once again the correlation of these errors are unknown.
We conservatively take this correlation to lie somewhere between 0\% and 100\% and include this in the same two-point systematic as the other continuum limit correlations.

Taking into account both two-point systematics, we obtain four correlation matrices, each with an associated statistical uncertainty,
\begin{align}
    C_0\equiv C_\text{lat}^{1\sigma,0\%} &= \begin{bmatrix}
        30.13(4.88) & -0.05(0.03) \\
        -0.05(0.03) &  1.95(0.47)
    \end{bmatrix} \,, \label{eq:latcov1}\\
    C_1\equiv C_\text{lat}^{2\sigma,0\%} &= \begin{bmatrix}
        34.04(16.80) &  0.32(0.05) \\
         0.32(0.05) &  1.12(0.07)
    \end{bmatrix} \,, \label{eq:latcov2}\\
    C_2\equiv C_\text{lat}^{1\sigma,100\%} &= \begin{bmatrix}
        30.13(4.88) &  1.56(0.03) \\
         1.56(0.03) &  1.95(0.47)
    \end{bmatrix} \,, \label{eq:latcov3}\\
    C_3\equiv C_\text{lat}^{2\sigma,100\%} &= \begin{bmatrix}
        34.04(16.80) &  1.94(0.05) \\
         1.94(0.05) &  1.12(0.07)
    \end{bmatrix} \,.\label{eq:latcov4}
\end{align}
By performing the fits described in the main text with each bootstrap sample 
for each of the four representative covariance matrices, we obtain the uncertainty in the rescaling factors and observed $\chi^2$ values arising from our uncertainty in the lattice covariance matrix.

It is important to note that the statistical covariance matrices, $C^{(1)}_{stat}$, used in the histogramming procedure, are determined from jackknifes over $N_J=48$ binned ``measurements''.
Therefore, the statistical uncertainty on these covariance matrices is determined by $N_J$, rather than by the number of independent samples used in the determination of the central values.
This number is smaller than the total number of independent samples available in the dataset, even accounting for autocorrelations.
This means that with the available data, a different sampling procedure (in particular, a larger number of bins) could produce an estimate of the statistical covariance with smaller statistical errors.
However, this conservative approach to binning ensures that the systematic errors arising from autocorrelations within the data are negligible compared to these statistical errors.
In addition, this choice provides consistency with the analysis of Ref.~\cite{Borsanyi:2020mff}. 
As the results of \sec{Sec:Results} show, the statistical error this conservative binning produces is still sufficiently small that it does not quantitatively affect the conclusions drawn in the present work. 

\section{Weighted average procedure in presence of uncertainties on uncertainties and correlations}
\label{Appendix:Chi2Methodology}

The minimum of the $\chi^2$ function in \eq{eq:chisqAveragefit}, with respect, to $\gamma$ is obtained for
\be
   \gamma = \frac{ \sum_{j,k} \left[ \left(C^{\tilde{\gamma}}_\text{lat} + C^{\tilde{\gamma}}_\text{R} \right)^{-1} \right]_{kj} \cdot \tilde{\gamma}_j }
   { \sum_{l,m} \left[ \left(C^{\tilde{\gamma}}_\text{lat} + C^{\tilde{\gamma}}_\text{R} \right)^{-1} \right]_{lm} },
   \label{Eq:chi2minimum}
\ee
which represents a weighted mean of the input $\tilde{\gamma}_j$ values. 
However, it is known that this approach can yield biased results~(see e.g.\ Ref.~\cite{DAgostini:1993arp}). 

According to the Gauss-Markov theorem, this $\chi^2$ minimization is equivalent to the minimization of the variance of a weighted average, with the sum of weights constrained to unity~(see e.g.\ Refs.~\cite{Lyons:1988rp,Cowan:1998ji,ParticleDataGroup:2010dbb}).
It is for this reason that this method is also called ``BLUE'', i.e.\ Best Linear Unbiased Estimate~(see e.g.\ Ref.~\cite{Lyons:1988rp} and references therein).
However, the word ``unbiased'' in the name above needs to be interpreted with care, as it implicitly involves several assumptions about the statistical model being employed.
Indeed, in presence of strong correlations among the inputs, the weights in \eq{Eq:chi2minimum} can be negative or larger than unity, yielding an average value that can be even outside the range of the input values, which may be problematic.
Chapter~7 of Ref.~\cite{Cowan:1998ji} points out that the measurements are likely to lie on the same side of the true value in the presence of strong positive correlations.
However, it is also indicated that this interpretation is subject to the implicit assumptions about the knowledge of the covariance matrix, as discussed in more detail below.

Ref.~\cite{DAgostini:1993arp} points out that a bias in this procedure is caused by the input covariance matrix because of ``the linearization on which the usual error propagation relies''.
While the $\chi^2$ definition of \eq{eq:chisqAveragefit} treats all the uncertainties as absolute~(i.e.\ corresponding to additive effects), at least some of them should be treated as relative~\footnote{That is the case e.g.\ for the luminosity uncertainty of the experimental measurements.}~(i.e.\ corresponding to multiplicative effects)~\cite{Blobel:2003wa}.

Furthermore, another problem of this approach is that 
\eq{eq:chisqAveragefit} involves an ill-behaved inversion of the covariance matrix.
Indeed, a small change in the covariance matrix can have an important impact on its inverse, hence on the weights and the resulting average.
Without special techniques, the inverse cannot be computed, nor can the $\chi^2$ minimization be performed~\cite{Blobel:2003wa}.
While modern numerical algorithms allow to address the technical aspect of this matter, some more fundamental problems related to the evaluation of the covariance matrix still persist~(see below).

In Ref.~\cite{DAgostini:1993arp} it is discussed that a possible solution to these problems consist in introducing, in the $\chi^2$ definition, one nuisance parameter for each correlated systematic uncertainty, together with the corresponding constraint terms.
Doing so, for a normalization uncertainty, a scaling factor is applied to both data and uncorrelated uncertainties.
An equivalent approach introduces the corresponding nuisance parameter as a scaling factor on the theoretical prediction~\cite{Blobel:2003wa,Pascaud:1995qs}.
An alternative method for the treatment of relative uncertainties consists in scaling them to the fitted value minimizing the $\chi^2$, re-deriving the corresponding covariance matrix and iterating~(see Refs.~\cite{Lyons:1989gh,DAgostini:1993arp} and references therein)~\footnote{However, for more general applications, such a procedure faces some important challenges when the rescaling is applied for the statistical uncertainties of unfolded measurements. Indeed, in such a case the migrations of events induce non-trivial correlations between the various bins, which imply that the statistical uncertainty in a given bin also depends on (i.e.\ scales as a function of) the expected numbers of events in other bins. Furthermore, the regularization procedures typically used in the unfolding also induce non-linear effects in the uncertainty propagation, making the uncertainty rescaling of such measurements even more challenging.}.
However we note that the rescaling of the uncertainties has little relevance in the current use-cases, when there's little difference between the $\tilde{\gamma}_j$ values to be combined or when one of them is more precisely determined than the others and dominates the average~(see \sec{sec:testing_R-ratio_results}).
It is also worth recalling that, if all the nuisance parameters correspond to~(are treated as) absolute systematic uncertainties, each of them acting as a coherent additive shift of the theoretical prediction, the approach using nuisance parameters is equivalent to the one of the ``standard'' $\chi^2$ with correlations from \eq{eq:chisqAveragefit} (see Refs.~\cite{Demortier:1999,Stump:2001gu,Botje:2001fx,Thorne:2002kn,Fogli:2002pt,ListTalk}).

The approach using nuisance parameters in the fit often leads to reduced systematic uncertainties, as they are constrained by the $\chi^2$ minimization.
However, this reduction assumes and strongly relies on a perfect knowledge of the phase-space dependence and correlations of the systematic uncertainties, an information explicitly used in the $\chi^2$ definition.
The same assumptions and features are implicit for the ``standard'' $\chi^2$ with correlations in \eqsss{eq:chisqwin}{eq:chisqAveragefit}{eq:chisqwinintResc}{eq:chisqwinint}, in particular for minimizing the variance of the average, as discussed above.
Indeed, the need for precisely known covariance matrices, mentioned in Refs.~\cite{DeGrand:2022lmc,Colangelo:2022vok}, becomes explicit in the equations above.

Building upon remarks made in the context of ATLAS jet performance and cross-section studies~\cite{ATLAS:2014hvo,ATLAS:2017kux,ATLAS:2017ble}~(see also the earlier discussion in Ref.~\cite{Malaescu:2012ts}), these strong assumptions were also questioned for the $\epem$ annihilation data used for the evaluation of $\amulohvp$, since there are clear indications that the amplitudes of systematic uncertainties and their correlations are generally impacted by uncertainties themselves~\cite{Davier:2019can,bogdan-Mainz-2018-DHMZ-UncOnUnc,Aoyama:2020ynm}.~\footnote{More recently, a similar remark has been independently made in Ref.~\cite{Cowan:2021sdy}. The remarks made in the context of the muon $g-2$ studies are also relevant when using the data of hadronic spectra for other applications, e.g.\ for the extraction of the strong coupling employing the Adler function~\cite{Davier:2023hhn}.}
Indeed, the size of the systematic uncertainties of the various measurements, the correlations of a systematic uncertainty component impacting several measurements, as well as the correlations between the systematic uncertainties impacting a given measurement are never really measured, but rather {\it estimated}.
For example, the tensions between the input measurements observed in several channels~(e.g.\ between BABAR and KLOE in the \pp channel) are a direct indication of underestimated uncertainties for the measurements and motivated the inclusion of an extra systematic uncertainty for the resulting dispersive integrals~\cite{Davier:2019can}~(see also \app{sec:R-ratio_obs}).
Such two-point systematic uncertainty, although necessary when using the currently available inputs, has unknown correlations between different phase-space regions.
Therefore, it comes at the price of bringing some fundamental limitations for the treatment of correlations in \eqsss{eq:chisqwin}{eq:chisqAveragefit}{eq:chisqwinintResc}{eq:chisqwinint}.
The recent CMD-3 measurement in the \pp channel~\cite{CMD-3:2023alj}, which shows some significant tension with the previous precise measurements of the same channel, emphasizes even further the existence and the need for a careful treatment of the uncertainties on uncertainties.
Furthermore, such limitations in the statistical treatment also originate from the uncertainties on uncertainties and on correlations for the lattice QCD calculations~(see \app{sec:lat_obs}).
Therefore, the statistical procedures employing such inputs~(or other quantities derived using them) should avoid overestimating the precision with which the uncertainties and their correlations are known, a conservative uncertainty treatment being preferable.
This is especially important since the uncertainties on the uncertainties and on their correlations are not available in the current publications of hadronic spectra.

In order to address the limitations related to the uncertainties on the systematic uncertainties and on their correlations, we use a procedure which consists in performing a weighted average, with the weights of the input $\tilde{\gamma}_j$ values being proportional to the inverse of their squared total uncertainties.
This is equivalent to using only the diagonal elements of the covariance matrices~(i.e.\ setting the other elements of the matrices to zero) in \eq{Eq:chi2minimum} when deriving the weights.
The weights used in this procedure are in the range between $0$ and $1$, and are on the larger end for more precise contributions.
At the same time, the full information of the input uncertainties, with their correlations, is propagated to the result of the weighted average and is used in the $\chi^2$ when assessing the level of agreement between the inputs and the average value.
A similar fitting procedure, using only the diagonal uncertainties in the $\chi^2$ followed by a propagation of the full set of uncertainties and their correlations, has been successfully used in Ref.~\cite{Davier:2019can} for combining the experimental measurements in the $\epem \rightarrow \pp$ channel.~\footnote{In Ref.~\cite{Malaescu:2012ts} one can find some further discussions on other unbiased combination procedures with realistic uncertainty estimates, taking into account the full information of the uncertainties and correlations of the inputs.}

\section{Stability of the results of \protect\sec{sec:testing_R-ratio_results} with respect to the averaging procedure}
\label{Appendix:alternate_averaging}

In this appendix we discuss the use of an alternative averaging procedure to the one employed in \sec{sec:testing_R-ratio_results}, to obtain the results for changes to the experimental R-ratio that could explain the differences between the lattice and data-driven results for $\amulohvp$, $\amulohvpwin$ and $\ddalpha$. 
In that section, we perform the average defined in \eq{Eq:chi2minimum}. That is, when we derive the averaging weights, we replace the full covariance matrix, $\left(C^{\tilde{\gamma}}_\text{lat} + C^{\tilde{\gamma}}_\text{R} \right)$, obtained by a linear propagation of lattice and experimental R-ratio uncertainties and correlations, by its diagonal form with no correlation. 
Here we consider what the effect of including these correlations is on our results.

Generically, the use of the full covariance matrices for determining the averaging weights in \eq{Eq:chi2minimum} changes these weights significantly, some becoming negative and others larger than unity~(see \sec{sec:testing_Rratio}).
However here, their deviations from zero or from unity are observed to be smaller than $0.06$.
Furthermore, in such cases either the $\tilde{\gamma}_j$ values from \eq{eq:rescaleRoneNormConstr} do not differ much~\footnote{The $\tilde{\gamma}_j$ values used for a given averaging differ by typically only $12\%$ or less, except for cases when the normalization fit is done in restricted $\sqrt{s}$ regions, below $0.63$~GeV or above $1.8$ or $3.0$~GeV. 
In that case, the shape differences among the kernels of the various moment integrals have a more important impact for \eq{eq:rescaleRoneNormConstr} and the differences between the $\tilde{\gamma}_j$ values can be as large as $30-80\%$.} or one of them is more precise and dominates the average. 
Thus, the values of the average $\gamma=\gamma_{1}$ are similar for the two averaging methods, within $12\%$ in the most extreme case, and in most cases within a few percent or less.

Even if this alternative approach minimizes the $\chi^{2}$ in \eq{eq:chisqAveragefit}~(and the uncertainty of the rescaling percentage $\delta_1$, which is propagated from the covariance matrices of the lattice QCD and dispersive results), it only reduces it by $15\%$~($9\%$) or less, compared to the values obtained with weights proportional to the inverse of the $\tilde{\gamma}_j$ uncertainties squared, which are used in the nominal approach to produce \tab{tab:rescalingAverage}.
Therefore, the $p$-values obtained with the two approaches are also similar, leading to similar conclusions about the level of compatibility between the lattice and dispersive moment integrals, for all the normalization shift scenarios considered in this paper.

The statistical variance and quantiles of the rescaling percentage $\delta_1$, obtained from the bootstrap variations of the lattice covariance matrices, are most often enhanced~(by up to $540\%$) in the alternative averaging method compared to the one used to obtain the results given in the main text, while in some rare cases they are reduced~(by up to $87\%$).
At the same time, the changes of $\delta_1$ under systematic variations of the lattice covariance matrix are also generally broader for the alternative averaging method.

Concerning the statistical variance and quantiles of the { \em uncertainty} on $\delta_1$~(propagated from the covariance matrices of the lattice QCD and dispersive results), they are most often enhanced~(by up to $133\%$) in the alternative averaging method compared to the nominal one, while in some rare cases they are reduced~(by up to $99\%$).
In all cases, as for the nominal method, the uncertainty of the rescaling percentage $\delta_1$ remains precisely determined when using the alternative averaging method.

On the other hand, the variance and quantiles of the $\chi^{2}$ values, due to the same bootstrap fluctuations as above, are generally reduced when using the alternative averaging method~(by up to $44\%$), while in some rare scenarios they are slightly enhanced~(by less than $3\%$).
The range covered by the $\chi^{2}$ values under systematic variations of the lattice covariance matrix is generally somewhat narrower for the alternative averaging method. 

Some of the features described above, for the comparison between the two averaging methods, are illustrated by comparisons between the plots in the first two rows of Figs.~\ref{Fig:ChiSqpValue_0_2_0}-\ref{Fig:AverageSigma_1_2_3}. These features are consistent with the fact that while both procedures are sensitive to the uncertainty on the uncertainty in the determination of the weights, only the alternative approach, which uses the full covariance matrices for determining the averaging weights, relies on correlations for computing the weights to minimize the $\chi^{2}$ and the uncertainty of the rescaling percentage $\delta_1$.

\section{Going beyond the rescaling in a single $\sqrt{s}$-interval}
\label{app:beyond_rescaling}

To go beyond the simple scenarios and interpretations of \app{Appendix:Chi2Methodology}, more window observables are required from the lattice, with the corresponding ones from the experimental R-ratio, split up into $\sqrt{s}$-intervals. These can be complemented by other quantities related to the HVP, such as the hadronic vacuum polarization function $\hat\Pi(Q^2)$ at various $Q^2$ accessible to the lattice. Again, all of these quantities must be obtained with a full description of the relevant covariances. 

\subsection{General formalism for comparing and possibly combining lattice and data-driven results}
\label{sec:general_model}

Consider a function $M_\text{R}(s,\alpha_i)$, that depends on $s$ and on parameters $\alpha_i$. 
It is chosen so as to provide a good description of the experimental R-ratio, or a re-binned version of it, once the $\alpha_i$ are appropriately adjusted.
Now, consider a modified model, $M_\text{lat}(s,\alpha_i,\beta_j)$, that depends on the additional parameters $\beta_j$.
This new model represents a guess at how the experimental R-ratio could be modified to reproduce the lattice observables of interest, once integrated over $s$ with the appropriate weights.
We then define the ratio of these two models as:
\be
\label{eq:LatticeReeModel}
F_\text{lat$\div$R}(s,\beta_j)\equiv M_\text{lat}(s,\alpha_i,\beta_j)/M_\text{R}(s,\alpha_i)
\ ,\ee
which is chosen not to depend on the $\alpha_i$ parameters.~\footnote{This factorization is used here just for an intuitive simplification of the explanation, while the methods discussed below can also be applied to more general $M_\text{lat}(s,\alpha_i,\beta_j)$ models.}
For instance, these functions can be physics-motivated models, spline-based parametrizations, orthogonal polynomials, histograms of the bins used for the experimental R-ratio or with some different binning, etc.
For either the R-ratio or lattice model functions, we define the integrals of interest as
\be
\label{eq:a_Rlattice_Model}
a_{jb}^{M_\text{R/lat}} \equiv \int_{I_b} {ds} \, \tilde{K}_j(s) \, M_\text{R/lat}(s)
\ ,\ee
where the $\alpha_i$ and $\beta_j$ model parameters are kept implicit.
As above, $I_b$ can be either some restricted $\sqrt{s}$-interval or it can cover the full range from threshold to $\infty$.
In the latter case we drop the index $b$ for the integral label, which also keeps the notations consistent with e.g.\ \eq{eq:amuwin_split}.
Note that if $M_\text{R/lat}$ is a histogram, then the integral in \eq{eq:a_Rlattice_Model} becomes a sum over bins.

For the lattice constraints of the model integrals one can write the $\chi^2$ function as:
\bea
\chi^2_\text{lat} = \sum_{j,k}\left[ \alat{j} - a_{j}^{M_\text{lat}} \right][C^{-1}_\text{lat}]_{jk}\left[ \alat{k} - a_{k}^{M_\text{lat}} \right] ,
\label{eq:chis_LatticeModel}
\eea
while, for the constraints from the experimental R-ratio, one can use approaches based on different $\chi^2$ definitions, such as
\bea
\chi^2_\text{R,1} &=& \sum_{j,k}\left[ \aR{j}-a_{j}^{M_\text{R}} \right][C^{-1}_\text{R}]_{jk}\left[ \aR{k}-a_{k}^{M_\text{R}} \right] ,
\label{eq:chis_RModel1} \\
\chi^2_\text{R,2} &=& \sum_{(jb),(kc)}\left[ \aR{jb}-a_{jb}^{M_\text{R}} \right][C^{-1}_\text{R}]_{(jb)(kc)}\nn\\
&& \qquad\qquad\qquad\qquad\times \left[ \aR{kc}-a_{kc}^{M_\text{R}} \right] ,
\label{eq:chis_RModel2} \\
\chi^2_\text{R,3} &=& \sum_{b,c}\left[ {\rm R}_{b} - {M_\text{R,b}} \right][C^{-1}_\text{R,bin}]_{bc}\left[ {\rm R}_{c} - {M_\text{R,c}} \right] ,
\label{eq:chis_RModel3}
\eea
where, as above, we indicate whether the covariance matrix of the $a_j$ or $a_{jb}$ moment integrals correspond to those of the lattice (data-driven) approach via the subscript ``lat'' (``R''). In addition, we call $C_\text{R,bin}$ the covariance matrix that corresponds to (a possibly rebinned version of) the experimental R-ratio.
Indeed, \eq{eq:chis_RModel1} concerns dispersive integrals on the full $s$ range, \eq{eq:chis_RModel2} includes further information on the shape of the R-ratio through the integrals computed on various $I_b$ intervals, while \eq{eq:chis_RModel3} uses directly the experimental R-ratio~(without any further convolution), either within its original bins or after merging the bins within larger intervals.
Based on these $\chi^2$ functions, one can consider several approaches for comparing and/or combining the R-ratio and lattice results.

A first approach can consist in constraining the parameters $\alpha_i$ of $M_\text{R}(s,\alpha_i)$ by minimizing one of the $\chi^2_\text{R}$ functions in  
\eqsint{eq:chis_RModel1}{eq:chis_RModel3},
followed by a determination of the $\beta_j$ parameters of $M_\text{lat}(s,\alpha_i,\beta_j)$ through a minimization of $\chi^2_\text{lat}$ from \eq{eq:chis_LatticeModel}.
In this two-step approach, the $\alpha_i$ parameters are held fixed when the second fit is performed and therefore they are not constrained by lattice results.
Still, their uncertainties and correlations are propagated from the first step to the second, e.g.\ by replacing $C_\text{lat}$ in \eq{eq:chis_LatticeModel} with the appropriate covariance matrix.

A second approach can consist in simultaneously constraining the $\alpha_i$ and $\beta_j$ parameters of the model, by minimizing $\chi^2_\text{lat} + \chi^2_\text{R}$, with either of the definitions in  
\eqsint{eq:chis_RModel1}{eq:chis_RModel3} for the contribution from the experimental R-ratio.
We note that this minimum is generally lower than the $\chi^2_\text{lat} + \chi^2_\text{R}$ value obtained for the models fitted in the two-step approach discussed above.
While the minimization of the $\chi^2$ defined in \eq{eq:chisqwin}, according to \eq{eq:amuwinmin} and \eq{eq:chisqwinMin}, represents a combination of the lattice and R-ratio results in the space of the $a_j$ integrals, without any extra constraints, the approach of fitting the $\alpha_i$ and $\beta_j$ parameters by minimizing $\chi^2_\text{lat} + \chi^2_\text{R}$ represents a similar combination in the R-ratio space, under the constraint of the $M_\text{lat}(s,\alpha_i,\beta_j)$ and $M_\text{R}(s,\alpha_i)$ models, respectively.

In a third approach, one can first minimize $\chi^2_\text{lat}$ from \eq{eq:chis_LatticeModel} with respect to the free parameters of the model $M_\text{lat}(s,\alpha_i,\beta_j)$, obtaining $\hat\alpha_i$ and $\hat\beta_j$.
One then considers a modified version of that lattice model which accounts for possible differences with the R-ratio.
Using the notation of \eq{eq:LatticeReeModel}, we define this model as $\hat M_\text{lat}(s,\hat\alpha_i) \equiv M_\text{lat}(s,\hat\alpha_i,\hat\beta_j) / F_\text{lat$\div$R}(s,\hat\beta_j)$.
This model is subsequently binned via averaging within chosen $\sqrt{s}$-intervals.
We call the resulting histogram $\hat M_\text{lat}^\text{bin}(s,\hat\alpha_i)$ and its covariance matrix $\hat C_\text{lat,bin}$.
In addition, we consider a model $M_\text{R}(s,\alpha_i)$ for the R-ratio and its version $\tilde M_\text{R}(s,\alpha_i)$, possibly re-binned to match the binning of $\hat M_\text{lat}^\text{bin}(s,\hat\alpha_i)$.
The parameters $\alpha_i$ are to be fitted simultaneously to the experimental R-ratio and $\hat M_\text{lat}^\text{bin}(s,\hat\alpha_i)$, by minimizing:
\bea
\chi^2 &=& \sum_{b,c}\left[ {\hat M_{\text{lat},b}^\text{bin}} - {\tilde M_\text{R,b}} \right][ \hat C_\text{lat,bin}^{-1} ]_{bc}\left[ {\hat M_{\text{lat},c}^\text{bin}} - {\tilde M_\text{R,c}} \right]\nn\\
&&+
\sum_{b,c}\left[ {\rm R}_{b} - {M_\text{R,b}} \right][C^{-1}_\text{R,bin}]_{bc}\left[ {\rm R}_{c} - {M_\text{R,c}} \right]
\label{eq:chi2_3rd_approach}
\eea
where the first sum runs over the lattice bins and the second, over the ones of the experimental R-ratio.
This effectively combines the lattice and experimental R-ratio results.

In this third approach, consider the case in which $M_\text{lat}$ is taken to be a binned histogram with the number of bins equal to the number of (linearly independent) lattice moment integrals, the count of each bin being a free parameter to be determined. 
Then the minimization of $\chi^2_\text{lat}$ of \eq{eq:chis_LatticeModel} is equivalent to an ``unregularized'' unfolding towards the space of the R-ratio~(in the sense that no other regularization besides the binning choice itself is being applied).
Indeed, this is equivalent to solving a system of constraints through a matrix inversion approach~(see e.g.\ Chapter 11 of Ref.~\cite{Cowan:1998ji}).
However, the parametrization of $M_\text{lat}(s,\alpha_i,\beta_j)$ can be used to inject an effective regularization into the inverse problem.
The unregularized approach avoids biases related to the matrix inversion.
However, it may induce large variances and strong (anti-)correlations between the determined quantities (i.e.\ fitted R-ratio values in various bins) and hence a large hierarchy among the eigenvalues of the corresponding covariance matrix.
In that case, any further use of such results would require a very precise determination of the covariance matrix which, in turn, would necessitate very precise knowledge of the input covariances.
Alternatively, it is generally preferable to use regularized unfolding methods, aiming for a trade off between moderate statistical variances and systematic biases induced through the regularization.~\footnote{Such regularization limits the number of free parameters that are to be determined, allowing us to also address the worries that are sometimes expressed concerning the ill-defined nature of the problem when using $\chi^2$-based approaches~(see e.g.\ Refs.~\cite{Nakahara:1999vy,Nakahara:1999bm,Asakawa:2000tr,Rothkopf:2011ef,Burnier:2013nla,Rothkopf:2016luz}).}

It is to be noted that the three approaches described above, with their possible variants, imply different hypotheses about the validity of the fitted models and the way in which their parameters are constrained.
The resulting $\chi^2/\ndof$ values represent tests of these hypotheses, providing information about the compatibility between these models and the input data, as well as about the compatibility between the lattice and R-ratio inputs themselves.
In the presence of tensions, one can consider enhancing the uncertainties of the combination result, following e.g.\ the approaches from Refs.~\cite{Davier:2010rnx,Davier:2010nc,Davier:2017zfy,Davier:2019can}, summarized in \app{sec:R-ratio_obs}.
This can be achieved by using the information from some (partial/local)~$\chi^2$ and/or adding some extra uncertainties to account for systematic deviations between the inputs.

Moreover, while  
\eqsint{eq:chis_LatticeModel}{eq:chis_RModel3}
are written using the full covariance matrices, the discussion of \app{Appendix:Chi2Methodology}, concerning the treatment of the uncertainties on the uncertainties and on the correlations is relevant here too and has to be implemented accordingly.
For instance, one can perform the $\chi^2$ fits for determining the model parameters using only the diagonal elements of the covariance matrices, while the full information on the uncertainties and their correlations is propagated to evaluate the uncertainties of the model parameters, their correlations and the fit quality.

\subsection{The case of rescaling factors in either a single or multiple $\sqrt{s}$-intervals}
\label{sec:single_param}

To make contact with what was done in the main body of the paper, we consider here the rescaling in a single $\sqrt{s}$-interval that was introduced in \sec{sec:testing_Rratio} and employed in \sec{sec:testing_R-ratio_results}. It can be retrieved as a special case of the general approaches discussed above.
Indeed, one can consider a ``model'' $M_\text{R}(s,\alpha_i)$ that consists in a histogram with the same binning as the experimental R-ratio itself.
The $\alpha_i$ are parameters to be determined and correspond to the bin counts.
Employing the first, two-step approach described below \eq{eq:chis_RModel3}, with $\chi^2_\text{R}$ given by $\chi^2_\text{R,3}$ from that equation, the first minimization step yields the bin counts with $\chi^2_\text{R} = 0$.
Moreover it propagates the covariance matrix of the experimental R-ratio to the histogram of $M_\text{R}(s,\alpha_i)$, i.e.\ to the parameters $\alpha_i$.
The model is completed with the function $F_\text{lat$\div$R}(s,\gamma)$, which is equal to a rescaling factor $\gamma$ in a subset of the $\sqrt{s}$-intervals, while it is equal to unity for the complementary subset of bins.
This yields a model $M_\text{lat}(s,\alpha_i,\gamma)$ that implements the prescription from \eq{eq:rescaleRoneNorm}. 
In particular, this model ensures that the ratios of integrals $ a_{jb}^{M_\text{R}} / a_{kb}^{M_\text{R}} $ and $ a_{jb}^{M_\text{lat}} / a_{kb}^{M_\text{lat}} $ are equal to $ \aR{jb} / \aR{kb} $. 
In turn, this guarantees that these ratios are independent of the parameter $\gamma$, for any $j$ and $k$ integrals and any interval $I_b$.
In other words, these ratios of integrals are driven by the corresponding kernels convoluted with~(i.e.\ averaged over) the shape of the experimental R-ratio, while the changes of normalization through the factor $\gamma$ impact coherently all the $a_{jb}^{M_\text{lat}}$ integrals on the rescaled $\sqrt{s}$-interval.
\eq{eq:chis_LatticeModel}, with the model $M_\text{lat}(s,\alpha_i,\gamma)$ and $C_\text{lat}$ completed with a propagation of the covariance matrix of the parameters $\alpha_i$, yields \eq{eq:chisqAveragefit}~(after a change of variable similar to \eq{eq:rescaleRoneNormConstr}, together with the corresponding linear uncertainty propagation).

The approach discussed above can be generalized by applying the constraint from \eq{eq:rescaleR} at the level of the corresponding R-ratio and lattice models, with multiple rescaling factors $\gamma_b$ applied in the $\sqrt{s}$-intervals $I_b$.
This corresponds to minimizing
\bea
\chi^2 & = & \sum_{j,k}\left[\alat{j}-\sum_b\gamma_b \aR{jb}\right]\left[\left( C_\text{lat} + \tilde{C}_\text{R} \right)^{-1}\right]_{jk}\nn\\
&& \qquad\qquad\qquad\qquad\times \left[\alat{k}-\sum_c\gamma_c \aR{kc}\right] \ ,
\label{eq:chisqwinintResc}
\eea
provided that the number of linearly independent moment integrals is greater or equal to the number of fitted $\gamma_b$ rescaling factors.
In this equation, $ \left( C_\text{lat} + \tilde{C}_\text{R} \right) $ is the covariance matrix corresponding to $\left[\alat{j}-\sum_b\gamma_b \aR{jb}\right]$, depending hence also on the $\gamma_b$ parameters.
As above, this implies assigning a pre-fit uncertainty to the $M_\text{R}(s,\alpha_i)$ model, which is then propagated via the corresponding covariance matrix.~\footnote{These methods can also be seen as fits of normalization factors for templates of moment integrals derived through the dispersive approach, used to describe the corresponding lattice results, as well as possible. For a discussion on the template fitting methods, with the various approaches for treating the uncertainties of the fitted distributions and of the templates themselves, see e.g.\ Refs.~\cite{Britzger:2021ocj,Cranmer:2021oxr,ATLAS:2022jbw} and references therein.}
In addition, if the $I_b$ intervals correspond to the original experimental R-ratio bins and their number is equal to that of linearly independent lattice observables, $\alat{j}$, then the rescaling hypothesis becomes exact.
In that case, one can have as many normalization rescaling factors as there are bins, allowing for a complete, unregularized reconstruction of the lattice R-ratio with the same granularity as that of the experimental R-ratio.
Please see the discussion after \eq{eq:chi2_3rd_approach} for more details.
It is also worth noting that in the case when all the $\gamma_b = 1$, the $\chi^2$ from \eq{eq:chisqwinintResc} matches the one of \eq{eq:chisqwinMin}, providing a direct compatibility check between the lattice and the dispersive results.

\subsection{Single or multiple rescaling factors for fitted moment integrals}
\label{sec:multiple_rescaling_fits}

An alternative to \eq{eq:chisqwinintResc}, 
which includes multiple rescaling factors, $\gamma_b$, is given by the model 
\be
\label{eq:rescaleRexpected}
\left\{
\begin{array}{rcl}
\alat{j} &=& \sum_{b} \gamma_b \,a_{jb}\\
\aR{jb} &=& a_{jb}
\end{array}
\right.
\ ,\ee
that is assumed to describe the inputs, $\alat{j}$ and $\aR{jb}$, within the uncertainties.
The second line provides a trivial model for the R-ratio contributions, $\aR{jb}$, from the $\sqrt{s}$-intervals, $I_b$, defined after \eq{eq:amuwin_split} and the first a rescaling model for the corresponding lattice observables defined in \sec{sec:observables}. 
Of course, the number of lattice observables, $\alat{j}$, must be larger than the number of free rescaling parameters. 
This is a special case of \eq{eq:LatticeReeModel}, as given in \eqs{eq:chis_LatticeModel}{eq:chis_RModel2} without, however, the constraints of the models, $M_\text{R/lat}$, for the lattice and R-ratio spectral functions.

Thus, to determine the parameters of this model, taking into account all correlations, one can minimize
\bea
&& \chi^2 = \sum_{j,k}\left[\alat{j}-\sum_b\gamma_b a_{jb}\right][C^{-1}_\text{lat}]_{jk}\left[\alat{k}-\sum_c\gamma_c a_{kc}\right]\nn\\
  && +\sum_{(jb),(kc)}\left[\aR{jb}-a_{jb}\right][C^{-1}_\text{R}]_{(jb)(kc)}\left[\aR{kc}-a_{kc}\right]\ ,
\label{eq:chisqwinint}
\eea
with respect to the $\gamma_b$ and $a_{jb}$. 
Again, lattice results are assumed to have no correlations with those obtained from the experimental R-ratio. 
The uncertainties on the parameters $\gamma_b$ and $a_{jb}$ can be obtained from the Hessian or by using pseudo-experiments. 
One can also consider the $p$-value obtained from the minimum value of the $\chi^2$ and the system's number of degrees of freedom, which is the number of lattice window observables minus the number of rescaling parameters. 
This $p$-value indicates the consistency of the available lattice and R-ratio input with the rescaling hypothesis above.

Now, suppose that the rescaling model of \eq{eq:rescaleRexpected} is well justified theoretically and that the $p$-value of the fit is acceptable. 
Then, the resulting parameter values provide a description of the lattice and R-ratio observables, via the right-hand sides of \eq{eq:rescaleRexpected}, that combines the information contained in the $\alat{i}$ and $\aR{jb}$.
When the number of observables and intervals increases, so will (anti-)correlations between them, leading to large variances on the results, well known in inverse problems.
Of course, if the number of rescaling parameters is the same as the number of lattice observables, then the system of equations of \eq{eq:rescaleRexpected} can be solved exactly, and one obtains $a_{jb}=\aR{jb}$ as well as the values of the $\gamma_b$ obtained by minimizing \eq{eq:chisqwinintResc}, in the same situation.

Beyond the possibility of combining lattice and R-ratio results for $\alat{j}$ and $\aR{jb}$, via the model of \eq{eq:rescaleRexpected}, the approaches of \sec{sec:testing_Rratio} and of \eqs{eq:rescaleRexpected}{eq:chisqwinint} differ in that the former is based on a model that is linear in the fitted parameters (\eq{eq:rescaleRoneNormConstr}), with a linear propagation of uncertainties, while the latter refers to a nonlinear model (\eq{eq:rescaleRexpected}), but with no linear propagation of uncertainties. 
They are expected to be equivalent in the limit that the rescaling factors are such that the $|\gamma_b-1|$ are small and as long as the relative uncertainties on the $\aR{jb}$ are also small.
We have checked this analytically and numerically, in the case of a single rescaling factor $\gamma$~(either fitted or fixed to $1$), for the results presented in \sec{Sec:Results}.
The non-linear effects in the uncertainty propagation mentioned above were also studied and found to be numerically small. 
The non-linearities in the deviation from $1$ of the rescaling parameter, $\gamma$, are also generally small, somewhat larger when the model does not describe the lattice and R-ratio results well.

The two approaches also answer slightly different questions. 
The one of \app{sec:single_param} corresponds to a rescaling of the dispersive integrals derived from the experimental R-ratio in different $\sqrt{s}$ sub-intervals $I_b$, while the one presented here, of a rescaling of the fitted $a_{jb}$, which may deviate from the input $a^R_{jb}$.

Moreover, the approach described here provides an unregularized reconstruction of both the lattice observables, via the first line of \eq{eq:rescaleRexpected}, and of the timelike contributions of $\sqrt{s}$-intervals, $I_b$, as given in the second line of \eq{eq:rescaleRexpected}. 
The resolution with which this sort of reconstruction can be performed is usually limited by (anti-)correlations between the various observables that generally leads to large variances on the results.

Note that, when all the $\gamma_{b}$ factors are set to unity, one can perform the change of variable $a_{j\bar{b}} = a_j - \sum_{b \ne \bar{b}} a_{jb}$ in \eq{eq:chisqwinint}, for an arbitrary index $\bar{b}$.
The covariance matrix $\bar{C}_{\rm R}$, resulting from the propagation of $C_{\rm R}$ through this change of variable, can be decomposed into a set of uncertainties that are either uncorrelated~($\sigma_{j(b)}$) or that correlate~($s^l_{j(b)}$) the $\aR{j(b)}$:~\footnote{In quantities such as $\aR{j(b)}$, the parentheses indicate that we are considering both $\aR{j}$ and $\aR{jb}$ with $b\neq\bar b$.} 
\bea
\left[ \bar{C}_{\rm R} \right]_{j(b),k(c)} &=& \sigma_{j(b)}^2 \cdot \delta_{j(b),k(c)}\nn\\ &&+ \sum_l { s^l_{j(b)} \cdot s^l_{k(c)} }~,
\eea
with the index $l$ running over the corresponding independent uncertainty components~\cite{ListTalk}.
This enables an equivalent re-writing of the $\chi^{2}$ function from \eq{eq:chisqwinint} as
\bea
&& \chi^2 = \sum_{j,k}\left[\alat{j}-a_j\right][C^{-1}_\text{lat}]_{jk}\left[\alat{k}-a_k\right]\nn\\
  && + \min\limits_{\beta_l} \Biggl\{ \sum_{j(b \ne \bar{b})} {\frac{\left[\aR{j(b)} + \sum_l { s^l_{j(b)} \cdot \beta_l } -a_{j(b)}\right]^2}{\sigma_{j(b)}^2}}\nn\\ &&\qquad\qquad\qquad\qquad\qquad\qquad+ \sum_l { \beta_l^2 }\Biggr\}\ ,\label{eq:chi2NPs}
\eea
using nuisance parameters, $s^l_{j(b)}$, to account for the correlated uncertainties~\cite{Demortier:1999,Stump:2001gu,Botje:2001fx,Thorne:2002kn,Fogli:2002pt,ListTalk}.~\footnote{In \eq{eq:chi2NPs}, the parentheses around the $b$ and $\bar b$ indices indicate that the sum in the second term of the equation runs over the $\aR{j}$ and $\aR{jb}$.}
While the $a_j$ parameters are constrained by both the lattice and the dispersive inputs, the $a_{j,b \ne \bar{b}}$ parameters remaining after the change of variable only enter the dispersive part of the $\chi^{2}$ function. 
The minimisation of this $\chi^{2}$ function with respect to $a_{j,b \ne \bar{b}}$ allows us to exactly cancel the corresponding contribution to the $\chi^{2}$, regardless of the contributions of the shifts of the nuisance parameters away from zero, which are parameterized by the $\beta_l$ and induced by the $a^{\rm R}_{j}$ and $a^{\rm lat}_{j}$ inputs.
Switching back from the representation in terms of nuisance parameters to the one with the covariance matrix, this time in the subspace of $a_j$ parameters, allows to show that the minimum of \eq{eq:chisqwinint} with respect to $a_{j,b \ne \bar{b}}$ reduces to \eq{eq:chisqwin}, while their global minimum is provided by \eq{eq:chisqwinMin}.
This also implies that for $\gamma_{b} = 1$ the minimum of the $\chi^{2}$ function from \eq{eq:chisqwinint} is stable with respect to the choice of the $I_b$ intervals, a feature also observed numerically.

\subsection{Parameterizing possible shape changes of the R-ratio suggested by lattice results}
\label{sec:legendre}

Once sufficient lattice input is available, one can refine the analysis in other ways. 
Since our results suggest that the $\rho$ peak could be responsible for the disagreement with the data-driven approach, one could imagine focusing on that region and subdividing it into smaller $\sqrt{s}$-intervals. 
Then, with the methods described in \apps{sec:single_param}{sec:multiple_rescaling_fits}, one could reconstruct, via rescaling parameters, the modifications to the experimental R-ratio suggested by the lattice observables.

\subsubsection{Using complete sets of polynomials}

Because we are interested in understanding how a tension between lattice and data-driven results could suggest a necessary alteration of the normalization, slope, curvature, etc.\ of the R-ratio in a given $\sqrt{s}$-interval, it also makes sense to parameterize the deviation from the experimental R-ratio as a sum of Legendre polynomials, $P_\ell$, that form a complete basis on that interval. 
Thus for simplicity, we choose the model $M_R(s,\alpha_i)$ to correspond to the histogram of the experimental R-ratio, with the original binning and the $\alpha_i$ parameters tuned to the corresponding bin counts.
We also choose the function $F_\text{lat$\div$R}$ of \eq{eq:LatticeReeModel} to have the following form:~\footnote{This function can be generalized to include Legendre polynomials over several complementary $\sqrt{s}$-intervals, with possible continuity and smoothness conditions at the edges of the intervals, obtained by imposing appropriate constraints on the Legendre coefficients. }
\be
\label{eq:legendre}
F_\text{lat$\div$R}(s,\beta_\ell) \equiv
\left\{
\begin{array}{lr}
     \frac2{s_2-s_1}\sum_{\ell\ge 0}\beta_\ell P_\ell\left(x(s)\right), &  s\in I_1\\
     1, &  s\in I_2
\end{array}
\right.
\ ,
\ee
where $I_1=[\sqrt{s_1},\sqrt{s_2}]$ and $I_2$ is the complementary interval on the real axis.

Then we write:
\be
a_j^{M_\text{lat}}=\sum_{b=1,2}a_{j,b}^{M_\text{lat}}\ ,
\ee
where $a_{jb}^{M_\text{lat}}$, $b=1,2$, are defined via \eq{eq:a_Rlattice_Model}, with the appropriate kernel $\tilde{K}_j(s)$.
The parameters of the model are determined via the first approach described in \app{sec:general_model}. 
The coefficients, $\beta_\ell$ of the Legendre polynomials are then determined by minimizing $\chi^2_\text{lat}$ of \eq{eq:chis_LatticeModel}. 
The resulting model, $M_\text{lat}(s,\alpha_i,\beta_\ell)$, serves as a regularization of the unfolding of the lattice R-ratio, in the sense discussed at the end of \app{sec:general_model}, provided that the number of fitted Legendre coefficients is smaller than the number of moment integrals being considered.

Note that the rescaling hypothesis of \sec{sec:testing_Rratio} is a special case of the one considered here, with $\ell$ restricted to the value $0$ and $\delta=\beta_0-1$. 

\subsubsection{Using physics-driven models}

Up until this point, none of the considered modifications to the experimental R-ratio incorporate theoretical constraints which the spectral function is known to obey. 
Parametrizations that do can be found in Refs.~\cite{Hanhart:2016pcd,DeTroconiz:2001rip,deTroconiz:2004yzs,Garcia-Martin:2011iqs,Colangelo:2018mtw,Ananthanarayan:2018nyx,Davier:2019can}.

Here, for concreteness, we focus on the parametrization given in Ref.~\cite{Colangelo:2018mtw}.
It relates the R-ratio to the $\pi^+$ electromagnetic form factor in the timelike region and it is the latter which is actually parametrized. 
This parametrization satisfies many requirements of analyticity and unitarity.
It also includes 2 and 3-pion channels, as well as inelastic corrections via a conformal polynomial, with a threshold dictated by phenomenology. 
The authors estimate that this parametrization is valid from the two-pion threshold to $\sqrt{s}\simeq1\,\gev$. To fit it to the measured R-ratio, the latter must be corrected for final-state radiation \cite{Colangelo:2018mtw}.

Because we do not actually analyze results with this parametrization here, we do not describe it in detail. 
Its expression is given in Ref.~\cite{Colangelo:2018mtw}.
However, we make a few comments on its parameters and how it may be used to describe a modification to the measured R-ratio compatible with lattice results.

The first set of parameters is the two-pion phase shift at $\sqrt{s_0}=0.8\,\gev$ and $\sqrt{s_1}=1.15\,\gev$. 
The second set of parameters describes $\rho$-$\omega$ mixing, in terms of the mass and width of the $\omega$, as well as of a mixing parameter. 
The last set of parameters describes inelasticity above $\sqrt{s_\text{in}}=M_\omega+M_{\pi^0}$, with typically 4 parameters. 
That constitutes a set of nine parameters that we label $p_k$. 

In Ref.~\cite{Colangelo:2020lcg}, the authors perform a study of the modifications to the measured R-ratio implied by a value of $\amulohvp$ that differs from its data-driven prediction.
However, the assumptions behind that study, and the questions which it answers are quite different from those of the approach proposed here.

In Ref.~\cite{Colangelo:2020lcg} the authors first perform a fit of the parametrization of Ref.~\cite{Colangelo:2018mtw} to all $e^+e^-\to\pi^+\pi^-$ measurements available at the time, fixing its nine parameters.
This allows them to predict the two-pion contribution to a number of quantities, including $\amulohvp$, in the data-driven approach.
Then they release some of the nine parameters, keeping the others fixed to their data-driven values, while performing the fit once more, with the constraint that the predicted value of $\amupipi$ takes on a precise, prescribed value.
This value is allowed to vary from its nominal, data-driven value to the value of $\amupipi$ that one obtains by assuming that the full difference between a lattice result for $\amulohvp$, close to that of Ref.~\cite{Borsanyi:2020mff}, and its data-driven counterpart can be ascribed to the two-pion contribution.
In particular, the approach allows the authors to determine the $\Delta\chi^2$ between a scenario in which $\amupipi$ takes on its ``lattice'' versus its nominal value.
In turn, the value of the $\Delta\chi^2$ indicates how compatible different values of $\amupipi$ are with $e^+e^-\to\pi^+\pi^-$ data and known constraints on the pion electromagnetic form factor.
The approach also allows the authors to present correlations between $\amupipi$ values and predictions for the two-pion contribution to other observables, such as the pion electromagnetic radius or the running of $\alpha$.

Here, in a sense that we now describe, we address the problem the other way around. As discussed throughout this article, we choose to consider the measured R-ratio, or quantities derived from it, and as many observables as possible that are computed in lattice QCD and that are related to HVP.
The latter include, for instance, the contributions to $\amulohvp$ in various time windows, the running of $\alpha$ in certain spacelike intervals, etc.
Concerning the spectral function aspect of the comparison, we not only consider the $\sqrt{s}$-interval $I_1=[2 M_{\pi^\pm},1\,\gev]$, but also its complementary interval $I_2$. 
As for all the comparisons between the two approaches that we have looked upon until now, knowledge of the correlations between all quantities considered must be determined as precisely as possible. 
Then, in the notation of \app{sec:general_model}, we call $M_\text{R}(s,\alpha_i)$ the model for the R-ratio.
To be more specific, in $I_1$ one can consider $M_\text{R}(s,\alpha_i)$ to be the model of Ref.~\cite{Colangelo:2018mtw}, so that nine of the $\alpha_i$ correspond to the parameters $p_k$ discussed above.
We continue calling $\alpha_i$ the other parameters of the model.
To describe the R-ratio in $I_1$, only the $p_k$ are needed.
The $\alpha_i$ are involved in describing the R-ratio in $I_2$.
In that interval, $M_\text{R}(s,\{p_k,\alpha_i\})$ can be chosen to be any of the models discussed above for quantities related to the R-ratio, e.g.\ the histogram of the measured spectral function, with parameters $\alpha_i$ corresponding to the counts in the original, measurement bins.

Now we consider a model for the candidate lattice R-ratio, $M_\text{lat}(s,\alpha_i,\beta_j)$.
Instead of defining it via \eq{eq:LatticeReeModel}, we take it to be the same model as $M_\text{R}(s,\{p_k,\alpha_i\})$ in $I_1$, but where some of the parameters $\beta_j$ correspond to possible shifts, $\delta p_l$, of a subset of the parameters $p_k$ of the model of Ref.~\cite{Colangelo:2018mtw}.
Thus, as for $M_\text{R}(s,\{p_k,\alpha_i\})$, we rewrite the lattice R-ratio model $M_\text{lat}(s,\{p_k,\alpha_i\},\{\delta p_l,\beta_j\})$, where the parameters $\alpha_i$ are those needed to describe the measured R-ratio in $I_2$, via $M_\text{R}$, and the $\beta_j$, those that are used to express the differences between the measured and lattice spectral functions in that same interval.
For instance, the $\beta_j$ could correspond to rescaling factors of the measured R-ratio in sub-intervals of $I_2$, discussed in \app{app:beyond_rescaling}, or Legendre polynomial modifications, as discussed in \app{sec:legendre}.

The next step consists in following the second approach described in \app{sec:general_model}, in which the parameters $p_k$, $\delta p_l$, $\alpha_i$ and $\beta_j$ are simultaneously determined by minimizing $\chi^2_\text{lat}+\chi^2_\text{R}$, where $\chi^2_\text{lat}$ is given by \eq{eq:chis_LatticeModel} and $\chi^2_\text{R}$ is one of the $\chi^2$ given in \eqs{eq:chis_RModel1} {eq:chis_RModel2}.
In this physics-motivated approach, both the measured and the candidate lattice R-ratios are described by a model that incorporates as many physical constraints as desired in interval $I_1$.
Moreover, the possible differences between the data-driven and lattice HVP are expressed in terms of parameters which have a physical meaning.
For instance, suppose that the $\delta p_l$ only correspond to modifications in the two-pion phase shift at $\sqrt{s_0}=0.8\,\gev$ and $\sqrt{s_1}=1.15\,\gev$.
If the minimization of $\chi^2_\text{lat}+\chi^2_\text{R}$ gives a good $\chi^2$ value and at least one of the $\delta p_l$ is not consistent with zero, then we can argue that, in the interval $I_1$, the lattice and data-driven approaches differ in their prediction for the phase of the pion form factor.
In addition, if one has reasons to believe that $M_\text{lat}$ correctly describes the differences between the lattice and data-driven approaches and one knows which of the two approaches correctly describes HVP, then the corresponding model can be viewed as a combination of the data-driven and lattice results, and can be used to make improved predictions for quantities that involve HVP.

We leave the comparison of lattice and data-driven results via physics-motivated models, as described here, for future work.

\section{Considering more observables in the data-driven approach}
\label{app:more_moments}

In Ref.~\cite{Colangelo:2022vok} it was stated that several window integrals, with possibly narrower $[ t_\text{min},\, t_\text{max} ]$ ranges compared to the ones studied up to now, could be employed to perform a more detailed comparison of the dispersive and lattice approaches.
While such a larger set of window moments generically enhances the amount of available information, this increase will eventually be limited by the (anti-)correlations among the windows, for both the dispersive and lattice determinations.

In the following studies we use a ``blinding'' approach, in the sense of only communicating the information on the dispersive uncertainties and their correlations for the moment integrals, but not the corresponding nominal values.
This is motivated by the fact that these moments have not yet been determined via lattice QCD and we wish to preserve the possibility of doing so without any risk of bias.
This is different to the approach followed in Ref.~\cite{Colangelo:2022vok} that quotes nominal values for moment integrals derived through the dispersive approach~(obtained with different data combination methodologies from those used here), even when these integrals have not yet been computed via lattice QCD.
We believe that going forward, all groups working on this problem should adopt our more conservative approach.
Employing ``blinding'' approaches is very important when working at the level of precision involved in these studies.
This is especially the case in the current context of the existing tensions between the dispersive approaches and the lattice calculations, as well as among the various experimental results themselves.

We consider first the seven window moments discussed in Ref.~\cite{Colangelo:2022vok}, then a set of nine window moments where the $[ 1.6, \infty[$~fm window is split into three, with extra thresholds at $2.6$ and $4$~fm, and finally a set of nineteen moments including ten extra \DaHadQsqFive{q^2} results, with spacelike $q^2$ varying from $-10~\gev^2$ to $-1~\gev^2$.
\tab{tab:Correl10Dalpha9windows} shows the correlation matrix of these nineteen moment integrals, computed using the dispersive approach.
It represents a first way of quantifying the amount of independent information available in these moments.
Indeed, one can notice that moment integrals with similar kernel shapes~(e.g.\ \DaHadQsqFive{q^2} moments computed for similar $q^2$ values) are strongly correlated, while other moments have weaker correlations.
This reflects the fact that differing kernel shapes supply a more effective way of exploiting the information provided by the timelike spectra.

\begin{table*}[htpb!]
\centering
\caption{ Correlation matrix of ten \DaHadQsqFive{q^2} and nine $\amulohvpwin$ dispersive moment integrals computed for various $q^2$ values and $[ t_\text{min},\,t_\text{max} ]$ intervals with $\Delta = 0.15$ fm, respectively. }
\label{tab:Correl10Dalpha9windows}
\vspace{0.2 cm}
\resizebox{\textwidth}{!}{
\begin{tabular}
{lccccccccccccccccccc}\hline\hline
 Moment integral &  \multicolumn{19}{c}{Correlation coefficients}  \\ \hline

 \DaHadQsqFive{ -10~{\rm GeV}^2} & 1     &       &       &       &       &       &       &       &       &       &       &        &       &       &       &       &       &       &       \\
 \DaHadQsqFive{ -9~{\rm GeV}^2}  & 0.999 & 1     &       &       &       &       &       &       &       &       &       &        &       &       &       &       &       &       &       \\
 \DaHadQsqFive{ -8~{\rm GeV}^2}  & 0.999 & 0.999 & 1     &       &       &       &       &       &       &       &       &        &       &       &       &       &       &       &       \\
 \DaHadQsqFive{ -7~{\rm GeV}^2}  & 0.996 & 0.998 & 0.999 & 1     &       &       &       &       &       &       &       &        &       &       &       &       &       &       &       \\
 \DaHadQsqFive{ -6~{\rm GeV}^2}  & 0.993 & 0.995 & 0.998 & 0.999 & 1     &       &       &       &       &       &       &        &       &       &       &       &       &       &       \\
 \DaHadQsqFive{ -5~{\rm GeV}^2}  & 0.986 & 0.990 & 0.994 & 0.997 & 0.999 & 1     &       &       &       &       &       &        &       &       &       &       &       &       &       \\
 \DaHadQsqFive{ -4~{\rm GeV}^2}  & 0.976 & 0.981 & 0.986 & 0.991 & 0.995 & 0.999 & 1     &       &       &       &       &        &       &       &       &       &       &       &       \\
 \DaHadQsqFive{ -3~{\rm GeV}^2}  & 0.960 & 0.966 & 0.973 & 0.980 & 0.986 & 0.993 & 0.998 & 1     &       &       &       &        &       &       &       &       &       &       &       \\
 \DaHadQsqFive{ -2~{\rm GeV}^2}  & 0.931 & 0.939 & 0.948 & 0.957 & 0.967 & 0.977 & 0.987 & 0.996 & 1     &       &       &        &       &       &       &       &       &       &       \\
 \DaHadQsqFive{ -1~{\rm GeV}^2}  & 0.874 & 0.885 & 0.896 & 0.909 & 0.923 & 0.938 & 0.955 & 0.973 & 0.990 & 1     &       &        &       &       &       &       &       &       &       \\
 $\amulohvpwin$ $[0, 0.1 ]$ fm    & 0.806 & 0.791 & 0.774 & 0.753 & 0.728 & 0.698 & 0.660 & 0.611 & 0.543 & 0.442 & 1     &        &       &       &       &       &       &       &       \\
 $\amulohvpwin$ $[ 0.1, 0.4 ]$ fm    & 0.959 & 0.955 & 0.949 & 0.942 & 0.931 & 0.916 & 0.895 & 0.864 & 0.813 & 0.723 & 0.864 & 1       &       &       &       &       &       &       &       \\
 $\amulohvpwin$ $[ 0.4, 0.7 ]$ fm    & 0.876 & 0.887 & 0.899 & 0.912 & 0.926 & 0.940 & 0.954 & 0.966 & 0.972 & 0.958 & 0.428 & 0.786 & 1     &       &       &       &       &       &       \\
 $\amulohvpwin$ $[ 0.7, 1   ]$ fm    & 0.711 & 0.726 & 0.743 & 0.762 & 0.784 & 0.809 & 0.838 & 0.873 & 0.91  & 0.961 & 0.206 & 0.509 & 0.893 & 1     &       &       &       &       &       \\
 $\amulohvpwin$ $[ 1  , 1.3 ]$ fm    & 0.604 & 0.619 & 0.636 & 0.656 & 0.678 & 0.705 & 0.738 & 0.778 & 0.831 & 0.901 & 0.123 & 0.365 & 0.775 & 0.973 & 1     &       &       &       &       \\
 $\amulohvpwin$ $[ 1.3, 1.6 ]$ fm    & 0.553 & 0.568 & 0.584 & 0.604 & 0.626 & 0.653 & 0.686 & 0.728 & 0.783 & 0.861 & 0.093 & 0.305 & 0.710 & 0.941 & 0.993 & 1     &       &       &       \\
 $\amulohvpwin$ $[ 1.6, 2.6 ]$ fm    & 0.508 & 0.522 & 0.537 & 0.556 & 0.577 & 0.604 & 0.636 & 0.677 & 0.733 & 0.814 & 0.074 & 0.260 & 0.647 & 0.891 & 0.963 & 0.987 & 1  	  &       &       \\
 $\amulohvpwin$ $[ 2.6, 4   ]$ fm    & 0.419 & 0.431 & 0.445 & 0.461 & 0.479 & 0.502 & 0.530 & 0.567 & 0.617 & 0.694 & 0.052 & 0.197 & 0.523 & 0.753 & 0.840 & 0.885 & 0.944 & 1     &       \\
 $\amulohvpwin$ $[ 4,   \infty [$ fm & 0.312 & 0.321 & 0.332 & 0.344 & 0.358 & 0.375 & 0.397 & 0.426 & 0.466 & 0.528 & 0.034 & 0.137 & 0.381 & 0.565 & 0.646 & 0.698 & 0.787 & 0.942 & 1     \\

\hline\hline
\end{tabular}
}
\end{table*}

A way of further quantifying the amount of independent information in the moment integrals is to evaluate the eigenvalues of the corresponding covariance matrix, since strong correlations typically imply the presence of small eigenvalues.
\tabss{tab:EV7windows}{tab:EV9windows}{tab:EV10Dalpha9windows} show the dispersive uncertainties for the three sets of moments, as well as the eigenvalues of the corresponding covariance matrices, which typically show a fast drop over several orders of magnitude.
It is worth noting that the uncertainties of the \DaHadQsqFive{q^2} and $\amulohvpwin$ integrals typically differ by several orders of magnitude, just because of the natural normalization of these moments, which is also reflected in the range covered by the corresponding eigenvalues of the covariance matrix.
This effect, also present to some extent when considering separate series of $\amulohvpwin$ and \DaHadQsqFive{q^2} integrals, interferes with the intent of quantifying the amount of independent information in the moments.

To take this into account, we also consider the eigenvalues of the correlation matrices of the moment integrals, as well as of the covariance matrices normalized by the nominal values of the moments~(i.e.\ $[C_\text{R}]_{ij} / \left( \aR{i} \cdot \aR{j} \right)$).
These are also displayed in \tabss{tab:EV7windows}{tab:EV9windows}{tab:EV10Dalpha9windows}~\footnote{It is worth noting that the rotation matrices of eigenvectors from the diagonalization of the covariance, correlation and normalized covariance matrices are different.}~\footnote{The presence of some small negative eigenvalues in \tab{tab:EV10Dalpha9windows} is related to numerical effects typically occurring in presence of strong correlations among moments.}.
For the set of seven $\amulohvpwin$ dispersive moment integrals in \tab{tab:EV7windows}, these eigenvalues cover seven orders of magnitude, which reflects the strong correlations among some of these moments.
For the three sets of moments, while the largest eigenvalues are enhanced by a factor of $2-3$, as more moments are added, the smallest eigenvalues drop by more and more orders of magnitude.
This indicates that the amount of independent information~(i.e.\ of independent degrees of freedom) is less than the number of added extra moments.
Indeed, when comparing the ratios of the various eigenvalues with the largest one, a reduction of the eigenvalues by two or four orders of magnitude for \tab{tab:EV7windows} allows us to accommodate typically one extra moment in \tab{tab:EV9windows}.
At the same time, when including the extra \DaHadQsqFive{q^2} moments in \tab{tab:EV10Dalpha9windows}, the potential gain in extra degrees of freedom seems rather marginal.
This is related to the fact that the \DaHadQsqFive{q^2} moments considered here are rather strongly correlated among themselves and have rather strong correlations with at least one of the $\amulohvpwin$ moments~(see \tab{tab:Correl10Dalpha9windows}).

One can attempt to provide more quantitative conclusions concerning the number of independent degrees of freedom that are available, in the data-driven approach, for a comparison with lattice results.
For this purpose, one can consider the eigenvalues of the correlation or normalized-covariance matrix, rescaled by the largest eigenvalue.
The number of independent degrees of freedom can then be obtained by counting the number of such ratios above a chosen lower bound.
This procedure is consistent with the idea that there is a limit to the precision with which we can obtain the uncertainty of moments of the measured R-ratio and linear combinations thereof. Such a limit approximately corresponds to that lower bound.

Here we choose this lower bound to be $10^{-6}$.
With this bound we find that the seven window observables listed in \tab{tab:EV7windows} contain a significant amount of linearly independent information. 
This number rises to eight when two more long-distance window observables are considered in \tab{tab:EV9windows}.
However, when ten values of \DaHadQsqFive{q^2} are added, in \tab{tab:Correl10Dalpha9windows}, to the nine window observables of \tab{tab:EV9windows}, no additional independent degrees of freedom appear.
Thus, assuming that a lower bound of around $10^{-6}$ is appropriate, it is reasonable to conclude that the number of linearly independent moments which can be obtained from the experimental R-ratio is less than ten for the types of observables considered here, which are the ones that should be computable, in the lattice approach, with sub-percent precision.

All these and other results presented in the paper point to the importance of determining the corresonding matrices in the lattice approach and performing studies similar to those presented here, when more moments become available in that approach.
\begin{table*}[htpb!]
\centering
\caption{ Total uncertainty of seven $\amulohvpwin$ dispersive moment integrals computed for various $[ t_\text{min},\,t_\text{max} ]$ intervals with $\Delta = 0.15$ fm. The eigenvalues of the corresponding covariance, correlation and normalized covariance matrices are also shown. }
\label{tab:EV7windows}
\vspace{0.2 cm}
\begin{tabular}{lccccc}\hline\hline
 Moment integral &  &  \multicolumn{3}{c}{Eigenvalues}  \\ 
 $\amulohvpwin$ $[ t_\text{min},\,t_\text{max} ]$     &  Total uncertainty     & Covariance  & Correlation  & Normalized covariance   \\ \hline

 $[0, 0.1 ]$ fm &     $ 8.18 \cdot 10^{-12} $  &  $ 3.12 \cdot 10^{-20} $ &  $ 4.85               $  &  $ 2.04 \cdot 10^{-4} $   \\
 $[ 0.1, 0.4 ]$ fm &     $ 3.86 \cdot 10^{-11} $  &  $ 3.52 \cdot 10^{-21} $ &  $ 1.76               $  &  $ 8.39 \cdot 10^{-5} $   \\
 $[ 0.4, 0.7 ]$ fm &     $ 6.43 \cdot 10^{-11} $  &  $ 6.38 \cdot 10^{-22} $ &  $ 3.32 \cdot 10^{-1} $  &  $ 1.45 \cdot 10^{-5} $   \\
 $[ 0.7, 1   ]$ fm &     $ 8.00 \cdot 10^{-11} $  &  $ 1.53 \cdot 10^{-22} $ &  $ 4.88 \cdot 10^{-2} $  &  $ 2.06 \cdot 10^{-6} $   \\
 $[ 1  , 1.3 ]$ fm &     $ 7.90 \cdot 10^{-11} $  &  $ 8.77 \cdot 10^{-24} $ &  $ 4.27 \cdot 10^{-3} $  &  $ 1.87 \cdot 10^{-7} $   \\
 $[ 1.3, 1.6 ]$ fm &     $ 6.32 \cdot 10^{-11} $  &  $ 4.64 \cdot 10^{-25} $ &  $ 4.47 \cdot 10^{-4} $  &  $ 1.88 \cdot 10^{-8} $   \\
 $[ 1.6, \infty [$ fm &  $ 1.15 \cdot 10^{-10} $  &  $ 7.89 \cdot 10^{-26} $ &  $ 1.51 \cdot 10^{-5} $  &  $ 6.34 \cdot 10^{-10} $  \\

\hline\hline
\end{tabular}
\end{table*}

\begin{table*}[htpb!]
\centering
\caption{ Total uncertainty of nine $\amulohvpwin$ dispersive moment integrals computed for various $[ t_\text{min},\,t_\text{max} ]$ intervals with $\Delta = 0.15$ fm. The eigenvalues of the corresponding covariance, correlation and normalized covariance matrices are also shown. }
\label{tab:EV9windows}
\vspace{0.2 cm}
\begin{tabular}{lccccc}\hline\hline
 Moment integral &    &  \multicolumn{3}{c}{Eigenvalues}  \\
 $\amulohvpwin$ $[ t_\text{min},\,t_\text{max} ]$     &   Total uncertainty   & Covariance  & Correlation  & Normalized covariance   \\ 
\hline
 $[0, 0.1 ]$ fm    &  $ 8.18 \cdot 10^{-12} $  &  $ 2.76 \cdot 10^{-20} $ &  $ 6.07                $  &  $ 2.49 \cdot 10^{-4}  $   \\
 $[ 0.1, 0.4 ]$ fm    &  $ 3.86 \cdot 10^{-11} $  &  $ 3.17 \cdot 10^{-21} $ &  $ 1.99                $  &  $ 9.28 \cdot 10^{-5}  $   \\
 $[ 0.4, 0.7 ]$ fm    &  $ 6.43 \cdot 10^{-11} $  &  $ 5.11 \cdot 10^{-22} $ &  $ 6.86 \cdot 10^{-1}  $  &  $ 2.72 \cdot 10^{-5}  $   \\
 $[ 0.7, 1   ]$ fm    &  $ 8.00 \cdot 10^{-11} $  &  $ 1.33 \cdot 10^{-22} $ &  $ 2.29 \cdot 10^{-1}  $  &  $ 9.71 \cdot 10^{-6}  $   \\
 $[ 1  , 1.3 ]$ fm    &  $ 7.90 \cdot 10^{-11} $  &  $ 1.52 \cdot 10^{-23} $ &  $ 2.12 \cdot 10^{-2}  $  &  $ 8.76 \cdot 10^{-7}  $   \\
 $[ 1.3, 1.6 ]$ fm    &  $ 6.32 \cdot 10^{-11} $  &  $ 1.20 \cdot 10^{-24} $ &  $ 2.36 \cdot 10^{-3}  $  &  $ 9.89 \cdot 10^{-8}  $   \\
 $[ 1.6, 2.6 ]$ fm    &  $ 9.32 \cdot 10^{-11} $  &  $ 2.91 \cdot 10^{-25} $ &  $ 3.90 \cdot 10^{-4}  $  &  $ 1.59 \cdot 10^{-8}  $   \\
 $[ 2.6, 4   ]$ fm    &  $ 2.01 \cdot 10^{-11} $  &  $ 2.15 \cdot 10^{-26} $ &  $ 3.43 \cdot 10^{-5}  $  &  $ 1.41 \cdot 10^{-9}  $   \\
 $[ 4,   \infty [$ fm &  $ 2.64 \cdot 10^{-12} $  &  $ 1.78 \cdot 10^{-27} $ &  $ 5.83 \cdot 10^{-7}  $  &  $ 2.50 \cdot 10^{-11} $   \\

\hline\hline
\end{tabular}
\end{table*}

\begin{table*}[htpb!]
\centering
\caption{ Total uncertainty of ten \DaHadQsqFive{q^2} and nine $\amulohvpwin$ dispersive moment integrals computed for various $q^2$ values and $[ t_\text{min},\,t_\text{max} ]$ intervals with $\Delta = 0.15$ fm, respectively. The eigenvalues of the corresponding covariance, correlation and normalized covariance matrices are also shown. }
\label{tab:EV10Dalpha9windows}
\vspace{0.2 cm}
\begin{tabular}{lccccc}\hline\hline
 &   &  \multicolumn{3}{c}{Eigenvalues}  \\ 
 Moment integral  & Total uncertainty & Covariance  & Correlation  & Normalized covariance   \\ 
                 \hline

 \DaHadQsqFive{ -10~{\rm GeV}^2} &  $ 4.80 \cdot 10^{ -5} $  &  $ 1.48 \cdot 10^{ -8}  $   &  $ 14.74                $   &  $ 5.21 \cdot 10^{ -4}  $   \\
 \DaHadQsqFive{ -9~{\rm GeV}^2}  &  $ 4.64 \cdot 10^{ -5} $  &  $ 2.17 \cdot 10^{-10}  $   &  $ 3.26                 $   &  $ 1.34 \cdot 10^{ -4}  $   \\
 \DaHadQsqFive{ -8~{\rm GeV}^2}  &  $ 4.48 \cdot 10^{ -5} $  &  $ 4.09 \cdot 10^{-12}  $   &  $ 7.36 \cdot 10^{ -1}  $   &  $ 2.92 \cdot 10^{ -5}  $   \\
 \DaHadQsqFive{ -7~{\rm GeV}^2}  &  $ 4.29 \cdot 10^{ -5} $  &  $ 1.94 \cdot 10^{-14}  $   &  $ 2.40 \cdot 10^{ -1}  $   &  $ 1.02 \cdot 10^{ -5}  $   \\
 \DaHadQsqFive{ -6~{\rm GeV}^2}  &  $ 4.09 \cdot 10^{ -5} $  &  $ 2.22 \cdot 10^{-16}  $   &  $ 2.16 \cdot 10^{ -2}  $   &  $ 8.90 \cdot 10^{ -7}  $   \\
 \DaHadQsqFive{ -5~{\rm GeV}^2}  &  $ 3.85 \cdot 10^{ -5} $  &  $ 3.40 \cdot 10^{-18}  $   &  $ 2.45 \cdot 10^{ -3}  $   &  $ 1.02 \cdot 10^{ -7}  $   \\
 \DaHadQsqFive{ -4~{\rm GeV}^2}  &  $ 3.57 \cdot 10^{ -5} $  &  $ 1.99 \cdot 10^{-20}  $   &  $ 4.17 \cdot 10^{ -4}  $   &  $ 1.68 \cdot 10^{ -8}  $   \\
 \DaHadQsqFive{ -3~{\rm GeV}^2}  &  $ 3.23 \cdot 10^{ -5} $  &  $ 9.27 \cdot 10^{-23}  $   &  $ 3.92 \cdot 10^{ -5}  $   &  $ 1.59 \cdot 10^{ -9}  $   \\
 \DaHadQsqFive{ -2~{\rm GeV}^2}  &  $ 2.78 \cdot 10^{ -5} $  &  $ 1.40 \cdot 10^{-23}  $   &  $ 2.33 \cdot 10^{ -6}  $   &  $ 8.94 \cdot 10^{-11}  $   \\
 \DaHadQsqFive{ -1~{\rm GeV}^2}  &  $ 2.07 \cdot 10^{ -5} $  &  $ 7.03 \cdot 10^{-25}  $   &  $ 1.15 \cdot 10^{ -7}  $   &  $ 4.78 \cdot 10^{-12}  $   \\
 $\amulohvpwin$ $[0, 0.1 ]$ fm    &  $ 8.18 \cdot 10^{-12} $  &  $ 1.47 \cdot 10^{-25}  $   &  $ 3.46 \cdot 10^{-10}  $   &  $ 1.22 \cdot 10^{-14}  $   \\
 $\amulohvpwin$ $[ 0.1, 0.4 ]$ fm    &  $ 3.86 \cdot 10^{-11} $  &  $ 7.96 \cdot 10^{-28}  $   &  $ 3.76 \cdot 10^{-13}  $   &  $ 1.19 \cdot 10^{-17}  $   \\
 $\amulohvpwin$ $[ 0.4, 0.7 ]$ fm    &  $ 6.43 \cdot 10^{-11} $  &  $ 1.66 \cdot 10^{-28}  $   &  $ 2.28 \cdot 10^{-13}  $   &  $ 7.40 \cdot 10^{-18}  $   \\
 $\amulohvpwin$ $[ 0.7, 1   ]$ fm    &  $ 8.00 \cdot 10^{-11} $  &  $ 4.50 \cdot 10^{-30}  $   &  $ 9.74 \cdot 10^{-14}  $   &  $ 3.10 \cdot 10^{-18}  $   \\
 $\amulohvpwin$ $[ 1  , 1.3 ]$ fm    &  $ 7.90 \cdot 10^{-11} $  &  $ 5.25 \cdot 10^{-31}  $   &  $ 3.07 \cdot 10^{-14}  $   &  $ 9.72 \cdot 10^{-19}  $   \\
 $\amulohvpwin$ $[ 1.3, 1.6 ]$ fm    &  $ 6.32 \cdot 10^{-11} $  &  $ 8.57 \cdot 10^{-32}  $   &  $ 1.40 \cdot 10^{-14}  $   &  $ 4.38 \cdot 10^{-19}  $   \\
 $\amulohvpwin$ $[ 1.6, 2.6 ]$ fm    &  $ 9.32 \cdot 10^{-11} $  &  $ -4.56 \cdot 10^{-27} $   &  $ -2.24 \cdot 10^{-14} $   &  $ -7.06 \cdot 10^{-19} $   \\
 $\amulohvpwin$ $[ 2.6, 4   ]$ fm    &  $ 2.01 \cdot 10^{-11} $  &  $ -2.01 \cdot 10^{-23} $   &  $ -3.79 \cdot 10^{-14} $   &  $ -1.19 \cdot 10^{-18} $   \\
 $\amulohvpwin$ $[ 4,   \infty [$ fm &  $ 2.64 \cdot 10^{-12} $  &  $ -1.12 \cdot 10^{-22} $   &  $ -1.22 \cdot 10^{-13} $   &  $ -3.83 \cdot 10^{-18} $   \\

\hline\hline
\end{tabular}
\end{table*}


\clearpage
\bibliographystyle{apsrev4-1}
\bibliography{./Bibliography.bib}

\begin{thebibliography}{123}%
\makeatletter
\providecommand \@ifxundefined [1]{%
 \@ifx{#1\undefined}
}%
\providecommand \@ifnum [1]{%
 \ifnum #1\expandafter \@firstoftwo
 \else \expandafter \@secondoftwo
 \fi
}%
\providecommand \@ifx [1]{%
 \ifx #1\expandafter \@firstoftwo
 \else \expandafter \@secondoftwo
 \fi
}%
\providecommand \natexlab [1]{#1}%
\providecommand \enquote  [1]{``#1''}%
\providecommand \bibnamefont  [1]{#1}%
\providecommand \bibfnamefont [1]{#1}%
\providecommand \citenamefont [1]{#1}%
\providecommand \href@noop [0]{\@secondoftwo}%
\providecommand \href [0]{\begingroup \@sanitize@url \@href}%
\providecommand \@href[1]{\@@startlink{#1}\@@href}%
\providecommand \@@href[1]{\endgroup#1\@@endlink}%
\providecommand \@sanitize@url [0]{\catcode `\\12\catcode `\$12\catcode
  `\&12\catcode `\#12\catcode `\^12\catcode `\_12\catcode `\%12\relax}%
\providecommand \@@startlink[1]{}%
\providecommand \@@endlink[0]{}%
\providecommand \url  [0]{\begingroup\@sanitize@url \@url }%
\providecommand \@url [1]{\endgroup\@href {#1}{\urlprefix }}%
\providecommand \urlprefix  [0]{URL }%
\providecommand \Eprint [0]{\href }%
\providecommand \doibase [0]{http://dx.doi.org/}%
\providecommand \selectlanguage [0]{\@gobble}%
\providecommand \bibinfo  [0]{\@secondoftwo}%
\providecommand \bibfield  [0]{\@secondoftwo}%
\providecommand \translation [1]{[#1]}%
\providecommand \BibitemOpen [0]{}%
\providecommand \bibitemStop [0]{}%
\providecommand \bibitemNoStop [0]{.\EOS\space}%
\providecommand \EOS [0]{\spacefactor3000\relax}%
\providecommand \BibitemShut  [1]{\csname bibitem#1\endcsname}%
\let\auto@bib@innerbib\@empty
\bibitem [{\citenamefont {Abi}\ \emph {et~al.}(2021)\citenamefont {Abi} \emph
  {et~al.}}]{Abi:2021gix}%
  \BibitemOpen
  \bibfield  {author} {\bibinfo {author} {\bibfnamefont {B.}~\bibnamefont
  {Abi}} \emph {et~al.} (\bibinfo {collaboration} {Muon g-2}),\ }\href
  {\doibase 10.1103/PhysRevLett.126.141801} {\bibfield  {journal} {\bibinfo
  {journal} {Phys. Rev. Lett.}\ }\textbf {\bibinfo {volume} {126}},\ \bibinfo
  {pages} {141801} (\bibinfo {year} {2021})},\ \Eprint
  {http://arxiv.org/abs/2104.03281} {arXiv:2104.03281 [hep-ex]} \BibitemShut
  {NoStop}%
\bibitem [{\citenamefont {Bennett}\ \emph {et~al.}(2006)\citenamefont {Bennett}
  \emph {et~al.}}]{Bennett:2006fi}%
  \BibitemOpen
  \bibfield  {author} {\bibinfo {author} {\bibfnamefont {G.~W.}\ \bibnamefont
  {Bennett}} \emph {et~al.} (\bibinfo {collaboration} {Muon g-2}),\ }\href
  {\doibase 10.1103/PhysRevD.73.072003} {\bibfield  {journal} {\bibinfo
  {journal} {Phys. Rev.}\ }\textbf {\bibinfo {volume} {D73}},\ \bibinfo {pages}
  {072003} (\bibinfo {year} {2006})},\ \Eprint
  {http://arxiv.org/abs/hep-ex/0602035} {arXiv:hep-ex/0602035 [hep-ex]}
  \BibitemShut {NoStop}%
\bibitem [{\citenamefont {Davier}\ \emph {et~al.}(2020)\citenamefont {Davier},
  \citenamefont {Hoecker}, \citenamefont {Malaescu},\ and\ \citenamefont
  {Zhang}}]{Davier:2019can}%
  \BibitemOpen
  \bibfield  {author} {\bibinfo {author} {\bibfnamefont {M.}~\bibnamefont
  {Davier}}, \bibinfo {author} {\bibfnamefont {A.}~\bibnamefont {Hoecker}},
  \bibinfo {author} {\bibfnamefont {B.}~\bibnamefont {Malaescu}}, \ and\
  \bibinfo {author} {\bibfnamefont {Z.}~\bibnamefont {Zhang}},\ }\href
  {\doibase 10.1140/epjc/s10052-020-7792-2} {\bibfield  {journal} {\bibinfo
  {journal} {Eur. Phys. J. C}\ }\textbf {\bibinfo {volume} {80}},\ \bibinfo
  {pages} {241} (\bibinfo {year} {2020})},\ \Eprint
  {http://arxiv.org/abs/1908.00921} {arXiv:1908.00921 [hep-ph]} \BibitemShut
  {NoStop}%
\bibitem [{\citenamefont {Keshavarzi}\ \emph
  {et~al.}(2020{\natexlab{a}})\citenamefont {Keshavarzi}, \citenamefont
  {Nomura},\ and\ \citenamefont {Teubner}}]{Keshavarzi:2019abf}%
  \BibitemOpen
  \bibfield  {author} {\bibinfo {author} {\bibfnamefont {A.}~\bibnamefont
  {Keshavarzi}}, \bibinfo {author} {\bibfnamefont {D.}~\bibnamefont {Nomura}},
  \ and\ \bibinfo {author} {\bibfnamefont {T.}~\bibnamefont {Teubner}},\ }\href
  {\doibase 10.1103/PhysRevD.101.014029} {\bibfield  {journal} {\bibinfo
  {journal} {Phys. Rev. D}\ }\textbf {\bibinfo {volume} {101}},\ \bibinfo
  {pages} {014029} (\bibinfo {year} {2020}{\natexlab{a}})},\ \Eprint
  {http://arxiv.org/abs/1911.00367} {arXiv:1911.00367 [hep-ph]} \BibitemShut
  {NoStop}%
\bibitem [{\citenamefont {Aoyama}\ \emph {et~al.}(2020)\citenamefont {Aoyama}
  \emph {et~al.}}]{Aoyama:2020ynm}%
  \BibitemOpen
  \bibfield  {author} {\bibinfo {author} {\bibfnamefont {T.}~\bibnamefont
  {Aoyama}} \emph {et~al.},\ }\href {\doibase 10.1016/j.physrep.2020.07.006}
  {\bibfield  {journal} {\bibinfo  {journal} {Phys. Rept.}\ }\textbf {\bibinfo
  {volume} {887}},\ \bibinfo {pages} {1} (\bibinfo {year} {2020})},\ \Eprint
  {http://arxiv.org/abs/2006.04822} {arXiv:2006.04822 [hep-ph]} \BibitemShut
  {NoStop}%
\bibitem [{\citenamefont {Borsanyi}\ \emph {et~al.}(2021)\citenamefont
  {Borsanyi} \emph {et~al.}}]{Borsanyi:2020mff}%
  \BibitemOpen
  \bibfield  {author} {\bibinfo {author} {\bibfnamefont {S.}~\bibnamefont
  {Borsanyi}} \emph {et~al.},\ }\href {\doibase 10.1038/s41586-021-03418-1}
  {\bibfield  {journal} {\bibinfo  {journal} {Nature}\ }\textbf {\bibinfo
  {volume} {593}},\ \bibinfo {pages} {51} (\bibinfo {year} {2021})},\ \Eprint
  {http://arxiv.org/abs/2002.12347} {arXiv:2002.12347 [hep-lat]} \BibitemShut
  {NoStop}%
\bibitem [{Vbw()}]{Vbwgm2_LL_202105}%
  \BibitemOpen
  \href
  {https://indico.ijclab.in2p3.fr/event/7183/contributions/22333/attachments/16546/21411/talk_short.pdf}
  {}\bibinfo {note} {See e.g.\ Laurent Lellouch, talk entitled "Leading
  hadronic contribution to the muon magnetic moment from lattice QCD", at the
  workshop "Virtual Breakfast with $g-2$", 19 May 2021, IJClab, Paris,
  France.}\BibitemShut {Stop}%
\bibitem [{\citenamefont {Blum}\ \emph {et~al.}(2018)\citenamefont {Blum},
  \citenamefont {Boyle}, \citenamefont {Gülpers}, \citenamefont {Izubuchi},
  \citenamefont {Jin}, \citenamefont {Jung}, \citenamefont {Jüttner},
  \citenamefont {Lehner}, \citenamefont {Portelli},\ and\ \citenamefont
  {Tsang}}]{Blum:2018mom}%
  \BibitemOpen
  \bibfield  {author} {\bibinfo {author} {\bibfnamefont {T.}~\bibnamefont
  {Blum}}, \bibinfo {author} {\bibfnamefont {P.~A.}\ \bibnamefont {Boyle}},
  \bibinfo {author} {\bibfnamefont {V.}~\bibnamefont {Gülpers}}, \bibinfo
  {author} {\bibfnamefont {T.}~\bibnamefont {Izubuchi}}, \bibinfo {author}
  {\bibfnamefont {L.}~\bibnamefont {Jin}}, \bibinfo {author} {\bibfnamefont
  {C.}~\bibnamefont {Jung}}, \bibinfo {author} {\bibfnamefont {A.}~\bibnamefont
  {Jüttner}}, \bibinfo {author} {\bibfnamefont {C.}~\bibnamefont {Lehner}},
  \bibinfo {author} {\bibfnamefont {A.}~\bibnamefont {Portelli}}, \ and\
  \bibinfo {author} {\bibfnamefont {J.~T.}\ \bibnamefont {Tsang}} (\bibinfo
  {collaboration} {RBC, UKQCD}),\ }\href {\doibase
  10.1103/PhysRevLett.121.022003} {\bibfield  {journal} {\bibinfo  {journal}
  {Phys. Rev. Lett.}\ }\textbf {\bibinfo {volume} {121}},\ \bibinfo {pages}
  {022003} (\bibinfo {year} {2018})},\ \Eprint
  {http://arxiv.org/abs/1801.07224} {arXiv:1801.07224 [hep-lat]} \BibitemShut
  {NoStop}%
\bibitem [{\citenamefont {C\`e}\ \emph
  {et~al.}(2022{\natexlab{a}})\citenamefont {C\`e} \emph
  {et~al.}}]{Ce:2022kxy}%
  \BibitemOpen
  \bibfield  {author} {\bibinfo {author} {\bibfnamefont {M.}~\bibnamefont
  {C\`e}} \emph {et~al.},\ }\href {\doibase 10.1103/PhysRevD.106.114502}
  {\bibfield  {journal} {\bibinfo  {journal} {Phys. Rev. D}\ }\textbf {\bibinfo
  {volume} {106}},\ \bibinfo {pages} {114502} (\bibinfo {year}
  {2022}{\natexlab{a}})},\ \Eprint {http://arxiv.org/abs/2206.06582}
  {arXiv:2206.06582 [hep-lat]} \BibitemShut {NoStop}%
\bibitem [{\citenamefont {Alexandrou}\ \emph {et~al.}(2023)\citenamefont
  {Alexandrou} \emph {et~al.}}]{ExtendedTwistedMass:2022jpw}%
  \BibitemOpen
  \bibfield  {author} {\bibinfo {author} {\bibfnamefont {C.}~\bibnamefont
  {Alexandrou}} \emph {et~al.} (\bibinfo {collaboration} {Extended Twisted
  Mass}),\ }\href {\doibase 10.1103/PhysRevD.107.074506} {\bibfield  {journal}
  {\bibinfo  {journal} {Phys. Rev. D}\ }\textbf {\bibinfo {volume} {107}},\
  \bibinfo {pages} {074506} (\bibinfo {year} {2023})},\ \Eprint
  {http://arxiv.org/abs/2206.15084} {arXiv:2206.15084 [hep-lat]} \BibitemShut
  {NoStop}%
\bibitem [{\citenamefont {Blum}\ \emph {et~al.}(2023)\citenamefont {Blum} \emph
  {et~al.}}]{Blum:2023qou}%
  \BibitemOpen
  \bibfield  {author} {\bibinfo {author} {\bibfnamefont {T.}~\bibnamefont
  {Blum}} \emph {et~al.},\ }\href@noop {} {\  (\bibinfo {year} {2023})},\
  \Eprint {http://arxiv.org/abs/2301.08696} {arXiv:2301.08696 [hep-lat]}
  \BibitemShut {NoStop}%
\bibitem [{\citenamefont {Bazavov}\ \emph {et~al.}(2023)\citenamefont {Bazavov}
  \emph {et~al.}}]{Bazavov:2023has}%
  \BibitemOpen
  \bibfield  {author} {\bibinfo {author} {\bibfnamefont {A.}~\bibnamefont
  {Bazavov}} \emph {et~al.},\ }\href@noop {} {\  (\bibinfo {year} {2023})},\
  \Eprint {http://arxiv.org/abs/2301.08274} {arXiv:2301.08274 [hep-lat]}
  \BibitemShut {NoStop}%
\bibitem [{\citenamefont {Ignatov}\ \emph {et~al.}(2023)\citenamefont {Ignatov}
  \emph {et~al.}}]{CMD-3:2023alj}%
  \BibitemOpen
  \bibfield  {author} {\bibinfo {author} {\bibfnamefont {F.~V.}\ \bibnamefont
  {Ignatov}} \emph {et~al.} (\bibinfo {collaboration} {CMD-3}),\ }\href@noop {}
  {\  (\bibinfo {year} {2023})},\ \Eprint {http://arxiv.org/abs/2302.08834}
  {arXiv:2302.08834 [hep-ex]} \BibitemShut {NoStop}%
\bibitem [{\citenamefont {Akhmetshin}\ \emph {et~al.}(2007)\citenamefont
  {Akhmetshin} \emph {et~al.}}]{CMD-2:2006gxt}%
  \BibitemOpen
  \bibfield  {author} {\bibinfo {author} {\bibfnamefont {R.~R.}\ \bibnamefont
  {Akhmetshin}} \emph {et~al.} (\bibinfo {collaboration} {CMD-2}),\ }\href
  {\doibase 10.1016/j.physletb.2007.01.073} {\bibfield  {journal} {\bibinfo
  {journal} {Phys. Lett. B}\ }\textbf {\bibinfo {volume} {648}},\ \bibinfo
  {pages} {28} (\bibinfo {year} {2007})},\ \Eprint
  {http://arxiv.org/abs/hep-ex/0610021} {arXiv:hep-ex/0610021} \BibitemShut
  {NoStop}%
\bibitem [{\citenamefont {Crivellin}\ \emph {et~al.}(2020)\citenamefont
  {Crivellin}, \citenamefont {Hoferichter}, \citenamefont {Manzari},\ and\
  \citenamefont {Montull}}]{Crivellin:2020zul}%
  \BibitemOpen
  \bibfield  {author} {\bibinfo {author} {\bibfnamefont {A.}~\bibnamefont
  {Crivellin}}, \bibinfo {author} {\bibfnamefont {M.}~\bibnamefont
  {Hoferichter}}, \bibinfo {author} {\bibfnamefont {C.~A.}\ \bibnamefont
  {Manzari}}, \ and\ \bibinfo {author} {\bibfnamefont {M.}~\bibnamefont
  {Montull}},\ }\href {\doibase 10.1103/PhysRevLett.125.091801} {\bibfield
  {journal} {\bibinfo  {journal} {Phys. Rev. Lett.}\ }\textbf {\bibinfo
  {volume} {125}},\ \bibinfo {pages} {091801} (\bibinfo {year} {2020})},\
  \Eprint {http://arxiv.org/abs/2003.04886} {arXiv:2003.04886 [hep-ph]}
  \BibitemShut {NoStop}%
\bibitem [{\citenamefont {Keshavarzi}\ \emph
  {et~al.}(2020{\natexlab{b}})\citenamefont {Keshavarzi}, \citenamefont
  {Marciano}, \citenamefont {Passera},\ and\ \citenamefont
  {Sirlin}}]{Keshavarzi:2020bfy}%
  \BibitemOpen
  \bibfield  {author} {\bibinfo {author} {\bibfnamefont {A.}~\bibnamefont
  {Keshavarzi}}, \bibinfo {author} {\bibfnamefont {W.~J.}\ \bibnamefont
  {Marciano}}, \bibinfo {author} {\bibfnamefont {M.}~\bibnamefont {Passera}}, \
  and\ \bibinfo {author} {\bibfnamefont {A.}~\bibnamefont {Sirlin}},\ }\href
  {\doibase 10.1103/PhysRevD.102.033002} {\bibfield  {journal} {\bibinfo
  {journal} {Phys. Rev. D}\ }\textbf {\bibinfo {volume} {102}},\ \bibinfo
  {pages} {033002} (\bibinfo {year} {2020}{\natexlab{b}})},\ \Eprint
  {http://arxiv.org/abs/2006.12666} {arXiv:2006.12666 [hep-ph]} \BibitemShut
  {NoStop}%
\bibitem [{\citenamefont {de~Rafael}(2020)}]{deRafael:2020uif}%
  \BibitemOpen
  \bibfield  {author} {\bibinfo {author} {\bibfnamefont {E.}~\bibnamefont
  {de~Rafael}},\ }\href {\doibase 10.1103/PhysRevD.102.056025} {\bibfield
  {journal} {\bibinfo  {journal} {Phys. Rev. D}\ }\textbf {\bibinfo {volume}
  {102}},\ \bibinfo {pages} {056025} (\bibinfo {year} {2020})},\ \Eprint
  {http://arxiv.org/abs/2006.13880} {arXiv:2006.13880 [hep-ph]} \BibitemShut
  {NoStop}%
\bibitem [{\citenamefont {Malaescu}\ and\ \citenamefont
  {Schott}(2021)}]{Malaescu:2020zuc}%
  \BibitemOpen
  \bibfield  {author} {\bibinfo {author} {\bibfnamefont {B.}~\bibnamefont
  {Malaescu}}\ and\ \bibinfo {author} {\bibfnamefont {M.}~\bibnamefont
  {Schott}},\ }\href {\doibase 10.1140/epjc/s10052-021-08848-9} {\bibfield
  {journal} {\bibinfo  {journal} {Eur. Phys. J. C}\ }\textbf {\bibinfo {volume}
  {81}},\ \bibinfo {pages} {46} (\bibinfo {year} {2021})},\ \Eprint
  {http://arxiv.org/abs/2008.08107} {arXiv:2008.08107 [hep-ph]} \BibitemShut
  {NoStop}%
\bibitem [{\citenamefont {Borsanyi}\ \emph
  {et~al.}(2018{\natexlab{a}})\citenamefont {Borsanyi} \emph
  {et~al.}}]{Borsanyi:2017zdw}%
  \BibitemOpen
  \bibfield  {author} {\bibinfo {author} {\bibfnamefont {S.}~\bibnamefont
  {Borsanyi}} \emph {et~al.} (\bibinfo {collaboration}
  {Budapest-Marseille-Wuppertal}),\ }\href {\doibase
  10.1103/PhysRevLett.121.022002} {\bibfield  {journal} {\bibinfo  {journal}
  {Phys. Rev. Lett.}\ }\textbf {\bibinfo {volume} {121}},\ \bibinfo {pages}
  {022002} (\bibinfo {year} {2018}{\natexlab{a}})},\ \Eprint
  {http://arxiv.org/abs/1711.04980} {arXiv:1711.04980 [hep-lat]} \BibitemShut
  {NoStop}%
\bibitem [{\citenamefont {C\`e}\ \emph
  {et~al.}(2022{\natexlab{b}})\citenamefont {C\`e}, \citenamefont {G\'erardin},
  \citenamefont {von Hippel}, \citenamefont {Meyer}, \citenamefont {Miura},
  \citenamefont {Ottnad}, \citenamefont {Risch}, \citenamefont {San~Jos\'e},
  \citenamefont {Wilhelm},\ and\ \citenamefont {Wittig}}]{Ce:2022eix}%
  \BibitemOpen
  \bibfield  {author} {\bibinfo {author} {\bibfnamefont {M.}~\bibnamefont
  {C\`e}}, \bibinfo {author} {\bibfnamefont {A.}~\bibnamefont {G\'erardin}},
  \bibinfo {author} {\bibfnamefont {G.}~\bibnamefont {von Hippel}}, \bibinfo
  {author} {\bibfnamefont {H.~B.}\ \bibnamefont {Meyer}}, \bibinfo {author}
  {\bibfnamefont {K.}~\bibnamefont {Miura}}, \bibinfo {author} {\bibfnamefont
  {K.}~\bibnamefont {Ottnad}}, \bibinfo {author} {\bibfnamefont
  {A.}~\bibnamefont {Risch}}, \bibinfo {author} {\bibfnamefont
  {T.}~\bibnamefont {San~Jos\'e}}, \bibinfo {author} {\bibfnamefont
  {J.}~\bibnamefont {Wilhelm}}, \ and\ \bibinfo {author} {\bibfnamefont
  {H.}~\bibnamefont {Wittig}},\ }\href {\doibase 10.1007/JHEP08(2022)220}
  {\bibfield  {journal} {\bibinfo  {journal} {JHEP}\ }\textbf {\bibinfo
  {volume} {08}},\ \bibinfo {pages} {220} (\bibinfo {year}
  {2022}{\natexlab{b}})},\ \Eprint {http://arxiv.org/abs/2203.08676}
  {arXiv:2203.08676 [hep-lat]} \BibitemShut {NoStop}%
\bibitem [{\citenamefont {Bernecker}\ and\ \citenamefont
  {Meyer}(2011)}]{Bernecker:2011gh}%
  \BibitemOpen
  \bibfield  {author} {\bibinfo {author} {\bibfnamefont {D.}~\bibnamefont
  {Bernecker}}\ and\ \bibinfo {author} {\bibfnamefont {H.~B.}\ \bibnamefont
  {Meyer}},\ }\href {\doibase 10.1140/epja/i2011-11148-6} {\bibfield  {journal}
  {\bibinfo  {journal} {Eur.Phys.J.}\ }\textbf {\bibinfo {volume} {A47}},\
  \bibinfo {pages} {148} (\bibinfo {year} {2011})},\ \Eprint
  {http://arxiv.org/abs/1107.4388} {1107.4388} \BibitemShut {NoStop}%
\bibitem [{\citenamefont {Hansen}\ \emph {et~al.}(2019)\citenamefont {Hansen},
  \citenamefont {Lupo},\ and\ \citenamefont {Tantalo}}]{Hansen:2019idp}%
  \BibitemOpen
  \bibfield  {author} {\bibinfo {author} {\bibfnamefont {M.}~\bibnamefont
  {Hansen}}, \bibinfo {author} {\bibfnamefont {A.}~\bibnamefont {Lupo}}, \ and\
  \bibinfo {author} {\bibfnamefont {N.}~\bibnamefont {Tantalo}},\ }\href
  {\doibase 10.1103/PhysRevD.99.094508} {\bibfield  {journal} {\bibinfo
  {journal} {Phys. Rev.}\ }\textbf {\bibinfo {volume} {D99}},\ \bibinfo {pages}
  {094508} (\bibinfo {year} {2019})},\ \Eprint
  {http://arxiv.org/abs/1903.06476} {arXiv:1903.06476 [hep-lat]} \BibitemShut
  {NoStop}%
\bibitem [{\citenamefont {Alexandrou}\ \emph {et~al.}(2022)\citenamefont
  {Alexandrou} \emph {et~al.}}]{Alexandrou:2022tyn}%
  \BibitemOpen
  \bibfield  {author} {\bibinfo {author} {\bibfnamefont {C.}~\bibnamefont
  {Alexandrou}} \emph {et~al.},\ }\href@noop {} {\  (\bibinfo {year} {2022})},\
  \Eprint {http://arxiv.org/abs/2212.08467} {arXiv:2212.08467 [hep-lat]}
  \BibitemShut {NoStop}%
\bibitem [{\citenamefont {Colangelo}\ \emph {et~al.}(2019)\citenamefont
  {Colangelo}, \citenamefont {Hoferichter},\ and\ \citenamefont
  {Stoffer}}]{Colangelo:2018mtw}%
  \BibitemOpen
  \bibfield  {author} {\bibinfo {author} {\bibfnamefont {G.}~\bibnamefont
  {Colangelo}}, \bibinfo {author} {\bibfnamefont {M.}~\bibnamefont
  {Hoferichter}}, \ and\ \bibinfo {author} {\bibfnamefont {P.}~\bibnamefont
  {Stoffer}},\ }\href {\doibase 10.1007/JHEP02(2019)006} {\bibfield  {journal}
  {\bibinfo  {journal} {JHEP}\ }\textbf {\bibinfo {volume} {02}},\ \bibinfo
  {pages} {006} (\bibinfo {year} {2019})},\ \Eprint
  {http://arxiv.org/abs/1810.00007} {arXiv:1810.00007 [hep-ph]} \BibitemShut
  {NoStop}%
\bibitem [{\citenamefont {Colangelo}\ \emph {et~al.}(2021)\citenamefont
  {Colangelo}, \citenamefont {Hoferichter},\ and\ \citenamefont
  {Stoffer}}]{Colangelo:2020lcg}%
  \BibitemOpen
  \bibfield  {author} {\bibinfo {author} {\bibfnamefont {G.}~\bibnamefont
  {Colangelo}}, \bibinfo {author} {\bibfnamefont {M.}~\bibnamefont
  {Hoferichter}}, \ and\ \bibinfo {author} {\bibfnamefont {P.}~\bibnamefont
  {Stoffer}},\ }\href {\doibase 10.1016/j.physletb.2021.136073} {\bibfield
  {journal} {\bibinfo  {journal} {Phys. Lett. B}\ }\textbf {\bibinfo {volume}
  {814}},\ \bibinfo {pages} {136073} (\bibinfo {year} {2021})},\ \Eprint
  {http://arxiv.org/abs/2010.07943} {arXiv:2010.07943 [hep-ph]} \BibitemShut
  {NoStop}%
\bibitem [{\citenamefont {Colangelo}\ \emph {et~al.}(2022)\citenamefont
  {Colangelo}, \citenamefont {El-Khadra}, \citenamefont {Hoferichter},
  \citenamefont {Keshavarzi}, \citenamefont {Lehner}, \citenamefont {Stoffer},\
  and\ \citenamefont {Teubner}}]{Colangelo:2022vok}%
  \BibitemOpen
  \bibfield  {author} {\bibinfo {author} {\bibfnamefont {G.}~\bibnamefont
  {Colangelo}}, \bibinfo {author} {\bibfnamefont {A.~X.}\ \bibnamefont
  {El-Khadra}}, \bibinfo {author} {\bibfnamefont {M.}~\bibnamefont
  {Hoferichter}}, \bibinfo {author} {\bibfnamefont {A.}~\bibnamefont
  {Keshavarzi}}, \bibinfo {author} {\bibfnamefont {C.}~\bibnamefont {Lehner}},
  \bibinfo {author} {\bibfnamefont {P.}~\bibnamefont {Stoffer}}, \ and\
  \bibinfo {author} {\bibfnamefont {T.}~\bibnamefont {Teubner}},\ }\href
  {\doibase 10.1016/j.physletb.2022.137313} {\bibfield  {journal} {\bibinfo
  {journal} {Phys. Lett. B}\ }\textbf {\bibinfo {volume} {833}},\ \bibinfo
  {pages} {137313} (\bibinfo {year} {2022})},\ \Eprint
  {http://arxiv.org/abs/2205.12963} {arXiv:2205.12963 [hep-ph]} \BibitemShut
  {NoStop}%
\bibitem [{\citenamefont {Boito}\ \emph {et~al.}(2023)\citenamefont {Boito},
  \citenamefont {Golterman}, \citenamefont {Maltman},\ and\ \citenamefont
  {Peris}}]{Boito:2022njs}%
  \BibitemOpen
  \bibfield  {author} {\bibinfo {author} {\bibfnamefont {D.}~\bibnamefont
  {Boito}}, \bibinfo {author} {\bibfnamefont {M.}~\bibnamefont {Golterman}},
  \bibinfo {author} {\bibfnamefont {K.}~\bibnamefont {Maltman}}, \ and\
  \bibinfo {author} {\bibfnamefont {S.}~\bibnamefont {Peris}},\ }\href
  {\doibase 10.1103/PhysRevD.107.034512} {\bibfield  {journal} {\bibinfo
  {journal} {Phys. Rev. D}\ }\textbf {\bibinfo {volume} {107}},\ \bibinfo
  {pages} {034512} (\bibinfo {year} {2023})},\ \Eprint
  {http://arxiv.org/abs/2210.13677} {arXiv:2210.13677 [hep-lat]} \BibitemShut
  {NoStop}%
\bibitem [{\citenamefont {Brandt}\ \emph {et~al.}(2015)\citenamefont {Brandt},
  \citenamefont {Francis}, \citenamefont {Meyer},\ and\ \citenamefont
  {Robaina}}]{Brandt:2015sxa}%
  \BibitemOpen
  \bibfield  {author} {\bibinfo {author} {\bibfnamefont {B.~B.}\ \bibnamefont
  {Brandt}}, \bibinfo {author} {\bibfnamefont {A.}~\bibnamefont {Francis}},
  \bibinfo {author} {\bibfnamefont {H.~B.}\ \bibnamefont {Meyer}}, \ and\
  \bibinfo {author} {\bibfnamefont {D.}~\bibnamefont {Robaina}},\ }\href
  {\doibase 10.1103/PhysRevD.92.094510} {\bibfield  {journal} {\bibinfo
  {journal} {Phys. Rev. D}\ }\textbf {\bibinfo {volume} {92}},\ \bibinfo
  {pages} {094510} (\bibinfo {year} {2015})},\ \Eprint
  {http://arxiv.org/abs/1506.05732} {arXiv:1506.05732 [hep-lat]} \BibitemShut
  {NoStop}%
\bibitem [{\citenamefont {Hansen}\ \emph {et~al.}(2017)\citenamefont {Hansen},
  \citenamefont {Meyer},\ and\ \citenamefont {Robaina}}]{Hansen:2017mnd}%
  \BibitemOpen
  \bibfield  {author} {\bibinfo {author} {\bibfnamefont {M.~T.}\ \bibnamefont
  {Hansen}}, \bibinfo {author} {\bibfnamefont {H.~B.}\ \bibnamefont {Meyer}}, \
  and\ \bibinfo {author} {\bibfnamefont {D.}~\bibnamefont {Robaina}},\ }\href
  {\doibase 10.1103/PhysRevD.96.094513} {\bibfield  {journal} {\bibinfo
  {journal} {Phys. Rev. D}\ }\textbf {\bibinfo {volume} {96}},\ \bibinfo
  {pages} {094513} (\bibinfo {year} {2017})},\ \Eprint
  {http://arxiv.org/abs/1704.08993} {arXiv:1704.08993 [hep-lat]} \BibitemShut
  {NoStop}%
\bibitem [{\citenamefont {Bulava}\ and\ \citenamefont
  {Hansen}(2019)}]{Bulava:2019kbi}%
  \BibitemOpen
  \bibfield  {author} {\bibinfo {author} {\bibfnamefont {J.}~\bibnamefont
  {Bulava}}\ and\ \bibinfo {author} {\bibfnamefont {M.~T.}\ \bibnamefont
  {Hansen}},\ }\href {\doibase 10.1103/PhysRevD.100.034521} {\bibfield
  {journal} {\bibinfo  {journal} {Phys. Rev. D}\ }\textbf {\bibinfo {volume}
  {100}},\ \bibinfo {pages} {034521} (\bibinfo {year} {2019})},\ \Eprint
  {http://arxiv.org/abs/1903.11735} {arXiv:1903.11735 [hep-lat]} \BibitemShut
  {NoStop}%
\bibitem [{\citenamefont {Bonanno}\ \emph {et~al.}(2023)\citenamefont
  {Bonanno}, \citenamefont {D'Angelo}, \citenamefont {D'Elia}, \citenamefont
  {Maio},\ and\ \citenamefont {Naviglio}}]{Bonanno:2023ljc}%
  \BibitemOpen
  \bibfield  {author} {\bibinfo {author} {\bibfnamefont {C.}~\bibnamefont
  {Bonanno}}, \bibinfo {author} {\bibfnamefont {F.}~\bibnamefont {D'Angelo}},
  \bibinfo {author} {\bibfnamefont {M.}~\bibnamefont {D'Elia}}, \bibinfo
  {author} {\bibfnamefont {L.}~\bibnamefont {Maio}}, \ and\ \bibinfo {author}
  {\bibfnamefont {M.}~\bibnamefont {Naviglio}},\ }\href@noop {} {\  (\bibinfo
  {year} {2023})},\ \Eprint {http://arxiv.org/abs/2305.17120} {arXiv:2305.17120
  [hep-lat]} \BibitemShut {NoStop}%
\bibitem [{\citenamefont {Frezzotti}\ \emph {et~al.}(2023)\citenamefont
  {Frezzotti}, \citenamefont {Gagliardi}, \citenamefont {Lubicz}, \citenamefont
  {Sanfilippo}, \citenamefont {Simula},\ and\ \citenamefont
  {Tantalo}}]{Frezzotti:2023nun}%
  \BibitemOpen
  \bibfield  {author} {\bibinfo {author} {\bibfnamefont {R.}~\bibnamefont
  {Frezzotti}}, \bibinfo {author} {\bibfnamefont {G.}~\bibnamefont
  {Gagliardi}}, \bibinfo {author} {\bibfnamefont {V.}~\bibnamefont {Lubicz}},
  \bibinfo {author} {\bibfnamefont {F.}~\bibnamefont {Sanfilippo}}, \bibinfo
  {author} {\bibfnamefont {S.}~\bibnamefont {Simula}}, \ and\ \bibinfo {author}
  {\bibfnamefont {N.}~\bibnamefont {Tantalo}},\ }\href@noop {} {\  (\bibinfo
  {year} {2023})},\ \Eprint {http://arxiv.org/abs/2306.07228} {arXiv:2306.07228
  [hep-lat]} \BibitemShut {NoStop}%
\bibitem [{\citenamefont {Bailas}\ \emph {et~al.}(2020)\citenamefont {Bailas},
  \citenamefont {Hashimoto},\ and\ \citenamefont {Ishikawa}}]{Bailas:2020qmv}%
  \BibitemOpen
  \bibfield  {author} {\bibinfo {author} {\bibfnamefont {G.}~\bibnamefont
  {Bailas}}, \bibinfo {author} {\bibfnamefont {S.}~\bibnamefont {Hashimoto}}, \
  and\ \bibinfo {author} {\bibfnamefont {T.}~\bibnamefont {Ishikawa}},\ }\href
  {\doibase 10.1093/ptep/ptaa044} {\bibfield  {journal} {\bibinfo  {journal}
  {PTEP}\ }\textbf {\bibinfo {volume} {2020}},\ \bibinfo {pages} {043B07}
  (\bibinfo {year} {2020})},\ \Eprint {http://arxiv.org/abs/2001.11779}
  {arXiv:2001.11779 [hep-lat]} \BibitemShut {NoStop}%
\bibitem [{\citenamefont {Nakahara}\ \emph {et~al.}(1999)\citenamefont
  {Nakahara}, \citenamefont {Asakawa},\ and\ \citenamefont
  {Hatsuda}}]{Nakahara:1999vy}%
  \BibitemOpen
  \bibfield  {author} {\bibinfo {author} {\bibfnamefont {Y.}~\bibnamefont
  {Nakahara}}, \bibinfo {author} {\bibfnamefont {M.}~\bibnamefont {Asakawa}}, \
  and\ \bibinfo {author} {\bibfnamefont {T.}~\bibnamefont {Hatsuda}},\ }\href
  {\doibase 10.1103/PhysRevD.60.091503} {\bibfield  {journal} {\bibinfo
  {journal} {Phys. Rev. D}\ }\textbf {\bibinfo {volume} {60}},\ \bibinfo
  {pages} {091503} (\bibinfo {year} {1999})},\ \Eprint
  {http://arxiv.org/abs/hep-lat/9905034} {arXiv:hep-lat/9905034} \BibitemShut
  {NoStop}%
\bibitem [{\citenamefont {Nakahara}\ \emph {et~al.}(2000)\citenamefont
  {Nakahara}, \citenamefont {Asakawa},\ and\ \citenamefont
  {Hatsuda}}]{Nakahara:1999bm}%
  \BibitemOpen
  \bibfield  {author} {\bibinfo {author} {\bibfnamefont {Y.}~\bibnamefont
  {Nakahara}}, \bibinfo {author} {\bibfnamefont {M.}~\bibnamefont {Asakawa}}, \
  and\ \bibinfo {author} {\bibfnamefont {T.}~\bibnamefont {Hatsuda}},\ }\href
  {\doibase 10.1016/S0920-5632(00)91620-7} {\bibfield  {journal} {\bibinfo
  {journal} {Nucl. Phys. B Proc. Suppl.}\ }\textbf {\bibinfo {volume} {83}},\
  \bibinfo {pages} {191} (\bibinfo {year} {2000})},\ \Eprint
  {http://arxiv.org/abs/hep-lat/9909137} {arXiv:hep-lat/9909137} \BibitemShut
  {NoStop}%
\bibitem [{\citenamefont {Asakawa}\ \emph {et~al.}(2001)\citenamefont
  {Asakawa}, \citenamefont {Hatsuda},\ and\ \citenamefont
  {Nakahara}}]{Asakawa:2000tr}%
  \BibitemOpen
  \bibfield  {author} {\bibinfo {author} {\bibfnamefont {M.}~\bibnamefont
  {Asakawa}}, \bibinfo {author} {\bibfnamefont {T.}~\bibnamefont {Hatsuda}}, \
  and\ \bibinfo {author} {\bibfnamefont {Y.}~\bibnamefont {Nakahara}},\ }\href
  {\doibase 10.1016/S0146-6410(01)00150-8} {\bibfield  {journal} {\bibinfo
  {journal} {Prog. Part. Nucl. Phys.}\ }\textbf {\bibinfo {volume} {46}},\
  \bibinfo {pages} {459} (\bibinfo {year} {2001})},\ \Eprint
  {http://arxiv.org/abs/hep-lat/0011040} {arXiv:hep-lat/0011040} \BibitemShut
  {NoStop}%
\bibitem [{\citenamefont {Rothkopf}\ \emph {et~al.}(2012)\citenamefont
  {Rothkopf}, \citenamefont {Hatsuda},\ and\ \citenamefont
  {Sasaki}}]{Rothkopf:2011db}%
  \BibitemOpen
  \bibfield  {author} {\bibinfo {author} {\bibfnamefont {A.}~\bibnamefont
  {Rothkopf}}, \bibinfo {author} {\bibfnamefont {T.}~\bibnamefont {Hatsuda}}, \
  and\ \bibinfo {author} {\bibfnamefont {S.}~\bibnamefont {Sasaki}},\ }\href
  {\doibase 10.1103/PhysRevLett.108.162001} {\bibfield  {journal} {\bibinfo
  {journal} {Phys. Rev. Lett.}\ }\textbf {\bibinfo {volume} {108}},\ \bibinfo
  {pages} {162001} (\bibinfo {year} {2012})},\ \Eprint
  {http://arxiv.org/abs/1108.1579} {arXiv:1108.1579 [hep-lat]} \BibitemShut
  {NoStop}%
\bibitem [{\citenamefont {Rothkopf}(2013)}]{Rothkopf:2011ef}%
  \BibitemOpen
  \bibfield  {author} {\bibinfo {author} {\bibfnamefont {A.}~\bibnamefont
  {Rothkopf}},\ }\href {\doibase 10.1016/j.jcp.2012.12.023} {\bibfield
  {journal} {\bibinfo  {journal} {J. Comput. Phys.}\ }\textbf {\bibinfo
  {volume} {238}},\ \bibinfo {pages} {106} (\bibinfo {year} {2013})},\ \Eprint
  {http://arxiv.org/abs/1110.6285} {arXiv:1110.6285 [physics.comp-ph]}
  \BibitemShut {NoStop}%
\bibitem [{\citenamefont {Asakawa}(2020)}]{Asakawa:2020hjs}%
  \BibitemOpen
  \bibfield  {author} {\bibinfo {author} {\bibfnamefont {M.}~\bibnamefont
  {Asakawa}},\ }\href@noop {} {\  (\bibinfo {year} {2020})},\ \Eprint
  {http://arxiv.org/abs/2001.10205} {arXiv:2001.10205 [hep-ph]} \BibitemShut
  {NoStop}%
\bibitem [{\citenamefont {Rothkopf}(2020)}]{Rothkopf:2020qqt}%
  \BibitemOpen
  \bibfield  {author} {\bibinfo {author} {\bibfnamefont {A.}~\bibnamefont
  {Rothkopf}},\ }\href {\doibase 10.3390/data5030085} {\  (\bibinfo {year}
  {2020}),\ 10.3390/data5030085},\ \Eprint {http://arxiv.org/abs/2002.09865}
  {arXiv:2002.09865 [physics.data-an]} \BibitemShut {NoStop}%
\bibitem [{\citenamefont {Burnier}\ and\ \citenamefont
  {Rothkopf}(2013)}]{Burnier:2013nla}%
  \BibitemOpen
  \bibfield  {author} {\bibinfo {author} {\bibfnamefont {Y.}~\bibnamefont
  {Burnier}}\ and\ \bibinfo {author} {\bibfnamefont {A.}~\bibnamefont
  {Rothkopf}},\ }\href {\doibase 10.1103/PhysRevLett.111.182003} {\bibfield
  {journal} {\bibinfo  {journal} {Phys. Rev. Lett.}\ }\textbf {\bibinfo
  {volume} {111}},\ \bibinfo {pages} {182003} (\bibinfo {year} {2013})},\
  \Eprint {http://arxiv.org/abs/1307.6106} {arXiv:1307.6106 [hep-lat]}
  \BibitemShut {NoStop}%
\bibitem [{\citenamefont {Rothkopf}(2017)}]{Rothkopf:2016luz}%
  \BibitemOpen
  \bibfield  {author} {\bibinfo {author} {\bibfnamefont {A.}~\bibnamefont
  {Rothkopf}},\ }\href {\doibase 10.1103/PhysRevD.95.056016} {\bibfield
  {journal} {\bibinfo  {journal} {Phys. Rev. D}\ }\textbf {\bibinfo {volume}
  {95}},\ \bibinfo {pages} {056016} (\bibinfo {year} {2017})},\ \Eprint
  {http://arxiv.org/abs/1611.00482} {arXiv:1611.00482 [hep-ph]} \BibitemShut
  {NoStop}%
\bibitem [{\citenamefont {Cyrol}\ \emph {et~al.}(2018)\citenamefont {Cyrol},
  \citenamefont {Pawlowski}, \citenamefont {Rothkopf},\ and\ \citenamefont
  {Wink}}]{Cyrol:2018xeq}%
  \BibitemOpen
  \bibfield  {author} {\bibinfo {author} {\bibfnamefont {A.~K.}\ \bibnamefont
  {Cyrol}}, \bibinfo {author} {\bibfnamefont {J.~M.}\ \bibnamefont
  {Pawlowski}}, \bibinfo {author} {\bibfnamefont {A.}~\bibnamefont {Rothkopf}},
  \ and\ \bibinfo {author} {\bibfnamefont {N.}~\bibnamefont {Wink}},\ }\href
  {\doibase 10.21468/SciPostPhys.5.6.065} {\bibfield  {journal} {\bibinfo
  {journal} {SciPost Phys.}\ }\textbf {\bibinfo {volume} {5}},\ \bibinfo
  {pages} {065} (\bibinfo {year} {2018})},\ \Eprint
  {http://arxiv.org/abs/1804.00945} {arXiv:1804.00945 [hep-ph]} \BibitemShut
  {NoStop}%
\bibitem [{\citenamefont {Li}\ \emph {et~al.}(2020)\citenamefont {Li},
  \citenamefont {Umeeda}, \citenamefont {Xu},\ and\ \citenamefont
  {Yu}}]{Li:2020xrz}%
  \BibitemOpen
  \bibfield  {author} {\bibinfo {author} {\bibfnamefont {H.-N.}\ \bibnamefont
  {Li}}, \bibinfo {author} {\bibfnamefont {H.}~\bibnamefont {Umeeda}}, \bibinfo
  {author} {\bibfnamefont {F.}~\bibnamefont {Xu}}, \ and\ \bibinfo {author}
  {\bibfnamefont {F.-S.}\ \bibnamefont {Yu}},\ }\href {\doibase
  10.1016/j.physletb.2020.135802} {\bibfield  {journal} {\bibinfo  {journal}
  {Phys. Lett. B}\ }\textbf {\bibinfo {volume} {810}},\ \bibinfo {pages}
  {135802} (\bibinfo {year} {2020})},\ \Eprint
  {http://arxiv.org/abs/2001.04079} {arXiv:2001.04079 [hep-ph]} \BibitemShut
  {NoStop}%
\bibitem [{\citenamefont {Li}\ and\ \citenamefont
  {Umeeda}(2020{\natexlab{a}})}]{Li:2020fiz}%
  \BibitemOpen
  \bibfield  {author} {\bibinfo {author} {\bibfnamefont {H.-n.}\ \bibnamefont
  {Li}}\ and\ \bibinfo {author} {\bibfnamefont {H.}~\bibnamefont {Umeeda}},\
  }\href {\doibase 10.1103/PhysRevD.102.094003} {\bibfield  {journal} {\bibinfo
   {journal} {Phys. Rev. D}\ }\textbf {\bibinfo {volume} {102}},\ \bibinfo
  {pages} {094003} (\bibinfo {year} {2020}{\natexlab{a}})},\ \Eprint
  {http://arxiv.org/abs/2004.06451} {arXiv:2004.06451 [hep-ph]} \BibitemShut
  {NoStop}%
\bibitem [{\citenamefont {Li}\ and\ \citenamefont
  {Umeeda}(2020{\natexlab{b}})}]{Li:2020ejs}%
  \BibitemOpen
  \bibfield  {author} {\bibinfo {author} {\bibfnamefont {H.-n.}\ \bibnamefont
  {Li}}\ and\ \bibinfo {author} {\bibfnamefont {H.}~\bibnamefont {Umeeda}},\
  }\href {\doibase 10.1103/PhysRevD.102.114014} {\bibfield  {journal} {\bibinfo
   {journal} {Phys. Rev. D}\ }\textbf {\bibinfo {volume} {102}},\ \bibinfo
  {pages} {114014} (\bibinfo {year} {2020}{\natexlab{b}})},\ \Eprint
  {http://arxiv.org/abs/2006.16593} {arXiv:2006.16593 [hep-ph]} \BibitemShut
  {NoStop}%
\bibitem [{\citenamefont {Xiong}\ \emph {et~al.}(2022)\citenamefont {Xiong},
  \citenamefont {Wei},\ and\ \citenamefont {Yu}}]{Xiong:2022uwj}%
  \BibitemOpen
  \bibfield  {author} {\bibinfo {author} {\bibfnamefont {A.-S.}\ \bibnamefont
  {Xiong}}, \bibinfo {author} {\bibfnamefont {T.}~\bibnamefont {Wei}}, \ and\
  \bibinfo {author} {\bibfnamefont {F.-S.}\ \bibnamefont {Yu}},\ }\href@noop {}
  {\  (\bibinfo {year} {2022})},\ \Eprint {http://arxiv.org/abs/2211.13753}
  {arXiv:2211.13753 [hep-th]} \BibitemShut {NoStop}%
\bibitem [{\citenamefont {Gubler}\ and\ \citenamefont
  {Oka}(2010)}]{Gubler:2010cf}%
  \BibitemOpen
  \bibfield  {author} {\bibinfo {author} {\bibfnamefont {P.}~\bibnamefont
  {Gubler}}\ and\ \bibinfo {author} {\bibfnamefont {M.}~\bibnamefont {Oka}},\
  }\href {\doibase 10.1143/PTP.124.995} {\bibfield  {journal} {\bibinfo
  {journal} {Prog. Theor. Phys.}\ }\textbf {\bibinfo {volume} {124}},\ \bibinfo
  {pages} {995} (\bibinfo {year} {2010})},\ \Eprint
  {http://arxiv.org/abs/1005.2459} {arXiv:1005.2459 [hep-ph]} \BibitemShut
  {NoStop}%
\bibitem [{\citenamefont {Bruno}\ and\ \citenamefont
  {Hansen}(2021)}]{Bruno:2020kyl}%
  \BibitemOpen
  \bibfield  {author} {\bibinfo {author} {\bibfnamefont {M.}~\bibnamefont
  {Bruno}}\ and\ \bibinfo {author} {\bibfnamefont {M.~T.}\ \bibnamefont
  {Hansen}},\ }\href {\doibase 10.1007/JHEP06(2021)043} {\bibfield  {journal}
  {\bibinfo  {journal} {JHEP}\ }\textbf {\bibinfo {volume} {06}},\ \bibinfo
  {pages} {043} (\bibinfo {year} {2021})},\ \Eprint
  {http://arxiv.org/abs/2012.11488} {arXiv:2012.11488 [hep-lat]} \BibitemShut
  {NoStop}%
\bibitem [{\citenamefont {Bergamaschi}\ \emph {et~al.}(2023)\citenamefont
  {Bergamaschi}, \citenamefont {Jay},\ and\ \citenamefont
  {Oare}}]{Bergamaschi:2023xzx}%
  \BibitemOpen
  \bibfield  {author} {\bibinfo {author} {\bibfnamefont {T.}~\bibnamefont
  {Bergamaschi}}, \bibinfo {author} {\bibfnamefont {W.~I.}\ \bibnamefont
  {Jay}}, \ and\ \bibinfo {author} {\bibfnamefont {P.~R.}\ \bibnamefont
  {Oare}},\ }\href@noop {} {\  (\bibinfo {year} {2023})},\ \Eprint
  {http://arxiv.org/abs/2305.16190} {arXiv:2305.16190 [hep-lat]} \BibitemShut
  {NoStop}%
\bibitem [{\citenamefont {Horak}\ \emph {et~al.}(2022)\citenamefont {Horak},
  \citenamefont {Pawlowski}, \citenamefont {Rodr\'\i{}guez-Quintero},
  \citenamefont {Turnwald}, \citenamefont {Urban}, \citenamefont {Wink},\ and\
  \citenamefont {Zafeiropoulos}}]{Horak:2021syv}%
  \BibitemOpen
  \bibfield  {author} {\bibinfo {author} {\bibfnamefont {J.}~\bibnamefont
  {Horak}}, \bibinfo {author} {\bibfnamefont {J.~M.}\ \bibnamefont
  {Pawlowski}}, \bibinfo {author} {\bibfnamefont {J.}~\bibnamefont
  {Rodr\'\i{}guez-Quintero}}, \bibinfo {author} {\bibfnamefont
  {J.}~\bibnamefont {Turnwald}}, \bibinfo {author} {\bibfnamefont {J.~M.}\
  \bibnamefont {Urban}}, \bibinfo {author} {\bibfnamefont {N.}~\bibnamefont
  {Wink}}, \ and\ \bibinfo {author} {\bibfnamefont {S.}~\bibnamefont
  {Zafeiropoulos}},\ }\href {\doibase 10.1103/PhysRevD.105.036014} {\bibfield
  {journal} {\bibinfo  {journal} {Phys. Rev. D}\ }\textbf {\bibinfo {volume}
  {105}},\ \bibinfo {pages} {036014} (\bibinfo {year} {2022})},\ \Eprint
  {http://arxiv.org/abs/2107.13464} {arXiv:2107.13464 [hep-ph]} \BibitemShut
  {NoStop}%
\bibitem [{\citenamefont {Offler}\ \emph {et~al.}(2019)\citenamefont {Offler},
  \citenamefont {Aarts}, \citenamefont {Allton}, \citenamefont {Glesaaen},
  \citenamefont {J\"ager}, \citenamefont {Kim}, \citenamefont {Lombardo},
  \citenamefont {Ryan},\ and\ \citenamefont {Skullerud}}]{Offler:2019eij}%
  \BibitemOpen
  \bibfield  {author} {\bibinfo {author} {\bibfnamefont {S.}~\bibnamefont
  {Offler}}, \bibinfo {author} {\bibfnamefont {G.}~\bibnamefont {Aarts}},
  \bibinfo {author} {\bibfnamefont {C.}~\bibnamefont {Allton}}, \bibinfo
  {author} {\bibfnamefont {J.}~\bibnamefont {Glesaaen}}, \bibinfo {author}
  {\bibfnamefont {B.}~\bibnamefont {J\"ager}}, \bibinfo {author} {\bibfnamefont
  {S.}~\bibnamefont {Kim}}, \bibinfo {author} {\bibfnamefont {M.~P.}\
  \bibnamefont {Lombardo}}, \bibinfo {author} {\bibfnamefont {S.~M.}\
  \bibnamefont {Ryan}}, \ and\ \bibinfo {author} {\bibfnamefont {J.-I.}\
  \bibnamefont {Skullerud}},\ }\href {\doibase 10.22323/1.363.0076} {\bibfield
  {journal} {\bibinfo  {journal} {PoS}\ }\textbf {\bibinfo {volume}
  {LATTICE2019}},\ \bibinfo {pages} {076} (\bibinfo {year} {2019})},\ \Eprint
  {http://arxiv.org/abs/1912.12900} {arXiv:1912.12900 [hep-lat]} \BibitemShut
  {NoStop}%
\bibitem [{\citenamefont {Offler}\ \emph {et~al.}(2022)\citenamefont {Offler},
  \citenamefont {Aarts}, \citenamefont {Allton}, \citenamefont {J\"ager},
  \citenamefont {Kim}, \citenamefont {Lombardo}, \citenamefont {Page},
  \citenamefont {Ryan}, \citenamefont {Skullerud},\ and\ \citenamefont
  {Spriggs}}]{Offler:2021fmg}%
  \BibitemOpen
  \bibfield  {author} {\bibinfo {author} {\bibfnamefont {S.}~\bibnamefont
  {Offler}}, \bibinfo {author} {\bibfnamefont {G.}~\bibnamefont {Aarts}},
  \bibinfo {author} {\bibfnamefont {C.}~\bibnamefont {Allton}}, \bibinfo
  {author} {\bibfnamefont {B.}~\bibnamefont {J\"ager}}, \bibinfo {author}
  {\bibfnamefont {S.}~\bibnamefont {Kim}}, \bibinfo {author} {\bibfnamefont
  {M.-P.}\ \bibnamefont {Lombardo}}, \bibinfo {author} {\bibfnamefont
  {B.}~\bibnamefont {Page}}, \bibinfo {author} {\bibfnamefont {S.~M.}\
  \bibnamefont {Ryan}}, \bibinfo {author} {\bibfnamefont {J.-I.}\ \bibnamefont
  {Skullerud}}, \ and\ \bibinfo {author} {\bibfnamefont {T.}~\bibnamefont
  {Spriggs}},\ }\href {\doibase 10.22323/1.396.0509} {\bibfield  {journal}
  {\bibinfo  {journal} {PoS}\ }\textbf {\bibinfo {volume} {LATTICE2021}},\
  \bibinfo {pages} {509} (\bibinfo {year} {2022})},\ \Eprint
  {http://arxiv.org/abs/2112.02116} {arXiv:2112.02116 [hep-lat]} \BibitemShut
  {NoStop}%
\bibitem [{\citenamefont {Kades}\ \emph {et~al.}(2020)\citenamefont {Kades},
  \citenamefont {Pawlowski}, \citenamefont {Rothkopf}, \citenamefont
  {Scherzer}, \citenamefont {Urban}, \citenamefont {Wetzel}, \citenamefont
  {Wink},\ and\ \citenamefont {Ziegler}}]{Kades:2019wtd}%
  \BibitemOpen
  \bibfield  {author} {\bibinfo {author} {\bibfnamefont {L.}~\bibnamefont
  {Kades}}, \bibinfo {author} {\bibfnamefont {J.~M.}\ \bibnamefont
  {Pawlowski}}, \bibinfo {author} {\bibfnamefont {A.}~\bibnamefont {Rothkopf}},
  \bibinfo {author} {\bibfnamefont {M.}~\bibnamefont {Scherzer}}, \bibinfo
  {author} {\bibfnamefont {J.~M.}\ \bibnamefont {Urban}}, \bibinfo {author}
  {\bibfnamefont {S.~J.}\ \bibnamefont {Wetzel}}, \bibinfo {author}
  {\bibfnamefont {N.}~\bibnamefont {Wink}}, \ and\ \bibinfo {author}
  {\bibfnamefont {F.~P.~G.}\ \bibnamefont {Ziegler}},\ }\href {\doibase
  10.1103/PhysRevD.102.096001} {\bibfield  {journal} {\bibinfo  {journal}
  {Phys. Rev. D}\ }\textbf {\bibinfo {volume} {102}},\ \bibinfo {pages}
  {096001} (\bibinfo {year} {2020})},\ \Eprint
  {http://arxiv.org/abs/1905.04305} {arXiv:1905.04305 [physics.comp-ph]}
  \BibitemShut {NoStop}%
\bibitem [{\citenamefont {Laanait}\ \emph {et~al.}(2019)\citenamefont
  {Laanait}, \citenamefont {Romero}, \citenamefont {Yin}, \citenamefont
  {Young}, \citenamefont {Treichler}, \citenamefont {Starchenko}, \citenamefont
  {Borisevich}, \citenamefont {Sergeev},\ and\ \citenamefont
  {Matheson}}]{laanait2019exascale}%
  \BibitemOpen
  \bibfield  {author} {\bibinfo {author} {\bibfnamefont {N.}~\bibnamefont
  {Laanait}}, \bibinfo {author} {\bibfnamefont {J.}~\bibnamefont {Romero}},
  \bibinfo {author} {\bibfnamefont {J.}~\bibnamefont {Yin}}, \bibinfo {author}
  {\bibfnamefont {M.~T.}\ \bibnamefont {Young}}, \bibinfo {author}
  {\bibfnamefont {S.}~\bibnamefont {Treichler}}, \bibinfo {author}
  {\bibfnamefont {V.}~\bibnamefont {Starchenko}}, \bibinfo {author}
  {\bibfnamefont {A.}~\bibnamefont {Borisevich}}, \bibinfo {author}
  {\bibfnamefont {A.}~\bibnamefont {Sergeev}}, \ and\ \bibinfo {author}
  {\bibfnamefont {M.}~\bibnamefont {Matheson}},\ }\href@noop {} {\enquote
  {\bibinfo {title} {Exascale deep learning for scientific inverse problems},}\
  } (\bibinfo {year} {2019}),\ \Eprint {http://arxiv.org/abs/1909.11150}
  {arXiv:1909.11150 [cs.LG]} \BibitemShut {NoStop}%
\bibitem [{\citenamefont {Chen}\ \emph {et~al.}(2021)\citenamefont {Chen},
  \citenamefont {Ding}, \citenamefont {Liu}, \citenamefont {Papp},\ and\
  \citenamefont {Yang}}]{Chen:2021giw}%
  \BibitemOpen
  \bibfield  {author} {\bibinfo {author} {\bibfnamefont {S.~Y.}\ \bibnamefont
  {Chen}}, \bibinfo {author} {\bibfnamefont {H.~T.}\ \bibnamefont {Ding}},
  \bibinfo {author} {\bibfnamefont {F.~Y.}\ \bibnamefont {Liu}}, \bibinfo
  {author} {\bibfnamefont {G.}~\bibnamefont {Papp}}, \ and\ \bibinfo {author}
  {\bibfnamefont {C.~B.}\ \bibnamefont {Yang}},\ }\href@noop {} {\  (\bibinfo
  {year} {2021})},\ \Eprint {http://arxiv.org/abs/2110.13521} {arXiv:2110.13521
  [hep-lat]} \BibitemShut {NoStop}%
\bibitem [{\citenamefont {Wang}\ \emph {et~al.}(2022)\citenamefont {Wang},
  \citenamefont {Shi},\ and\ \citenamefont {Zhou}}]{Wang:2021jou}%
  \BibitemOpen
  \bibfield  {author} {\bibinfo {author} {\bibfnamefont {L.}~\bibnamefont
  {Wang}}, \bibinfo {author} {\bibfnamefont {S.}~\bibnamefont {Shi}}, \ and\
  \bibinfo {author} {\bibfnamefont {K.}~\bibnamefont {Zhou}},\ }\href {\doibase
  10.1103/PhysRevD.106.L051502} {\bibfield  {journal} {\bibinfo  {journal}
  {Phys. Rev. D}\ }\textbf {\bibinfo {volume} {106}},\ \bibinfo {pages}
  {L051502} (\bibinfo {year} {2022})},\ \Eprint
  {http://arxiv.org/abs/2111.14760} {arXiv:2111.14760 [hep-ph]} \BibitemShut
  {NoStop}%
\bibitem [{\citenamefont {Wang}\ \emph {et~al.}(2021)\citenamefont {Wang},
  \citenamefont {Shi},\ and\ \citenamefont {Zhou}}]{Wang:2021cqw}%
  \BibitemOpen
  \bibfield  {author} {\bibinfo {author} {\bibfnamefont {L.}~\bibnamefont
  {Wang}}, \bibinfo {author} {\bibfnamefont {S.}~\bibnamefont {Shi}}, \ and\
  \bibinfo {author} {\bibfnamefont {K.}~\bibnamefont {Zhou}},\ }in\ \href@noop
  {} {\emph {\bibinfo {booktitle} {{35th Conference on Neural Information
  Processing Systems}}}}\ (\bibinfo {year} {2021})\ \Eprint
  {http://arxiv.org/abs/2112.06206} {arXiv:2112.06206 [physics.comp-ph]}
  \BibitemShut {NoStop}%
\bibitem [{\citenamefont {Shi}\ \emph {et~al.}(2023)\citenamefont {Shi},
  \citenamefont {Wang},\ and\ \citenamefont {Zhou}}]{Shi:2022yqw}%
  \BibitemOpen
  \bibfield  {author} {\bibinfo {author} {\bibfnamefont {S.}~\bibnamefont
  {Shi}}, \bibinfo {author} {\bibfnamefont {L.}~\bibnamefont {Wang}}, \ and\
  \bibinfo {author} {\bibfnamefont {K.}~\bibnamefont {Zhou}},\ }\href {\doibase
  10.1016/j.cpc.2022.108547} {\bibfield  {journal} {\bibinfo  {journal}
  {Comput. Phys. Commun.}\ }\textbf {\bibinfo {volume} {282}},\ \bibinfo
  {pages} {108547} (\bibinfo {year} {2023})},\ \Eprint
  {http://arxiv.org/abs/2201.02564} {arXiv:2201.02564 [hep-ph]} \BibitemShut
  {NoStop}%
\bibitem [{\citenamefont {Lechien}\ and\ \citenamefont
  {Dudal}(2022)}]{Lechien:2022ieg}%
  \BibitemOpen
  \bibfield  {author} {\bibinfo {author} {\bibfnamefont {T.}~\bibnamefont
  {Lechien}}\ and\ \bibinfo {author} {\bibfnamefont {D.}~\bibnamefont
  {Dudal}},\ }\href {\doibase 10.21468/SciPostPhys.13.4.097} {\bibfield
  {journal} {\bibinfo  {journal} {SciPost Phys.}\ }\textbf {\bibinfo {volume}
  {13}},\ \bibinfo {pages} {097} (\bibinfo {year} {2022})},\ \Eprint
  {http://arxiv.org/abs/2203.03293} {arXiv:2203.03293 [hep-lat]} \BibitemShut
  {NoStop}%
\bibitem [{\citenamefont {Boyda}\ \emph {et~al.}(2022)\citenamefont {Boyda}
  \emph {et~al.}}]{Boyda:2022nmh}%
  \BibitemOpen
  \bibfield  {author} {\bibinfo {author} {\bibfnamefont {D.}~\bibnamefont
  {Boyda}} \emph {et~al.},\ }in\ \href@noop {} {\emph {\bibinfo {booktitle}
  {{Snowmass 2021}}}}\ (\bibinfo {year} {2022})\ \Eprint
  {http://arxiv.org/abs/2202.05838} {arXiv:2202.05838 [hep-lat]} \BibitemShut
  {NoStop}%
\bibitem [{\citenamefont {Buzzicotti}\ \emph {et~al.}(2023)\citenamefont
  {Buzzicotti}, \citenamefont {De~Santis},\ and\ \citenamefont
  {Tantalo}}]{Buzzicotti:2023qdv}%
  \BibitemOpen
  \bibfield  {author} {\bibinfo {author} {\bibfnamefont {M.}~\bibnamefont
  {Buzzicotti}}, \bibinfo {author} {\bibfnamefont {A.}~\bibnamefont
  {De~Santis}}, \ and\ \bibinfo {author} {\bibfnamefont {N.}~\bibnamefont
  {Tantalo}},\ }\href@noop {} {\  (\bibinfo {year} {2023})},\ \Eprint
  {http://arxiv.org/abs/2307.00808} {arXiv:2307.00808 [hep-lat]} \BibitemShut
  {NoStop}%
\bibitem [{\citenamefont {Chambers}\ \emph {et~al.}(2017)\citenamefont
  {Chambers}, \citenamefont {Horsley}, \citenamefont {Nakamura}, \citenamefont
  {Perlt}, \citenamefont {Rakow}, \citenamefont {Schierholz}, \citenamefont
  {Schiller}, \citenamefont {Somfleth}, \citenamefont {Young},\ and\
  \citenamefont {Zanotti}}]{Chambers:2017dov}%
  \BibitemOpen
  \bibfield  {author} {\bibinfo {author} {\bibfnamefont {A.~J.}\ \bibnamefont
  {Chambers}}, \bibinfo {author} {\bibfnamefont {R.}~\bibnamefont {Horsley}},
  \bibinfo {author} {\bibfnamefont {Y.}~\bibnamefont {Nakamura}}, \bibinfo
  {author} {\bibfnamefont {H.}~\bibnamefont {Perlt}}, \bibinfo {author}
  {\bibfnamefont {P.~E.~L.}\ \bibnamefont {Rakow}}, \bibinfo {author}
  {\bibfnamefont {G.}~\bibnamefont {Schierholz}}, \bibinfo {author}
  {\bibfnamefont {A.}~\bibnamefont {Schiller}}, \bibinfo {author}
  {\bibfnamefont {K.}~\bibnamefont {Somfleth}}, \bibinfo {author}
  {\bibfnamefont {R.~D.}\ \bibnamefont {Young}}, \ and\ \bibinfo {author}
  {\bibfnamefont {J.~M.}\ \bibnamefont {Zanotti}},\ }\href {\doibase
  10.1103/PhysRevLett.118.242001} {\bibfield  {journal} {\bibinfo  {journal}
  {Phys. Rev. Lett.}\ }\textbf {\bibinfo {volume} {118}},\ \bibinfo {pages}
  {242001} (\bibinfo {year} {2017})},\ \Eprint
  {http://arxiv.org/abs/1703.01153} {arXiv:1703.01153 [hep-lat]} \BibitemShut
  {NoStop}%
\bibitem [{\citenamefont {Karpie}\ \emph {et~al.}(2019)\citenamefont {Karpie},
  \citenamefont {Orginos}, \citenamefont {Rothkopf},\ and\ \citenamefont
  {Zafeiropoulos}}]{Karpie:2019eiq}%
  \BibitemOpen
  \bibfield  {author} {\bibinfo {author} {\bibfnamefont {J.}~\bibnamefont
  {Karpie}}, \bibinfo {author} {\bibfnamefont {K.}~\bibnamefont {Orginos}},
  \bibinfo {author} {\bibfnamefont {A.}~\bibnamefont {Rothkopf}}, \ and\
  \bibinfo {author} {\bibfnamefont {S.}~\bibnamefont {Zafeiropoulos}},\ }\href
  {\doibase 10.1007/JHEP04(2019)057} {\bibfield  {journal} {\bibinfo  {journal}
  {JHEP}\ }\textbf {\bibinfo {volume} {04}},\ \bibinfo {pages} {057} (\bibinfo
  {year} {2019})},\ \Eprint {http://arxiv.org/abs/1901.05408} {arXiv:1901.05408
  [hep-lat]} \BibitemShut {NoStop}%
\bibitem [{\citenamefont {Candido}\ \emph {et~al.}(2023)\citenamefont
  {Candido}, \citenamefont {Del~Debbio}, \citenamefont {Giani},\ and\
  \citenamefont {Petrillo}}]{Candido:2023nnb}%
  \BibitemOpen
  \bibfield  {author} {\bibinfo {author} {\bibfnamefont {A.}~\bibnamefont
  {Candido}}, \bibinfo {author} {\bibfnamefont {L.}~\bibnamefont {Del~Debbio}},
  \bibinfo {author} {\bibfnamefont {T.}~\bibnamefont {Giani}}, \ and\ \bibinfo
  {author} {\bibfnamefont {G.}~\bibnamefont {Petrillo}},\ }\href {\doibase
  10.22323/1.430.0098} {\bibfield  {journal} {\bibinfo  {journal} {PoS}\
  }\textbf {\bibinfo {volume} {LATTICE2022}},\ \bibinfo {pages} {098} (\bibinfo
  {year} {2023})},\ \Eprint {http://arxiv.org/abs/2302.14731} {arXiv:2302.14731
  [hep-lat]} \BibitemShut {NoStop}%
\bibitem [{\citenamefont {Dudal}\ \emph {et~al.}(2014)\citenamefont {Dudal},
  \citenamefont {Oliveira},\ and\ \citenamefont {Silva}}]{Dudal:2013yva}%
  \BibitemOpen
  \bibfield  {author} {\bibinfo {author} {\bibfnamefont {D.}~\bibnamefont
  {Dudal}}, \bibinfo {author} {\bibfnamefont {O.}~\bibnamefont {Oliveira}}, \
  and\ \bibinfo {author} {\bibfnamefont {P.~J.}\ \bibnamefont {Silva}},\ }\href
  {\doibase 10.1103/PhysRevD.89.014010} {\bibfield  {journal} {\bibinfo
  {journal} {Phys. Rev. D}\ }\textbf {\bibinfo {volume} {89}},\ \bibinfo
  {pages} {014010} (\bibinfo {year} {2014})},\ \Eprint
  {http://arxiv.org/abs/1310.4069} {arXiv:1310.4069 [hep-lat]} \BibitemShut
  {NoStop}%
\bibitem [{\citenamefont {Tripolt}\ \emph {et~al.}(2019)\citenamefont
  {Tripolt}, \citenamefont {Gubler}, \citenamefont {Ulybyshev},\ and\
  \citenamefont {Von~Smekal}}]{Tripolt:2018xeo}%
  \BibitemOpen
  \bibfield  {author} {\bibinfo {author} {\bibfnamefont {R.-A.}\ \bibnamefont
  {Tripolt}}, \bibinfo {author} {\bibfnamefont {P.}~\bibnamefont {Gubler}},
  \bibinfo {author} {\bibfnamefont {M.}~\bibnamefont {Ulybyshev}}, \ and\
  \bibinfo {author} {\bibfnamefont {L.}~\bibnamefont {Von~Smekal}},\ }\href
  {\doibase 10.1016/j.cpc.2018.11.012} {\bibfield  {journal} {\bibinfo
  {journal} {Comput. Phys. Commun.}\ }\textbf {\bibinfo {volume} {237}},\
  \bibinfo {pages} {129} (\bibinfo {year} {2019})},\ \Eprint
  {http://arxiv.org/abs/1801.10348} {arXiv:1801.10348 [hep-ph]} \BibitemShut
  {NoStop}%
\bibitem [{\citenamefont {Rothkopf}(2022{\natexlab{a}})}]{Rothkopf:2022ctl}%
  \BibitemOpen
  \bibfield  {author} {\bibinfo {author} {\bibfnamefont {A.}~\bibnamefont
  {Rothkopf}},\ }\href@noop {} {\  (\bibinfo {year} {2022}{\natexlab{a}})},\
  \Eprint {http://arxiv.org/abs/2208.13590} {arXiv:2208.13590 [hep-lat]}
  \BibitemShut {NoStop}%
\bibitem [{\citenamefont {Rothkopf}(2022{\natexlab{b}})}]{Rothkopf:2022fyo}%
  \BibitemOpen
  \bibfield  {author} {\bibinfo {author} {\bibfnamefont {A.}~\bibnamefont
  {Rothkopf}},\ }\href {\doibase 10.1051/epjconf/202227401004} {\bibfield
  {journal} {\bibinfo  {journal} {EPJ Web Conf.}\ }\textbf {\bibinfo {volume}
  {274}},\ \bibinfo {pages} {01004} (\bibinfo {year} {2022}{\natexlab{b}})},\
  \Eprint {http://arxiv.org/abs/2211.10680} {arXiv:2211.10680 [hep-lat]}
  \BibitemShut {NoStop}%
\bibitem [{\citenamefont {Davier}\ \emph {et~al.}(2010)\citenamefont {Davier},
  \citenamefont {Hoecker}, \citenamefont {Malaescu}, \citenamefont {Yuan},\
  and\ \citenamefont {Zhang}}]{Davier:2010rnx}%
  \BibitemOpen
  \bibfield  {author} {\bibinfo {author} {\bibfnamefont {M.}~\bibnamefont
  {Davier}}, \bibinfo {author} {\bibfnamefont {A.}~\bibnamefont {Hoecker}},
  \bibinfo {author} {\bibfnamefont {B.}~\bibnamefont {Malaescu}}, \bibinfo
  {author} {\bibfnamefont {C.~Z.}\ \bibnamefont {Yuan}}, \ and\ \bibinfo
  {author} {\bibfnamefont {Z.}~\bibnamefont {Zhang}},\ }\href {\doibase
  10.1140/epjc/s10052-010-1246-1} {\bibfield  {journal} {\bibinfo  {journal}
  {Eur. Phys. J. C}\ }\textbf {\bibinfo {volume} {66}},\ \bibinfo {pages} {1}
  (\bibinfo {year} {2010})},\ \Eprint {http://arxiv.org/abs/0908.4300}
  {arXiv:0908.4300 [hep-ph]} \BibitemShut {NoStop}%
\bibitem [{\citenamefont {Davier}\ \emph {et~al.}(2011)\citenamefont {Davier},
  \citenamefont {Hoecker}, \citenamefont {Malaescu},\ and\ \citenamefont
  {Zhang}}]{Davier:2010nc}%
  \BibitemOpen
  \bibfield  {author} {\bibinfo {author} {\bibfnamefont {M.}~\bibnamefont
  {Davier}}, \bibinfo {author} {\bibfnamefont {A.}~\bibnamefont {Hoecker}},
  \bibinfo {author} {\bibfnamefont {B.}~\bibnamefont {Malaescu}}, \ and\
  \bibinfo {author} {\bibfnamefont {Z.}~\bibnamefont {Zhang}},\ }\href
  {\doibase 10.1140/epjc/s10052-012-1874-8} {\bibfield  {journal} {\bibinfo
  {journal} {Eur. Phys. J. C}\ }\textbf {\bibinfo {volume} {71}},\ \bibinfo
  {pages} {1515} (\bibinfo {year} {2011})},\ \bibinfo {note} {[Erratum:
  Eur.Phys.J.C 72, 1874 (2012)]},\ \Eprint {http://arxiv.org/abs/1010.4180}
  {arXiv:1010.4180 [hep-ph]} \BibitemShut {NoStop}%
\bibitem [{\citenamefont {Davier}\ \emph {et~al.}(2017)\citenamefont {Davier},
  \citenamefont {Hoecker}, \citenamefont {Malaescu},\ and\ \citenamefont
  {Zhang}}]{Davier:2017zfy}%
  \BibitemOpen
  \bibfield  {author} {\bibinfo {author} {\bibfnamefont {M.}~\bibnamefont
  {Davier}}, \bibinfo {author} {\bibfnamefont {A.}~\bibnamefont {Hoecker}},
  \bibinfo {author} {\bibfnamefont {B.}~\bibnamefont {Malaescu}}, \ and\
  \bibinfo {author} {\bibfnamefont {Z.}~\bibnamefont {Zhang}},\ }\href
  {\doibase 10.1140/epjc/s10052-017-5161-6} {\bibfield  {journal} {\bibinfo
  {journal} {Eur. Phys. J. C}\ }\textbf {\bibinfo {volume} {77}},\ \bibinfo
  {pages} {827} (\bibinfo {year} {2017})},\ \Eprint
  {http://arxiv.org/abs/1706.09436} {arXiv:1706.09436 [hep-ph]} \BibitemShut
  {NoStop}%
\bibitem [{\citenamefont {D'Agostini}(1994)}]{DAgostini:1993arp}%
  \BibitemOpen
  \bibfield  {author} {\bibinfo {author} {\bibfnamefont {G.}~\bibnamefont
  {D'Agostini}},\ }\href {\doibase 10.1016/0168-9002(94)90719-6} {\bibfield
  {journal} {\bibinfo  {journal} {Nucl. Instrum. Meth. A}\ }\textbf {\bibinfo
  {volume} {346}},\ \bibinfo {pages} {306} (\bibinfo {year}
  {1994})}\BibitemShut {NoStop}%
\bibitem [{\citenamefont {Blobel}(2003)}]{Blobel:2003wa}%
  \BibitemOpen
  \bibfield  {author} {\bibinfo {author} {\bibfnamefont {V.}~\bibnamefont
  {Blobel}},\ }\href@noop {} {\bibfield  {journal} {\bibinfo  {journal}
  {eConf}\ }\textbf {\bibinfo {volume} {C030908}},\ \bibinfo {pages} {MOET002}
  (\bibinfo {year} {2003})}\BibitemShut {NoStop}%
\bibitem [{\citenamefont {Chao}\ \emph {et~al.}(2023)\citenamefont {Chao},
  \citenamefont {Meyer},\ and\ \citenamefont {Parrino}}]{Chao:2022ycy}%
  \BibitemOpen
  \bibfield  {author} {\bibinfo {author} {\bibfnamefont {E.-H.}\ \bibnamefont
  {Chao}}, \bibinfo {author} {\bibfnamefont {H.~B.}\ \bibnamefont {Meyer}}, \
  and\ \bibinfo {author} {\bibfnamefont {J.}~\bibnamefont {Parrino}},\ }\href
  {\doibase 10.1103/PhysRevD.107.054505} {\bibfield  {journal} {\bibinfo
  {journal} {Phys. Rev. D}\ }\textbf {\bibinfo {volume} {107}},\ \bibinfo
  {pages} {054505} (\bibinfo {year} {2023})},\ \Eprint
  {http://arxiv.org/abs/2211.15581} {arXiv:2211.15581 [hep-lat]} \BibitemShut
  {NoStop}%
\bibitem [{\citenamefont {Lehner}\ and\ \citenamefont
  {Meyer}(2020)}]{Lehner:2020crt}%
  \BibitemOpen
  \bibfield  {author} {\bibinfo {author} {\bibfnamefont {C.}~\bibnamefont
  {Lehner}}\ and\ \bibinfo {author} {\bibfnamefont {A.~S.}\ \bibnamefont
  {Meyer}},\ }\href {\doibase 10.1103/PhysRevD.101.074515} {\bibfield
  {journal} {\bibinfo  {journal} {Phys. Rev. D}\ }\textbf {\bibinfo {volume}
  {101}},\ \bibinfo {pages} {074515} (\bibinfo {year} {2020})},\ \Eprint
  {http://arxiv.org/abs/2003.04177} {arXiv:2003.04177 [hep-lat]} \BibitemShut
  {NoStop}%
\bibitem [{\citenamefont {Wang}\ \emph {et~al.}(2023)\citenamefont {Wang},
  \citenamefont {Draper}, \citenamefont {Liu},\ and\ \citenamefont
  {Yang}}]{Wang:2022lkq}%
  \BibitemOpen
  \bibfield  {author} {\bibinfo {author} {\bibfnamefont {G.}~\bibnamefont
  {Wang}}, \bibinfo {author} {\bibfnamefont {T.}~\bibnamefont {Draper}},
  \bibinfo {author} {\bibfnamefont {K.-F.}\ \bibnamefont {Liu}}, \ and\
  \bibinfo {author} {\bibfnamefont {Y.-B.}\ \bibnamefont {Yang}} (\bibinfo
  {collaboration} {chiQCD}),\ }\href {\doibase 10.1103/PhysRevD.107.034513}
  {\bibfield  {journal} {\bibinfo  {journal} {Phys. Rev. D}\ }\textbf {\bibinfo
  {volume} {107}},\ \bibinfo {pages} {034513} (\bibinfo {year} {2023})},\
  \Eprint {http://arxiv.org/abs/2204.01280} {arXiv:2204.01280 [hep-lat]}
  \BibitemShut {NoStop}%
\bibitem [{\citenamefont {Aubin}\ \emph {et~al.}(2022)\citenamefont {Aubin},
  \citenamefont {Blum}, \citenamefont {Golterman},\ and\ \citenamefont
  {Peris}}]{Aubin:2022hgm}%
  \BibitemOpen
  \bibfield  {author} {\bibinfo {author} {\bibfnamefont {C.}~\bibnamefont
  {Aubin}}, \bibinfo {author} {\bibfnamefont {T.}~\bibnamefont {Blum}},
  \bibinfo {author} {\bibfnamefont {M.}~\bibnamefont {Golterman}}, \ and\
  \bibinfo {author} {\bibfnamefont {S.}~\bibnamefont {Peris}},\ }\href
  {\doibase 10.1103/PhysRevD.106.054503} {\bibfield  {journal} {\bibinfo
  {journal} {Phys. Rev. D}\ }\textbf {\bibinfo {volume} {106}},\ \bibinfo
  {pages} {054503} (\bibinfo {year} {2022})},\ \Eprint
  {http://arxiv.org/abs/2204.12256} {arXiv:2204.12256 [hep-lat]} \BibitemShut
  {NoStop}%
\bibitem [{\citenamefont {C\`e}\ \emph {et~al.}(2021)\citenamefont {C\`e},
  \citenamefont {Harris}, \citenamefont {Meyer}, \citenamefont {Toniato},\ and\
  \citenamefont {T\"or\"ok}}]{Ce:2021xgd}%
  \BibitemOpen
  \bibfield  {author} {\bibinfo {author} {\bibfnamefont {M.}~\bibnamefont
  {C\`e}}, \bibinfo {author} {\bibfnamefont {T.}~\bibnamefont {Harris}},
  \bibinfo {author} {\bibfnamefont {H.~B.}\ \bibnamefont {Meyer}}, \bibinfo
  {author} {\bibfnamefont {A.}~\bibnamefont {Toniato}}, \ and\ \bibinfo
  {author} {\bibfnamefont {C.}~\bibnamefont {T\"or\"ok}},\ }\href {\doibase
  10.1007/JHEP12(2021)215} {\bibfield  {journal} {\bibinfo  {journal} {JHEP}\
  }\textbf {\bibinfo {volume} {12}},\ \bibinfo {pages} {215} (\bibinfo {year}
  {2021})},\ \Eprint {http://arxiv.org/abs/2106.15293} {arXiv:2106.15293
  [hep-lat]} \BibitemShut {NoStop}%
\bibitem [{\citenamefont {Ambrosino}\ \emph {et~al.}(2009)\citenamefont
  {Ambrosino} \emph {et~al.}}]{KLOE:2008fmq}%
  \BibitemOpen
  \bibfield  {author} {\bibinfo {author} {\bibfnamefont {F.}~\bibnamefont
  {Ambrosino}} \emph {et~al.} (\bibinfo {collaboration} {KLOE}),\ }\href
  {\doibase 10.1016/j.physletb.2008.10.060} {\bibfield  {journal} {\bibinfo
  {journal} {Phys. Lett. B}\ }\textbf {\bibinfo {volume} {670}},\ \bibinfo
  {pages} {285} (\bibinfo {year} {2009})},\ \Eprint
  {http://arxiv.org/abs/0809.3950} {arXiv:0809.3950 [hep-ex]} \BibitemShut
  {NoStop}%
\bibitem [{\citenamefont {Ambrosino}\ \emph {et~al.}(2011)\citenamefont
  {Ambrosino} \emph {et~al.}}]{KLOE:2010qei}%
  \BibitemOpen
  \bibfield  {author} {\bibinfo {author} {\bibfnamefont {F.}~\bibnamefont
  {Ambrosino}} \emph {et~al.} (\bibinfo {collaboration} {KLOE}),\ }\href
  {\doibase 10.1016/j.physletb.2011.04.055} {\bibfield  {journal} {\bibinfo
  {journal} {Phys. Lett. B}\ }\textbf {\bibinfo {volume} {700}},\ \bibinfo
  {pages} {102} (\bibinfo {year} {2011})},\ \Eprint
  {http://arxiv.org/abs/1006.5313} {arXiv:1006.5313 [hep-ex]} \BibitemShut
  {NoStop}%
\bibitem [{\citenamefont {Babusci}\ \emph {et~al.}(2013)\citenamefont {Babusci}
  \emph {et~al.}}]{KLOE:2012anl}%
  \BibitemOpen
  \bibfield  {author} {\bibinfo {author} {\bibfnamefont {D.}~\bibnamefont
  {Babusci}} \emph {et~al.} (\bibinfo {collaboration} {KLOE}),\ }\href
  {\doibase 10.1016/j.physletb.2013.02.029} {\bibfield  {journal} {\bibinfo
  {journal} {Phys. Lett. B}\ }\textbf {\bibinfo {volume} {720}},\ \bibinfo
  {pages} {336} (\bibinfo {year} {2013})},\ \Eprint
  {http://arxiv.org/abs/1212.4524} {arXiv:1212.4524 [hep-ex]} \BibitemShut
  {NoStop}%
\bibitem [{\citenamefont {Lees}\ \emph {et~al.}(2012)\citenamefont {Lees} \emph
  {et~al.}}]{BaBar:2012bdw}%
  \BibitemOpen
  \bibfield  {author} {\bibinfo {author} {\bibfnamefont {J.~P.}\ \bibnamefont
  {Lees}} \emph {et~al.} (\bibinfo {collaboration} {BaBar}),\ }\href {\doibase
  10.1103/PhysRevD.86.032013} {\bibfield  {journal} {\bibinfo  {journal} {Phys.
  Rev. D}\ }\textbf {\bibinfo {volume} {86}},\ \bibinfo {pages} {032013}
  (\bibinfo {year} {2012})},\ \Eprint {http://arxiv.org/abs/1205.2228}
  {arXiv:1205.2228 [hep-ex]} \BibitemShut {NoStop}%
\bibitem [{\citenamefont {Keshavarzi}\ \emph {et~al.}(2018)\citenamefont
  {Keshavarzi}, \citenamefont {Nomura},\ and\ \citenamefont
  {Teubner}}]{Keshavarzi:2018mgv}%
  \BibitemOpen
  \bibfield  {author} {\bibinfo {author} {\bibfnamefont {A.}~\bibnamefont
  {Keshavarzi}}, \bibinfo {author} {\bibfnamefont {D.}~\bibnamefont {Nomura}},
  \ and\ \bibinfo {author} {\bibfnamefont {T.}~\bibnamefont {Teubner}},\ }\href
  {\doibase 10.1103/PhysRevD.97.114025} {\bibfield  {journal} {\bibinfo
  {journal} {Phys. Rev.}\ }\textbf {\bibinfo {volume} {D97}},\ \bibinfo {pages}
  {114025} (\bibinfo {year} {2018})},\ \Eprint
  {http://arxiv.org/abs/1802.02995} {arXiv:1802.02995 [hep-ph]} \BibitemShut
  {NoStop}%
\bibitem [{\citenamefont {Colangelo}(2022)}]{Colangelo:2022xfy}%
  \BibitemOpen
  \bibfield  {author} {\bibinfo {author} {\bibfnamefont {G.}~\bibnamefont
  {Colangelo}},\ }\href {\doibase 10.1051/epjconf/202225801004} {\bibfield
  {journal} {\bibinfo  {journal} {EPJ Web Conf.}\ }\textbf {\bibinfo {volume}
  {258}},\ \bibinfo {pages} {01004} (\bibinfo {year} {2022})}\BibitemShut
  {NoStop}%
\bibitem [{\citenamefont {Aul'chenko}\ \emph {et~al.}(2006)\citenamefont
  {Aul'chenko} \emph {et~al.}}]{Aulchenko:2006dxz}%
  \BibitemOpen
  \bibfield  {author} {\bibinfo {author} {\bibfnamefont {V.~M.}\ \bibnamefont
  {Aul'chenko}} \emph {et~al.},\ }\href {\doibase 10.1134/S0021364006200021}
  {\bibfield  {journal} {\bibinfo  {journal} {JETP Lett.}\ }\textbf {\bibinfo
  {volume} {84}},\ \bibinfo {pages} {413} (\bibinfo {year} {2006})},\ \Eprint
  {http://arxiv.org/abs/hep-ex/0610016} {arXiv:hep-ex/0610016} \BibitemShut
  {NoStop}%
\bibitem [{\citenamefont {Achasov}\ \emph {et~al.}(2006)\citenamefont {Achasov}
  \emph {et~al.}}]{Achasov:2006vp}%
  \BibitemOpen
  \bibfield  {author} {\bibinfo {author} {\bibfnamefont {M.~N.}\ \bibnamefont
  {Achasov}} \emph {et~al.},\ }\href {\doibase 10.1134/S106377610609007X}
  {\bibfield  {journal} {\bibinfo  {journal} {J. Exp. Theor. Phys.}\ }\textbf
  {\bibinfo {volume} {103}},\ \bibinfo {pages} {380} (\bibinfo {year}
  {2006})},\ \Eprint {http://arxiv.org/abs/hep-ex/0605013}
  {arXiv:hep-ex/0605013} \BibitemShut {NoStop}%
\bibitem [{\citenamefont {Aubert}\ \emph {et~al.}(2009)\citenamefont {Aubert}
  \emph {et~al.}}]{BaBar:2009wpw}%
  \BibitemOpen
  \bibfield  {author} {\bibinfo {author} {\bibfnamefont {B.}~\bibnamefont
  {Aubert}} \emph {et~al.} (\bibinfo {collaboration} {BaBar}),\ }\href
  {\doibase 10.1103/PhysRevLett.103.231801} {\bibfield  {journal} {\bibinfo
  {journal} {Phys. Rev. Lett.}\ }\textbf {\bibinfo {volume} {103}},\ \bibinfo
  {pages} {231801} (\bibinfo {year} {2009})},\ \Eprint
  {http://arxiv.org/abs/0908.3589} {arXiv:0908.3589 [hep-ex]} \BibitemShut
  {NoStop}%
\bibitem [{\citenamefont {Husung}\ \emph {et~al.}(2020)\citenamefont {Husung},
  \citenamefont {Marquard},\ and\ \citenamefont {Sommer}}]{Husung:2019ytz}%
  \BibitemOpen
  \bibfield  {author} {\bibinfo {author} {\bibfnamefont {N.}~\bibnamefont
  {Husung}}, \bibinfo {author} {\bibfnamefont {P.}~\bibnamefont {Marquard}}, \
  and\ \bibinfo {author} {\bibfnamefont {R.}~\bibnamefont {Sommer}},\ }\href
  {\doibase 10.1140/epjc/s10052-020-7685-4} {\bibfield  {journal} {\bibinfo
  {journal} {Eur. Phys. J. C}\ }\textbf {\bibinfo {volume} {80}},\ \bibinfo
  {pages} {200} (\bibinfo {year} {2020})},\ \Eprint
  {http://arxiv.org/abs/1912.08498} {arXiv:1912.08498 [hep-lat]} \BibitemShut
  {NoStop}%
\bibitem [{Hus()}]{Husung:2020}%
  \BibitemOpen
  \href@noop {} {}\bibinfo {howpublished}
  {\url{https://indico.cern.ch/event/956699/contributions/4108682/attachments/2145724/3616679/SymanzikCompressed.pdf}}\BibitemShut
  {NoStop}%
\bibitem [{\citenamefont {Zyla}\ \emph {et~al.}(2020)\citenamefont {Zyla} \emph
  {et~al.}}]{ParticleDataGroup:2020ssz}%
  \BibitemOpen
  \bibfield  {author} {\bibinfo {author} {\bibfnamefont {P.~A.}\ \bibnamefont
  {Zyla}} \emph {et~al.} (\bibinfo {collaboration} {Particle Data Group}),\
  }\href {\doibase 10.1093/ptep/ptaa104} {\bibfield  {journal} {\bibinfo
  {journal} {PTEP}\ }\textbf {\bibinfo {volume} {2020}},\ \bibinfo {pages}
  {083C01} (\bibinfo {year} {2020})}\BibitemShut {NoStop}%
\bibitem [{\citenamefont {Herzog}\ \emph {et~al.}(2017)\citenamefont {Herzog},
  \citenamefont {Ruijl}, \citenamefont {Ueda}, \citenamefont {Vermaseren},\
  and\ \citenamefont {Vogt}}]{Herzog:2017ohr}%
  \BibitemOpen
  \bibfield  {author} {\bibinfo {author} {\bibfnamefont {F.}~\bibnamefont
  {Herzog}}, \bibinfo {author} {\bibfnamefont {B.}~\bibnamefont {Ruijl}},
  \bibinfo {author} {\bibfnamefont {T.}~\bibnamefont {Ueda}}, \bibinfo {author}
  {\bibfnamefont {J.~A.~M.}\ \bibnamefont {Vermaseren}}, \ and\ \bibinfo
  {author} {\bibfnamefont {A.}~\bibnamefont {Vogt}},\ }\href {\doibase
  10.1007/JHEP02(2017)090} {\bibfield  {journal} {\bibinfo  {journal} {JHEP}\
  }\textbf {\bibinfo {volume} {02}},\ \bibinfo {pages} {090} (\bibinfo {year}
  {2017})},\ \Eprint {http://arxiv.org/abs/1701.01404} {arXiv:1701.01404
  [hep-ph]} \BibitemShut {NoStop}%
\bibitem [{\citenamefont {Schroder}\ and\ \citenamefont
  {Steinhauser}(2006)}]{Schroder:2005hy}%
  \BibitemOpen
  \bibfield  {author} {\bibinfo {author} {\bibfnamefont {Y.}~\bibnamefont
  {Schroder}}\ and\ \bibinfo {author} {\bibfnamefont {M.}~\bibnamefont
  {Steinhauser}},\ }\href {\doibase 10.1088/1126-6708/2006/01/051} {\bibfield
  {journal} {\bibinfo  {journal} {JHEP}\ }\textbf {\bibinfo {volume} {01}},\
  \bibinfo {pages} {051} (\bibinfo {year} {2006})},\ \Eprint
  {http://arxiv.org/abs/hep-ph/0512058} {arXiv:hep-ph/0512058} \BibitemShut
  {NoStop}%
\bibitem [{\citenamefont {Dürr}\ \emph {et~al.}(2008)\citenamefont {Dürr}
  \emph {et~al.}}]{Durr:2008zz}%
  \BibitemOpen
  \bibfield  {author} {\bibinfo {author} {\bibfnamefont {S.}~\bibnamefont
  {Dürr}} \emph {et~al.} (\bibinfo {collaboration}
  {Budapest-Marseille-Wuppertal}),\ }\href {\doibase 10.1126/science.1163233}
  {\bibfield  {journal} {\bibinfo  {journal} {Science}\ }\textbf {\bibinfo
  {volume} {322}},\ \bibinfo {pages} {1224} (\bibinfo {year} {2008})},\ \Eprint
  {http://arxiv.org/abs/0906.3599} {arXiv:0906.3599 [hep-lat]} \BibitemShut
  {NoStop}%
\bibitem [{\citenamefont {Borsanyi}\ \emph {et~al.}(2015)\citenamefont
  {Borsanyi} \emph {et~al.}}]{Borsanyi:2014jba}%
  \BibitemOpen
  \bibfield  {author} {\bibinfo {author} {\bibfnamefont {S.}~\bibnamefont
  {Borsanyi}} \emph {et~al.},\ }\href {\doibase 10.1126/science.1257050}
  {\bibfield  {journal} {\bibinfo  {journal} {Science}\ }\textbf {\bibinfo
  {volume} {347}},\ \bibinfo {pages} {1452} (\bibinfo {year} {2015})},\ \Eprint
  {http://arxiv.org/abs/1406.4088} {arXiv:1406.4088 [hep-lat]} \BibitemShut
  {NoStop}%
\bibitem [{\citenamefont {Borsanyi}\ \emph
  {et~al.}(2018{\natexlab{b}})\citenamefont {Borsanyi} \emph
  {et~al.}}]{Budapest-Marseille-Wuppertal:2017okr}%
  \BibitemOpen
  \bibfield  {author} {\bibinfo {author} {\bibfnamefont {S.}~\bibnamefont
  {Borsanyi}} \emph {et~al.} (\bibinfo {collaboration}
  {Budapest-Marseille-Wuppertal}),\ }\href {\doibase
  10.1103/PhysRevLett.121.022002} {\bibfield  {journal} {\bibinfo  {journal}
  {Phys. Rev. Lett.}\ }\textbf {\bibinfo {volume} {121}},\ \bibinfo {pages}
  {022002} (\bibinfo {year} {2018}{\natexlab{b}})},\ \Eprint
  {http://arxiv.org/abs/1711.04980} {arXiv:1711.04980 [hep-lat]} \BibitemShut
  {NoStop}%
\bibitem [{\citenamefont {Lyons}\ \emph {et~al.}(1988)\citenamefont {Lyons},
  \citenamefont {Gibaut},\ and\ \citenamefont {Clifford}}]{Lyons:1988rp}%
  \BibitemOpen
  \bibfield  {author} {\bibinfo {author} {\bibfnamefont {L.}~\bibnamefont
  {Lyons}}, \bibinfo {author} {\bibfnamefont {D.}~\bibnamefont {Gibaut}}, \
  and\ \bibinfo {author} {\bibfnamefont {P.}~\bibnamefont {Clifford}},\ }\href
  {\doibase 10.1016/0168-9002(88)90018-6} {\bibfield  {journal} {\bibinfo
  {journal} {Nucl. Instrum. Meth. A}\ }\textbf {\bibinfo {volume} {270}},\
  \bibinfo {pages} {110} (\bibinfo {year} {1988})}\BibitemShut {NoStop}%
\bibitem [{\citenamefont {Cowan}(1998)}]{Cowan:1998ji}%
  \BibitemOpen
  \bibfield  {author} {\bibinfo {author} {\bibfnamefont {G.}~\bibnamefont
  {Cowan}},\ }\href@noop {} {\emph {\bibinfo {title} {{Statistical data
  analysis}}}}\ (\bibinfo {year} {1998})\BibitemShut {NoStop}%
\bibitem [{\citenamefont {Nakamura}\ \emph {et~al.}(2010)\citenamefont
  {Nakamura} \emph {et~al.}}]{ParticleDataGroup:2010dbb}%
  \BibitemOpen
  \bibfield  {author} {\bibinfo {author} {\bibfnamefont {K.}~\bibnamefont
  {Nakamura}} \emph {et~al.} (\bibinfo {collaboration} {Particle Data Group}),\
  }\href {\doibase 10.1088/0954-3899/37/7A/075021} {\bibfield  {journal}
  {\bibinfo  {journal} {J. Phys. G}\ }\textbf {\bibinfo {volume} {37}},\
  \bibinfo {pages} {075021} (\bibinfo {year} {2010})}\BibitemShut {NoStop}%
\bibitem [{\citenamefont {Pascaud}\ and\ \citenamefont
  {Zomer}(1995)}]{Pascaud:1995qs}%
  \BibitemOpen
  \bibfield  {author} {\bibinfo {author} {\bibfnamefont {C.}~\bibnamefont
  {Pascaud}}\ and\ \bibinfo {author} {\bibfnamefont {F.}~\bibnamefont
  {Zomer}},\ }\href@noop {} {\bibfield  {journal} {\bibinfo  {journal}
  {LAL-95-05}\ } (\bibinfo {year} {1995})}\BibitemShut {NoStop}%
\bibitem [{\citenamefont {Lyons}\ \emph {et~al.}(1990)\citenamefont {Lyons},
  \citenamefont {Martin},\ and\ \citenamefont {Saxon}}]{Lyons:1989gh}%
  \BibitemOpen
  \bibfield  {author} {\bibinfo {author} {\bibfnamefont {L.}~\bibnamefont
  {Lyons}}, \bibinfo {author} {\bibfnamefont {A.~J.}\ \bibnamefont {Martin}}, \
  and\ \bibinfo {author} {\bibfnamefont {D.~H.}\ \bibnamefont {Saxon}},\ }\href
  {\doibase 10.1103/PhysRevD.41.982} {\bibfield  {journal} {\bibinfo  {journal}
  {Phys. Rev. D}\ }\textbf {\bibinfo {volume} {41}},\ \bibinfo {pages} {982}
  (\bibinfo {year} {1990})}\BibitemShut {NoStop}%
\bibitem [{\citenamefont {Demortier}(1999)}]{Demortier:1999}%
  \BibitemOpen
  \bibfield  {author} {\bibinfo {author} {\bibfnamefont {L.}~\bibnamefont
  {Demortier}},\ }\href
  {https://www-cdf.fnal.gov/physics/statistics/notes/cdf8661_chi2fit_w_corr_syst.pdf}
  {\emph {\bibinfo {title} {{Equivalence of the best-fit and covariance-matrix
  methods for comparing binned data with a model in the presence of correlated
  systematic uncertainties}}}},\ \bibinfo {type} {CDF Note}\ \bibinfo {number}
  {8661}\ (\bibinfo {year} {1999})\BibitemShut {NoStop}%
\bibitem [{\citenamefont {Stump}\ \emph {et~al.}(2001)\citenamefont {Stump},
  \citenamefont {Pumplin}, \citenamefont {Brock}, \citenamefont {Casey},
  \citenamefont {Huston}, \citenamefont {Kalk}, \citenamefont {Lai},\ and\
  \citenamefont {Tung}}]{Stump:2001gu}%
  \BibitemOpen
  \bibfield  {author} {\bibinfo {author} {\bibfnamefont {D.}~\bibnamefont
  {Stump}}, \bibinfo {author} {\bibfnamefont {J.}~\bibnamefont {Pumplin}},
  \bibinfo {author} {\bibfnamefont {R.}~\bibnamefont {Brock}}, \bibinfo
  {author} {\bibfnamefont {D.}~\bibnamefont {Casey}}, \bibinfo {author}
  {\bibfnamefont {J.}~\bibnamefont {Huston}}, \bibinfo {author} {\bibfnamefont
  {J.}~\bibnamefont {Kalk}}, \bibinfo {author} {\bibfnamefont {H.~L.}\
  \bibnamefont {Lai}}, \ and\ \bibinfo {author} {\bibfnamefont {W.~K.}\
  \bibnamefont {Tung}},\ }\href {\doibase 10.1103/PhysRevD.65.014012}
  {\bibfield  {journal} {\bibinfo  {journal} {Phys. Rev. D}\ }\textbf {\bibinfo
  {volume} {65}},\ \bibinfo {pages} {014012} (\bibinfo {year} {2001})},\
  \Eprint {http://arxiv.org/abs/hep-ph/0101051} {arXiv:hep-ph/0101051}
  \BibitemShut {NoStop}%
\bibitem [{\citenamefont {Botje}(2002)}]{Botje:2001fx}%
  \BibitemOpen
  \bibfield  {author} {\bibinfo {author} {\bibfnamefont {M.}~\bibnamefont
  {Botje}},\ }\href {\doibase 10.1088/0954-3899/28/5/305} {\bibfield  {journal}
  {\bibinfo  {journal} {J. Phys. G}\ }\textbf {\bibinfo {volume} {28}},\
  \bibinfo {pages} {779} (\bibinfo {year} {2002})},\ \Eprint
  {http://arxiv.org/abs/hep-ph/0110123} {arXiv:hep-ph/0110123} \BibitemShut
  {NoStop}%
\bibitem [{\citenamefont {Thorne}(2002)}]{Thorne:2002kn}%
  \BibitemOpen
  \bibfield  {author} {\bibinfo {author} {\bibfnamefont {R.~S.}\ \bibnamefont
  {Thorne}},\ }\href {\doibase 10.1088/0954-3899/28/10/314} {\bibfield
  {journal} {\bibinfo  {journal} {J. Phys. G}\ }\textbf {\bibinfo {volume}
  {28}},\ \bibinfo {pages} {2705} (\bibinfo {year} {2002})},\ \Eprint
  {http://arxiv.org/abs/hep-ph/0205235} {arXiv:hep-ph/0205235} \BibitemShut
  {NoStop}%
\bibitem [{\citenamefont {Fogli}\ \emph {et~al.}(2002)\citenamefont {Fogli},
  \citenamefont {Lisi}, \citenamefont {Marrone}, \citenamefont {Montanino},\
  and\ \citenamefont {Palazzo}}]{Fogli:2002pt}%
  \BibitemOpen
  \bibfield  {author} {\bibinfo {author} {\bibfnamefont {G.~L.}\ \bibnamefont
  {Fogli}}, \bibinfo {author} {\bibfnamefont {E.}~\bibnamefont {Lisi}},
  \bibinfo {author} {\bibfnamefont {A.}~\bibnamefont {Marrone}}, \bibinfo
  {author} {\bibfnamefont {D.}~\bibnamefont {Montanino}}, \ and\ \bibinfo
  {author} {\bibfnamefont {A.}~\bibnamefont {Palazzo}},\ }\href {\doibase
  10.1103/PhysRevD.66.053010} {\bibfield  {journal} {\bibinfo  {journal} {Phys.
  Rev. D}\ }\textbf {\bibinfo {volume} {66}},\ \bibinfo {pages} {053010}
  (\bibinfo {year} {2002})},\ \Eprint {http://arxiv.org/abs/hep-ph/0206162}
  {arXiv:hep-ph/0206162} \BibitemShut {NoStop}%
\bibitem [{\citenamefont {List}(2010)}]{ListTalk}%
  \BibitemOpen
  \bibfield  {author} {\bibinfo {author} {\bibfnamefont {B.}~\bibnamefont
  {List}},\ }\href@noop {} {\enquote {\bibinfo {title} {Decomposition of a
  covariance matrix into uncorrelated and correlated errors},}\ }\bibinfo
  {howpublished} {\url{https://indico.desy.de/event/3009/contributions/64704/}}
  (\bibinfo {year} {2010}),\ \bibinfo {note} {alliance Workshop on Unfolding
  and Data Correction, DESY}\BibitemShut {NoStop}%
\bibitem [{\citenamefont {DeGrand}(2022)}]{DeGrand:2022lmc}%
  \BibitemOpen
  \bibfield  {author} {\bibinfo {author} {\bibfnamefont {T.}~\bibnamefont
  {DeGrand}},\ }\href {\doibase 10.1103/PhysRevD.106.014504} {\bibfield
  {journal} {\bibinfo  {journal} {Phys. Rev. D}\ }\textbf {\bibinfo {volume}
  {106}},\ \bibinfo {pages} {014504} (\bibinfo {year} {2022})},\ \Eprint
  {http://arxiv.org/abs/2203.04393} {arXiv:2203.04393 [hep-lat]} \BibitemShut
  {NoStop}%
\bibitem [{\citenamefont {Aad}\ \emph {et~al.}(2015)\citenamefont {Aad} \emph
  {et~al.}}]{ATLAS:2014hvo}%
  \BibitemOpen
  \bibfield  {author} {\bibinfo {author} {\bibfnamefont {G.}~\bibnamefont
  {Aad}} \emph {et~al.} (\bibinfo {collaboration} {ATLAS}),\ }\href {\doibase
  10.1140/epjc/s10052-014-3190-y} {\bibfield  {journal} {\bibinfo  {journal}
  {Eur. Phys. J. C}\ }\textbf {\bibinfo {volume} {75}},\ \bibinfo {pages} {17}
  (\bibinfo {year} {2015})},\ \Eprint {http://arxiv.org/abs/1406.0076}
  {arXiv:1406.0076 [hep-ex]} \BibitemShut {NoStop}%
\bibitem [{\citenamefont {Aaboud}\ \emph {et~al.}(2017)\citenamefont {Aaboud}
  \emph {et~al.}}]{ATLAS:2017kux}%
  \BibitemOpen
  \bibfield  {author} {\bibinfo {author} {\bibfnamefont {M.}~\bibnamefont
  {Aaboud}} \emph {et~al.} (\bibinfo {collaboration} {ATLAS}),\ }\href
  {\doibase 10.1007/JHEP09(2017)020} {\bibfield  {journal} {\bibinfo  {journal}
  {JHEP}\ }\textbf {\bibinfo {volume} {09}},\ \bibinfo {pages} {020} (\bibinfo
  {year} {2017})},\ \Eprint {http://arxiv.org/abs/1706.03192} {arXiv:1706.03192
  [hep-ex]} \BibitemShut {NoStop}%
\bibitem [{\citenamefont {Aaboud}\ \emph {et~al.}(2018)\citenamefont {Aaboud}
  \emph {et~al.}}]{ATLAS:2017ble}%
  \BibitemOpen
  \bibfield  {author} {\bibinfo {author} {\bibfnamefont {M.}~\bibnamefont
  {Aaboud}} \emph {et~al.} (\bibinfo {collaboration} {ATLAS}),\ }\href
  {\doibase 10.1007/JHEP05(2018)195} {\bibfield  {journal} {\bibinfo  {journal}
  {JHEP}\ }\textbf {\bibinfo {volume} {05}},\ \bibinfo {pages} {195} (\bibinfo
  {year} {2018})},\ \Eprint {http://arxiv.org/abs/1711.02692} {arXiv:1711.02692
  [hep-ex]} \BibitemShut {NoStop}%
\bibitem [{\citenamefont {Malaescu}\ and\ \citenamefont
  {Starovoitov}(2012)}]{Malaescu:2012ts}%
  \BibitemOpen
  \bibfield  {author} {\bibinfo {author} {\bibfnamefont {B.}~\bibnamefont
  {Malaescu}}\ and\ \bibinfo {author} {\bibfnamefont {P.}~\bibnamefont
  {Starovoitov}},\ }\href {\doibase 10.1140/epjc/s10052-012-2041-y} {\bibfield
  {journal} {\bibinfo  {journal} {Eur. Phys. J. C}\ }\textbf {\bibinfo {volume}
  {72}},\ \bibinfo {pages} {2041} (\bibinfo {year} {2012})},\ \Eprint
  {http://arxiv.org/abs/1203.5416} {arXiv:1203.5416 [hep-ph]} \BibitemShut
  {NoStop}%
\bibitem [{\citenamefont {Malaescu}(2018)}]{bogdan-Mainz-2018-DHMZ-UncOnUnc}%
  \BibitemOpen
  \bibfield  {author} {\bibinfo {author} {\bibfnamefont {B.}~\bibnamefont
  {Malaescu}},\ }\href@noop {} {\enquote {\bibinfo {title} {{Treatment of
  uncertainties and correlations in combinations of $e^+e^-$ annihilation
  data}},}\ }\bibinfo {howpublished}
  {\url{https://indico.him.uni-mainz.de/event/11/session/1/contribution/42/material/slides/0.pdf}}
  (\bibinfo {year} {2018}),\ \bibinfo {note} {{Muon $g-2$ Theory Initiative
  workshop Mainz}}\BibitemShut {NoStop}%
\bibitem [{\citenamefont {Cowan}(2022)}]{Cowan:2021sdy}%
  \BibitemOpen
  \bibfield  {author} {\bibinfo {author} {\bibfnamefont {G.}~\bibnamefont
  {Cowan}},\ }\href {\doibase 10.1051/epjconf/202225809002} {\bibfield
  {journal} {\bibinfo  {journal} {EPJ Web Conf.}\ }\textbf {\bibinfo {volume}
  {258}},\ \bibinfo {pages} {09002} (\bibinfo {year} {2022})},\ \Eprint
  {http://arxiv.org/abs/2107.02652} {arXiv:2107.02652 [hep-ph]} \BibitemShut
  {NoStop}%
\bibitem [{\citenamefont {Davier}\ \emph {et~al.}(2023)\citenamefont {Davier},
  \citenamefont {D\'\i{}az-Calder\'on}, \citenamefont {Malaescu}, \citenamefont
  {Pich}, \citenamefont {Rodr\'\i{}guez-S\'anchez},\ and\ \citenamefont
  {Zhang}}]{Davier:2023hhn}%
  \BibitemOpen
  \bibfield  {author} {\bibinfo {author} {\bibfnamefont {M.}~\bibnamefont
  {Davier}}, \bibinfo {author} {\bibfnamefont {D.}~\bibnamefont
  {D\'\i{}az-Calder\'on}}, \bibinfo {author} {\bibfnamefont {B.}~\bibnamefont
  {Malaescu}}, \bibinfo {author} {\bibfnamefont {A.}~\bibnamefont {Pich}},
  \bibinfo {author} {\bibfnamefont {A.}~\bibnamefont
  {Rodr\'\i{}guez-S\'anchez}}, \ and\ \bibinfo {author} {\bibfnamefont
  {Z.}~\bibnamefont {Zhang}},\ }\href {\doibase 10.1007/JHEP04(2023)067}
  {\bibfield  {journal} {\bibinfo  {journal} {JHEP}\ }\textbf {\bibinfo
  {volume} {04}},\ \bibinfo {pages} {067} (\bibinfo {year} {2023})},\ \Eprint
  {http://arxiv.org/abs/2302.01359} {arXiv:2302.01359 [hep-ph]} \BibitemShut
  {NoStop}%
\bibitem [{\citenamefont {Britzger}(2022)}]{Britzger:2021ocj}%
  \BibitemOpen
  \bibfield  {author} {\bibinfo {author} {\bibfnamefont {D.}~\bibnamefont
  {Britzger}},\ }\href {\doibase 10.1140/epjc/s10052-022-10581-w} {\bibfield
  {journal} {\bibinfo  {journal} {Eur. Phys. J. C}\ }\textbf {\bibinfo {volume}
  {82}},\ \bibinfo {pages} {731} (\bibinfo {year} {2022})},\ \Eprint
  {http://arxiv.org/abs/2112.01548} {arXiv:2112.01548 [physics.data-an]}
  \BibitemShut {NoStop}%
\bibitem [{\citenamefont {Cranmer}\ and\ \citenamefont
  {Held}(2021)}]{Cranmer:2021oxr}%
  \BibitemOpen
  \bibfield  {author} {\bibinfo {author} {\bibfnamefont {K.}~\bibnamefont
  {Cranmer}}\ and\ \bibinfo {author} {\bibfnamefont {A.}~\bibnamefont {Held}},\
  }\href {\doibase 10.1051/epjconf/202125103067} {\bibfield  {journal}
  {\bibinfo  {journal} {EPJ Web Conf.}\ }\textbf {\bibinfo {volume} {251}},\
  \bibinfo {pages} {03067} (\bibinfo {year} {2021})}\BibitemShut {NoStop}%
\bibitem [{\citenamefont {Aad}\ \emph {et~al.}(2023)\citenamefont {Aad} \emph
  {et~al.}}]{ATLAS:2022jbw}%
  \BibitemOpen
  \bibfield  {author} {\bibinfo {author} {\bibfnamefont {G.}~\bibnamefont
  {Aad}} \emph {et~al.} (\bibinfo {collaboration} {ATLAS}),\ }\href {\doibase
  10.1007/JHEP06(2023)019} {\bibfield  {journal} {\bibinfo  {journal} {JHEP}\
  }\textbf {\bibinfo {volume} {06}},\ \bibinfo {pages} {019} (\bibinfo {year}
  {2023})},\ \Eprint {http://arxiv.org/abs/2209.00583} {arXiv:2209.00583
  [hep-ex]} \BibitemShut {NoStop}%
\bibitem [{\citenamefont {Hanhart}\ \emph {et~al.}(2017)\citenamefont
  {Hanhart}, \citenamefont {Holz}, \citenamefont {Kubis}, \citenamefont
  {Kup\'s\'c}, \citenamefont {Wirzba},\ and\ \citenamefont
  {Xiao}}]{Hanhart:2016pcd}%
  \BibitemOpen
  \bibfield  {author} {\bibinfo {author} {\bibfnamefont {C.}~\bibnamefont
  {Hanhart}}, \bibinfo {author} {\bibfnamefont {S.}~\bibnamefont {Holz}},
  \bibinfo {author} {\bibfnamefont {B.}~\bibnamefont {Kubis}}, \bibinfo
  {author} {\bibfnamefont {A.}~\bibnamefont {Kup\'s\'c}}, \bibinfo {author}
  {\bibfnamefont {A.}~\bibnamefont {Wirzba}}, \ and\ \bibinfo {author}
  {\bibfnamefont {C.~W.}\ \bibnamefont {Xiao}},\ }\href {\doibase
  10.1140/epjc/s10052-017-4651-x} {\bibfield  {journal} {\bibinfo  {journal}
  {Eur. Phys. J. C}\ }\textbf {\bibinfo {volume} {77}},\ \bibinfo {pages} {98}
  (\bibinfo {year} {2017})},\ \bibinfo {note} {[Erratum: Eur.Phys.J.C 78, 450
  (2018)]},\ \Eprint {http://arxiv.org/abs/1611.09359} {arXiv:1611.09359
  [hep-ph]} \BibitemShut {NoStop}%
\bibitem [{\citenamefont {De~Troconiz}\ and\ \citenamefont
  {Yndurain}(2002)}]{DeTroconiz:2001rip}%
  \BibitemOpen
  \bibfield  {author} {\bibinfo {author} {\bibfnamefont {J.~F.}\ \bibnamefont
  {De~Troconiz}}\ and\ \bibinfo {author} {\bibfnamefont {F.~J.}\ \bibnamefont
  {Yndurain}},\ }\href {\doibase 10.1103/PhysRevD.65.093001} {\bibfield
  {journal} {\bibinfo  {journal} {Phys. Rev. D}\ }\textbf {\bibinfo {volume}
  {65}},\ \bibinfo {pages} {093001} (\bibinfo {year} {2002})},\ \Eprint
  {http://arxiv.org/abs/hep-ph/0106025} {arXiv:hep-ph/0106025} \BibitemShut
  {NoStop}%
\bibitem [{\citenamefont {de~Troconiz}\ and\ \citenamefont
  {Yndurain}(2005)}]{deTroconiz:2004yzs}%
  \BibitemOpen
  \bibfield  {author} {\bibinfo {author} {\bibfnamefont {J.~F.}\ \bibnamefont
  {de~Troconiz}}\ and\ \bibinfo {author} {\bibfnamefont {F.~J.}\ \bibnamefont
  {Yndurain}},\ }\href {\doibase 10.1103/PhysRevD.71.073008} {\bibfield
  {journal} {\bibinfo  {journal} {Phys. Rev. D}\ }\textbf {\bibinfo {volume}
  {71}},\ \bibinfo {pages} {073008} (\bibinfo {year} {2005})},\ \Eprint
  {http://arxiv.org/abs/hep-ph/0402285} {arXiv:hep-ph/0402285} \BibitemShut
  {NoStop}%
\bibitem [{\citenamefont {Garcia-Martin}\ \emph {et~al.}(2011)\citenamefont
  {Garcia-Martin}, \citenamefont {Kaminski}, \citenamefont {Pelaez},
  \citenamefont {Ruiz~de Elvira},\ and\ \citenamefont
  {Yndurain}}]{Garcia-Martin:2011iqs}%
  \BibitemOpen
  \bibfield  {author} {\bibinfo {author} {\bibfnamefont {R.}~\bibnamefont
  {Garcia-Martin}}, \bibinfo {author} {\bibfnamefont {R.}~\bibnamefont
  {Kaminski}}, \bibinfo {author} {\bibfnamefont {J.~R.}\ \bibnamefont
  {Pelaez}}, \bibinfo {author} {\bibfnamefont {J.}~\bibnamefont {Ruiz~de
  Elvira}}, \ and\ \bibinfo {author} {\bibfnamefont {F.~J.}\ \bibnamefont
  {Yndurain}},\ }\href {\doibase 10.1103/PhysRevD.83.074004} {\bibfield
  {journal} {\bibinfo  {journal} {Phys. Rev. D}\ }\textbf {\bibinfo {volume}
  {83}},\ \bibinfo {pages} {074004} (\bibinfo {year} {2011})},\ \Eprint
  {http://arxiv.org/abs/1102.2183} {arXiv:1102.2183 [hep-ph]} \BibitemShut
  {NoStop}%
\bibitem [{\citenamefont {Ananthanarayan}\ \emph {et~al.}(2018)\citenamefont
  {Ananthanarayan}, \citenamefont {Caprini},\ and\ \citenamefont
  {Das}}]{Ananthanarayan:2018nyx}%
  \BibitemOpen
  \bibfield  {author} {\bibinfo {author} {\bibfnamefont {B.}~\bibnamefont
  {Ananthanarayan}}, \bibinfo {author} {\bibfnamefont {I.}~\bibnamefont
  {Caprini}}, \ and\ \bibinfo {author} {\bibfnamefont {D.}~\bibnamefont
  {Das}},\ }\href {\doibase 10.1103/PhysRevD.98.114015} {\bibfield  {journal}
  {\bibinfo  {journal} {Phys. Rev. D}\ }\textbf {\bibinfo {volume} {98}},\
  \bibinfo {pages} {114015} (\bibinfo {year} {2018})},\ \Eprint
  {http://arxiv.org/abs/1810.09265} {arXiv:1810.09265 [hep-ph]} \BibitemShut
  {NoStop}%
\end{thebibliography}%

\end{document}